\documentclass[a4paper,11pt]{article}
\usepackage{amsmath}
\numberwithin{equation}{section}
\numberwithin{equation}{subsection}

\usepackage{amssymb}
\usepackage{amsfonts}
\usepackage{color}
\usepackage{graphicx}
\usepackage{wrapfig}
\usepackage{slashed}

\def\circlet{\rlap{\raise0.20truecm\hbox{\small$\circ$}}}
\def\à{\`a}
\def\ò{\`o}
\def\ì{\`\i}
\def\ù{\`u}
\def\à{\`a}
\def\è{\`e}
\def\é{\'e}
\def\È{\`E}

\begin{document}

\title{{THE CONFORMAL UNIVERSE II: \\
Conformal Symmetry, its Spontaneous Breakdown\\
and Higgs Fields in Conformally Flat Spacetime}}

\author{{\large \em FINAL VERSION}\\
     \\
     Renato Nobili\\
    {\normalsize    E-mail: renato.nobili@unipd.it}\\}
\date{Padova, 26 March 2016}
\maketitle
\pagestyle{myheadings} \markright{R.Nobili, The Conformal Universe II}

\begin{abstract}
\noindent
This is the second of three papers on Conformal General Relativity (CGR). The conformal group is introduced here
as the invariance group of the partial order of causal events in $n$D spacetime. Its general structure, discrete
symmetries and field representations are described in detail. The spontaneous breakdown of conformal symmetry is
then discussed and the role played by a ghost scalar field and a physical scalar field in 4D spacetime 
are evidenced. Kinematic--, conformal-- and proper--time hyperbolic coordinates are introduced in a negatively curved Milne
spacetime for the purpose of providing three different but equivalent representations of CGR. The first of these is
grounded in a Riemannian manifold and is manifestly conformal invariant, the second is grounded in a conformally connected
Cartan manifold but its conformal invariance is hidden, the third is grounded in the Riemannian manifold of the Milne spacetime   
and has the formal structure of General Relativity (GR). The relation between CGR and standard inflationary
cosmology is also briefly discussed. Lastly, in view of the detailed study of Higgs--field dynamics carried out in
the third paper, the action integrals, motion equations and total energy--momentum tensors of the Higgs field
interacting with the dilation field are described in the three representations mentioned above.
\end{abstract}

{\bf keywords}: {\em causality, conformal symmetry, orthochronous inversion, dilation, elation}

\tableofcontents

\section{The conformal group and its representations}
As delineated in \S\,2 of Part I \cite{PART1}, both General Relativity (GR) and Conformal General Relativity
(CGR) describe the universe as grounded in a differentiable manifold enveloped in a continuum of local
tangent spaces, all of which support isomorphic representations of a finite continuous group, called the
{\em fundamental group}. The fundamental group of GR is the Poincar\'e group and that of CGR is its
conformal extension.

The importance of the finite group of conformal transformations is due to the fact that it is the largest
invariance--group of the partial order of causal events in flat spacetime. We can infer from this that
the infinite group of conformal diffeomorphisms is the widest invariance
group of the partial ordering of causal events in a differentiable spacetime manifold.

In the next three subsections, the structure and most important representations of the conformal
group in $n$D spacetime are described in detail. In the further subsections, only the conformal group
in 4D spacetime will be considered.

\subsection{The conformal group as the invariance group of causality}
\label{confgroupas}
The relationship between causality and group theoretical properties of spacetime geometry was investigated by
Alexandrov in 1953 \cite{ALEXANDROV1} \cite{ALEXANDROV2}) and Zeeman in 1964. I briefly summarize the
approach of the latter.

Let us define an {\em $n$D Minkowski spacetime ${\cal M}_n$} as the Cartesian product
of an ($n$-1)--dimensional Euclidean space ${\cal R}^{n-1}$ by a real time axis $T$. The former is
intended to represent the set of possible inertial observers at rest in ${\cal R}^{n-1}$ and equipped
with perfectly synchronized clocks, and the latter is intended to represent common time $x^0 \in T$
marked by the clocks. All observers are allowed to communicate with one another by signals of limited
speed, the upper limit of which is conventionally assumed not to exceed 1 (the speed of light in ${\cal
R}^{n-1}$). Let us indicate by $x^1, x^2,..., x^{n-1}$ the coordinates of ${\cal R}^{n-1}$. Hence,
regardless of any metrical considerations, we can write ${\cal M}_n = {\cal R}^{n-1}\times T$, which
means that ${\cal M}_n$ is equivalent to the $n$--dimensional affine space. Clearly, points $x = \{x^0,
x^1, x^2,..., x^{n-1}\}\in {\cal M}_n$ also represent the set of all possible point--like events,
observable in ${\cal R}^{n-1}$ at time $x^0$, partially ordered by the relation $x\leq y$, defined as follows:

\smallskip

\centerline{\em $x$ can influence $y$ if and only if $y^0-x^0\le
\sqrt{(y^1-x^1)^2+\dots+(y^{n-1}-x^{n-1})^2}$.}

\smallskip

\noindent In contrast to causality in Newtonian spacetime, this partial ordering equips ${\cal R}^{n-1}$ with
a natural topology, the basis of which may be formed by the open sets of points $y\in {\cal R}^{n-1}$,
satisfying the inequalities
$$\sqrt{(y^1-x^1)^2+\dots+(y^{n-1}-x^{n-1})^2}< \varepsilon $$ for arbitrarily small $\varepsilon$. The set
of points $y\ge x$ with fixed $x$ and variable $y$ defines the {\em future cone} of $x$; the set of points $y\ge
x$ with fixed $y$ and variable $x$ defines the {\it past cone} of $y$. Equation $$(y-x)^2\equiv (y^0-x^0)^2-
(y^1-x^1)^2-\dots-(y^{n-1}-x^{n-1})^2=0$$ defines the family of (double) light--cones of  ${\cal M}_n$.
The following theorem then holds:

{\it Let $Z$ be a one--to--one mapping of $n$D Minkowski spacetime ${\cal M}_n$ on to itself -- no assumption being
made about whether $Z$ is linear or continuous. If $Z$ preserves the causal order of events and $n> 2$, then
$Z$ maps light--cones onto light--cones and belongs to a group that is the direct product of the $n$D Poincar\'e
group and the dilation group}.

The same result was also obtained by other authors \cite{BORCHERS} \cite{GOEBEL} on the basis of weaker topological
assumptions, and it is conceivable that it may be directly obtained by pure lattice--theoretical methods and suitable
automorphism conditions, regardless of any embedding of causal events in an affine space.

To be specific, the invariance group of causality acts on $x^\mu$, $(\mu = 0,
1, \dots, n-1)$, as follows
\begin{eqnarray}
\label{transl}
T(a):  x^\mu  & \rightarrow & x^{\mu} + a^{\mu} \quad \,\hbox{(translations)\,;} \\
\label{dilat} S(\alpha):  x^\mu  & \rightarrow &  e^{\alpha} x^{\mu} \qquad
\,\,\,\hbox{(dilations)}\,;\\
\label{lorentz} \Lambda(\omega):  x^\mu & \rightarrow &
\Lambda^\mu_\nu(\omega)\,x^{\nu} \quad \hbox{(Lorentz rotations)}\,.
\end{eqnarray}
Here $a^\mu$, $\alpha$ and the tensor $\omega\equiv\omega^{\rho\sigma} =-\omega^{\sigma\rho}$ are respectively
the parameters of translations, dilation and Lorentz rotations. However, the invariance group of causality is
somewhat larger \cite{HAWKING1} \cite{POPOVICI}, since the partial ordering of causal events is also preserved, for instance,
by the following map
$$
I_0: x^\mu \rightarrow - \frac{x^\mu}{x^2}\,,
$$
where $x^2 = x^\mu x_\mu, x_\mu=\eta_{\mu\nu}x^\nu$,  $\eta_{\mu\nu} = \hbox{diag}\{1, -1,\dots, -1\}$, which we
call the {\em orthochronous inversion} with respect to event $x=0\in {\cal M}_n$ \cite{NOBILI}. The equalities
$(I_0)^2 = 1$, $I_0 \Lambda(\omega)I_0 = \Lambda(\omega)$  and $I_0S(\alpha)I_0 = S(-\alpha)$ are manifest.
By translation, we obtain the orthochronous inversion with respect to any point $a\in {\cal M}_n$, which acts on
$x^\mu$ as follows:
$$
I_a: x^\mu \rightarrow - \frac{x^\mu-a^\mu}{(x-a)^2}\,.
$$

Clearly, the causal order is also preserved by the transformations $E(b) = I_0T(b)I_0$, which manifestly form an
Abelian group and act on $x^\mu$ as follows
\begin{equation}
\label{elat}E(b): \,\, x^\mu  \rightarrow \frac{x^\mu - b^\mu x^2}{1-2bx+b^2x^2}\,,
\end{equation}
where $b x$ stands for $b^\mu x_\mu$. These are commonly known as {\em special conformal transformations}, but we call them
{\em elations}, since this is the name coined for them by Cartan in 1922 \cite{CARTAN1}. In conclusion, the topologically
connected component of the complete invariance group of causal order in the $n$D Minkow\-ski spacetime is the
{\em $n$--dimensional conformal group} ${C}(1, n-1)$ formed of transformations (\ref{transl})--(\ref{elat}),
comprehensively depending on $n(n+3)/2+1$ real parameters.

From an invariantive point of view, ${C}(1, n-1)$ may be regarded as the more general continuous group generated
by infinitesimal transformations of the form $ x^\mu \rightarrow x^\mu + \varepsilon u^\mu(x)$,  where $\varepsilon$
is an infinitesimal parameter and $u_\mu(x)$ are such that squared line element
$ds^2 =g_{\mu\nu} dx^\mu dx^\nu$ undergoes the transformation $ds^2 \rightarrow [1+\varepsilon \lambda(x)]\, ds^2$,
where $\lambda(x)$ is an arbitrary real function. This requires $u^\mu(x) = a^\mu + d\,\, x^\mu +
g^{\mu\nu}\omega_{\nu\lambda}\,x^\lambda + 2\,c_\lambda x^\lambda x^\mu - c_\mu x^\lambda x_\lambda$,
where $a^\mu,  b, c_\lambda$ are arbitrary constant and $\omega_{\nu\lambda}$ is antisymmetric  \cite{HAAG92}.

Indicating by $P_\mu$, $M_{\mu\nu}$, $D$ and $K_\mu$ the generators of $T(a), \Lambda(\omega), S(\alpha)$ and $E(b)$,
respectively, we can easily determine their actions on $x^\mu$
\begin{eqnarray}
& & P_\mu x^\nu = -i\delta_\mu^\nu\,,\quad M_{\mu\nu}x^\lambda = i\big(\delta_\nu^\lambda x^\mu-\delta_\mu^\lambda x^\nu\big)
\,,\nonumber \\
& & D x^\mu = -i x^\mu\,, \quad K_\mu x^\nu = i\big(x^2\delta_\mu^\nu - 2x_\mu x^\nu\big) \nonumber\,,
\end{eqnarray}
where $\delta_\mu^\nu$ is the Kronecker delta. Indicating by $\partial_\mu$ the partial derivative with
respect to $x^\mu$, their actions on arbitrary differentiable functions $f$ of $x$ are
\begin{eqnarray}
\label{dergen1} & & P_\mu f(x) = -i\partial_\mu f(x) \,;\quad  M_{\mu\nu} f(x)
= i\big(x_\mu\partial_\nu -x_\nu\partial_\mu\big)f(x)\,; \nonumber\\
\label{dergen2} & & D f(x) = -i x^\mu \partial_\mu f(x) \,;\quad K_\mu f(x) =
i\big(x^2\partial_\mu -2x_\mu x^\nu \partial_\nu\big)f(x) \,; \nonumber
\end{eqnarray}
the Lie algebra of which satisfies the following commutation relations
\begin{eqnarray}
\label{commut1}
& &[P_\mu, P_\nu]  = [K_\mu, K_\nu] =0\,;\quad [P_\mu, K_\nu] = 2i\big(g_{\mu\nu}D+M_{\mu\nu}\big) \,;\\
\label{commut2}
& &[D, P_\mu]  = i P_\mu\,; \quad [D, K_\mu] =-i K_\mu\,; \quad [D, M_{\mu\nu}] =0\,;\\
\label{commut3} & &[M_{\mu\nu}, P_\rho]  = i \big(g_{\nu\rho}P_\mu
-g_{\mu\rho}P_\nu\big) \,; \quad [M_{\mu\nu}, K_\rho]  = i \big(g_{\nu\rho}K_\mu-g_{\mu\rho}K_\nu\big) \,;\\
\label{commut4} & &[M_{\mu\nu}, M_{\rho\sigma}]  = i\big(g_{\mu\sigma}M_{\nu\rho} +
g_{\nu\rho}M_{\mu\sigma}-g_{\mu\rho}M_{\nu\sigma} -g_{\nu\sigma}M_{\mu\rho}\big)\,,
\end{eqnarray}
which form the prototype of the abstract Lie algebra of ${C}(1, n-1)$
\cite{MACKSALAM}.

For the sake of clarity and completeness, we add to Eqs (\ref{commut1})--(\ref{commut4}) the discrete
operations
\begin{equation}
\label{inverse} I_0 P_\mu I_0  = K_\mu\,; \quad\!\! I_0 K_\mu I_0  =
P_\mu\,;\quad\!\! I_0 D I_0 = -D \,; \quad\!\! I_0 M_{\mu\nu} I_0 =
M_{\mu\nu}\,.
\end{equation}

Note that $I_0$ and $P_\mu$ alone suffice to generate ${C}(1, n-1)$. In fact, using Eqs  (\ref{commut1})--(\ref{commut3})
and the first of Eqs (\ref{inverse}), we can define all other group generators as follows:
$$K_\mu = I_0 P_\mu I_0\,,\quad D = \frac{i}{8}g^{\mu\nu}[K_\mu, P_\nu]\,, \quad M_{\mu\nu} =
\frac{i}{2}[K_\nu,P_\mu] - g_{\mu\nu}D\,.$$
This tells us many things about the partial ordering of causal events in ${\cal M}_n$. In fact, we may think of $I_0$
as representing the operator which performs the partial ordering of causally related events as seen by a point--like
observer located at $x = 0$, which receives signals from its own past and sends signals to its own future, of $T(a)$ as the operator
which shifts the observer from $x = 0$ to $x = a$ in ${\cal M}_n$, and of $I_a$ as a continuous set of involutions which
impart a symmetric--space structure to the lattice of causal event.

Lastly note that, provided $n$ is even, we can include parity transformation $P: \{x^0, \vec x\} \rightarrow \{x^0, -\vec x\}$
as a second discrete element of the conformal group. Instead, time--reversal must be excluded, since it does not preserve the
causal order of events. This makes a difference between GR and CGR. In fact, time reversal, so familiar to GR, is replaced by
orthochronous inversion $I_0$ conventionally centered at an arbitrary point $x=0$.

\subsection{Remarkable geometric properties of orthochronous inversions}
\label{OnIa} Orthochronous inversion $I_a$, where $a$ is any point of ${\cal
M}_n$, has the following properties:

1) It leaves invariant the double cone centered at $a$, swapping the interiors
of the future and past cones so as to preserve the time arrow and the
collineation of all points on straight lines through $a$, as shown in Fig.\,1.

\begin{figure}[!ht]
\centering
\mbox{%
\begin{minipage}{.35\textwidth}
\includegraphics[scale=0.6]{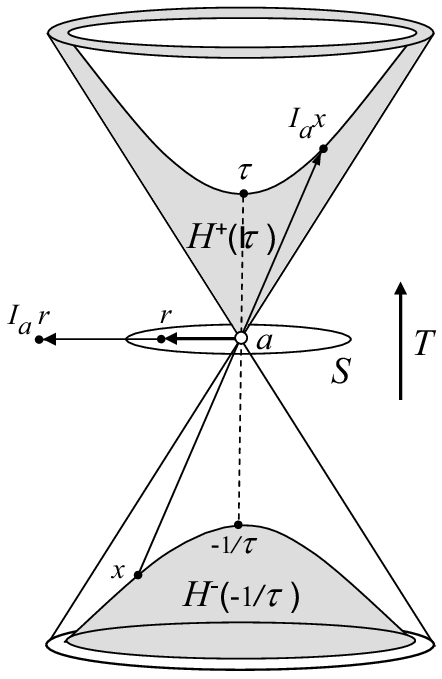}
\end{minipage}%
\quad
\begin{minipage}[c]{.45\textwidth}
\caption{\small The orthochronous inversion centered at a point $a$ of the Minkowski
spacetime interchanges the events lying in the interior of the double cone stemming
from $a$ in such a way that spacetime regions of the past--cone close to $a$ are mapped
onto regions of the future--cone far from $a$ (and vice versa). $T = $ time axis; $S = $
unit $(n-1)$--D sphere centered at $a$ and orthogonal to $T$.}
\end{minipage}%
}
\end{figure}

2) It partitions the events of the past and future cones stemming from $a$ into
a two--fold family of $(n-1)$-D hyperboloids parameterized by kinematic time
$$
\tau =\pm \sqrt{ (x^0-a^0)^2+\dots +(x^{n-1}-a^{n-1})^2}\,.
$$

3) It maps future region $H^+_a(\tau)$, extending from cone--vertex $a$ to the
$(n-1)$--D hyperboloid at $\tau$, into region  $H^-_a(-1/\tau)$ of the past
cone defined by the  $(n-1)$--D  hyperboloid at $-\infty$ and the hyperboloid
at $\tau' = -1/\tau$, and vice versa (Fig.\,1, gray areas). Similarly, it maps
the set--theoretical complements of $H^+_a(\tau), H^-_a(-1/\tau)$ onto each other.

4) It performs the polar inversion of points $r$ internal to the space--like unit $(n-1)$--D sphere $S$ centered
at $a$ and orthogonal to time axis $T$ into points $r' = I_ar$ external to $S$, and vice versa.

5) Functions which are invariant under $I_a$ depend upon kinematic time $\tau$ only. Thus, if they vanish near the
vertex of the past cone, they also vanish at the infinite kinematic time of the future cone, and vice versa. This
property has an important implication in that, if $I_0$ is invariant under the spontaneous breakdown of conformal symmetry,
for very large $\tau$ the universe reaches the same physical conditions as those immediately beforethe symmetry--breaking event.
In other words, the action--integral invariance under $I_0$ is compatible with the assumption that the time course of the
universe may be described as a transition from the state of an unstable initial vacuum to that of a stable final vacuum
with the same physical properties.

\subsection{Conformal transformations of local fields}
\label{fieldtransfs} When a differential operator $g$ is applied to a differentiable function $f$ of $x$,
the function changes as $gf(x) = f(gx)$, which may be interpreted as the form taken by $f$ in the reference
frame of coordinates $x'=gx$. When a second differential operator $g'$ acts on $f(gx)$, we obtain
$g'f(gx) = f(gg'x)$, i.e., we have $g'gf(x) =  f(gg'x)$, showing that $g'$ and $g$ act on the reference
frame in reverse order.

The action of $g$ on a local quantum field $\Psi(x)$ of dimension, or weight, $w_\Psi$, bearing a spin
subscript $\rho$, has the general form  $g\Psi_\rho(x)= {\cal F}_\rho^\sigma(g^{-1}, x) \Psi_\sigma(gx)$,
where ${\cal F}(g^{-1}, x)$ is a matrix obeying the composition law
$$
{\cal F}(g^{-1}_2,x)\,{\cal F}(g^{-1}_1, g_2 x)= {\cal F}(g^{-1}_2g^{-1}_1, x)\,.
$$
These equations are consistent with coordinate transformations, since the product of two
transformations $g_1, g_2$ yields
$$
g_2g_1 \Psi_\rho(x)  = \, {\cal F}^\sigma_\rho[(g_1 g_2)^{-1},
x]\Psi_\sigma(g_1g_2 x)\,,
$$
with $g_2, g_1$ always appearing in reverse order on the right--hand member.

According to these rules, the action of ${C}(1, n-1)$ generators on an
irreducible unitary representation  $\Psi_\rho(x)$ of the Poincar\'e group,
describing a field of dimension  $w_\Psi$ and spin subscript $\rho$,  may be
summarized as follows
\begin{eqnarray}
\label{dergenf1} & & [P_\mu, \Psi_\rho]  = -i \partial_\mu \Psi_\rho  \,; \\
\label{dergenf2} & & [K_\mu, \Psi_\rho] = i \big[x^2
\partial_\mu - 2x_\mu \big(x^\rho \partial_\rho - w_\Psi\big)\big] \Psi_\rho +
i x^\nu\big(\Sigma_{\mu\nu}\big)_\rho^\sigma\Psi_\sigma\,;\\
\label{dergenf3} & & [D, \Psi_\rho] = -i \big(x^\mu\partial_\mu -  w_\Psi\big)
\Psi_\rho \,;\\
\label{dergenf4} & & [M_{\mu\nu}, \Psi_\rho] = i \big(x_\mu \partial_\nu -  x_\nu
\partial_\mu\big) \Psi_\rho - i\big(\Sigma_{\mu\nu}\big)_\rho^\sigma  \Psi_\sigma\,;
\end{eqnarray}
where $ \Sigma_{\mu\nu}$ are the spin matrices, i.e., the generators of Lorenz
rotations on the spin space.  The corresponding set of finite conformal
transformations are
\begin{eqnarray}
\label{conf1} & & T(a): \Psi_\rho(x)\rightarrow \Psi_\rho(x+a)\,;\\
\label{conf2} & & E(b): \Psi_\rho(x)\rightarrow {\cal E}(-b,x)^\sigma_\rho
\Psi_\sigma\bigg(\frac{x - bx^2}{1-2bx+b^2x^2}\bigg)\,;\\
\label{conf3} & & S(\alpha): \Psi_\rho(x)\rightarrow
e^{-w_\Psi\alpha}\,\Psi_\rho(e^{\alpha} x)\,;\\
\label{conf4} & &  \Lambda(\omega): \Psi_\rho(x)\rightarrow {\cal
L}^\sigma_\rho(-\omega) \Psi_\sigma[\Lambda(\omega)\, x]\,;
\end{eqnarray}
where  ${\cal E}(-b,x), {\cal L}(-\omega)$ are suitable matrices which perform
the conformal transformations of spin components, respectively for elations and
Lorentz rotations.

As regards the orthochronous inversion, we generally have
\begin{equation}
\label{conf5}  I_0: \Psi_\rho(x)\rightarrow {\cal I}_0(x)^\sigma_\rho
\Psi_\sigma (-x/x^2)\,,
\end{equation}
where matrix ${\cal I}_0(x)$ obeys the equation
\begin{equation}
\label{IoIo}  {\cal I}_0(x){\cal I}_0(-x/x^2)=1\,.
\end{equation}
For consistency with (\ref{conf1}), (\ref{conf2}) and Eqs  $E(b) =
I_0(x)\,T(b)\,I_0(x)$, we also have
\begin{equation}
\label{ITI} {\cal E}^\sigma_\rho(-b,x)= {\cal I}_0(x)\,{\cal I}_0(x-b)\,.
\end{equation}

For the needs of a Langrangian theory, the adjoint representation of $\Psi_\alpha$ must also be defined. It is indicated
by $\bar\Psi= \Psi^{\dag} {\cal B}$, where ${\cal B}$ is a suitable matrix, or complex number, chosen so as to satisfy
equation $\bar{\bar\Psi}=\Psi$ -- which implies ${\cal B}\,{\cal B}^\dag=1$ -- and the self--adjointness condition of
the Hamiltonian. Therefore, under the action of a group element $g$, the adjoint representation $\bar \Psi^\rho(x)$
is subject to the transformation
$$
g: \bar \Psi^\rho(x)\rightarrow \bar\Psi^\sigma(gx)\,\bar{\cal F}(g^{-1},x)^\rho_\sigma\,,
$$
where $\bar{\cal F}(g^{-1}, x)= {\cal B}^\dag {\cal F}^\dag(g^{-1}, x){\cal B}$.

\subsection{Discrete symmetries of the conformal group in 4D}
In this and the next two subsections we focus on the transformation properties of a spinor field $\psi(x)$, since
those of all other fields can be obtained by reducing direct products of spinor field representations.

As is well--known in standard field theory, the algebra of spinor representation contains the discrete group formed
by {\em parity operator} $P$, {\em charge conjugation} $C$ and {\em time reversal} $T$. The last
commutes with $P$, and $C$, and {\em elicity projectors} $P_\pm$, with $P_++P_-=1$, which are defined by equations
$\psi_R =P_+\psi$ and $\psi_L=P_-\psi$, where $R$ and $L$ stand respectively for the right--handed and the
left--handed spinor elicities. However, as already pointed out at the end of \S\,\ref{confgroupas}, passing from
the Poincar\'e to the conformal group, we must exclude $T$, which violates the causal order, and transfer the role of
this operator to $I_0$.

Let us normalize Dirac matrices $\gamma^\mu$ in such a way that $\gamma^0 = (\gamma^0)^\dag =\widetilde{\gamma^0}$
and $\gamma^2 = (\gamma^2)^\dag=\widetilde{\gamma^2}$, where the tilde superscript indicates matrix transposition,
and $\gamma^5= i\gamma^0\gamma^1\gamma^2\gamma^3$. Then, the equalities $\psi_L =\frac{1}{2}(1+\gamma^5)\,\psi$
and $\psi_R =\frac{1}{2}(1-\gamma^5)\,\psi$ hold. As is well--known in basic Quantum Mechanics, in this
representation $P$ and $C$ act on $\psi_{L,R}$ as follows:
\begin{eqnarray}
\label{PCRops} & & P: \psi_{R,L}(x) \rightarrow {\cal P}\psi_{L,R}(Px)\,;\quad
C: \psi_{R,L}(x) \rightarrow {\cal C}\psi_{L,R}^{*}(x)\,;\nonumber\\ & & CP:
\psi_{R,L}(x) \rightarrow {\cal CP} \psi^*_{R,L}(Px)\,;
\end{eqnarray}
where $Px\equiv P\{ x^0, \vec x\} = \{ x^0, -\vec x\}$, ${\cal P}=\gamma^0$,
${\cal C}=i\gamma^2\gamma^0$, ${\cal CP}=i\gamma^2$ and $\psi_{R,L}^*$ are the
complex conjugates of $\psi_{R,L}$. We can easily verify the equations
$$
{\cal P}{\cal C}=-{\cal C}{\cal P}\,,\quad {\cal
C}^{-1}= {\cal C}^\dag =-{\cal C}\,,\quad {\cal C}\gamma^5 = -\gamma^5{\cal C}
\,,\quad {\cal P}\gamma^5 =-\gamma^5{\cal P}\,,
$$
the last two of which show that both $\cal P$ and $\cal C$ interchange
$L$ with $R$ (whereas $\cal PC$ leaves them unchanged).

Note that, despite their anticommutativity, $\cal P$ and $\cal C$  act commutatively on fermion
bilinears, of which all spinor observables are made.

The general form of ${\cal I}_0(x)$ introduced in Eq (\ref{conf5}) is determined by the requirement
that $I_0$ commutes with $P, C, P_\pm$, like time reversal $T$, as the heuristic principle of preservation of
formal laws suggests, and that it satisfies Eq (\ref{IoIo}). In summary,
\begin{eqnarray}
\label{[I0,P, C]} & & \gamma^0{\cal I}_0(Px)= {\cal I}_0(x)\,\gamma^0\,;\quad
\gamma^2\gamma^0{\cal I}_0(x)= {\cal
I}_0(x)\,\gamma^2\gamma^0\,;\nonumber\\
\label{[I0gamma5, IoIo]} & & \gamma^5{\cal I}_0(x)= {\cal
I}_0(x)\,\gamma^5\,;\quad {\cal I}_0(x^\mu)\,{\cal
I}_0(-x^\mu/x^2)=1\,.\nonumber
\end{eqnarray}
Of course,  we also require that $I_0$ is not equivalent to 1 and $CP$. We can
easily verify that all these conditions lead to the general formula
\begin{equation}
\label{I0delta} {\cal I}_0(x)=\pm (x^2)^\delta \frac{\slashed x}{\vert x
\vert}\,\gamma^2\gamma^0=\pm (x^2)^{\delta-1/2}\gamma^2\gamma^0  {\slashed
x}\,,
\end{equation}
where $\delta$ is a real number,  $\slashed{x}= \eta_{\mu\nu}\gamma^\mu x^\nu$, with
$$
\eta_{\mu\nu}=\mbox{diag}(1,-1,-1,-1)\,,
$$
and $\vert x\vert = \sqrt{x^2}=\sqrt{\slashed{x}^2}$. Since the sign on the left of Eq
(\ref{I0delta}) is arbitrary, we make it equal to $-1$ for the sake of convenience.

Note that ${\slashed x}$ is a pseudo--scalar, since ${\cal P}{\slashed x}\, {\cal P}^{-1}=
\gamma^0\gamma_\mu\gamma^0 (Px)^\mu = - {\slashed x}^\dag$.

To find $\delta$, we impose the condition that the conformal invariant free--field Feynman propagator
\begin{equation}
\langle 0\vert T\big\{\psi_\alpha(x_1), \overline
\psi_\beta(x_2)\big\}\vert 0\rangle= \frac{i}{2\pi^2}\,\frac{\slashed{x}_1-
\slashed{x}_2}{[(x_1-x_2)^2+i\epsilon]^2} \,,\nonumber
\end{equation} where $T\big\{\cdots \big\}$ indicates time ordering, is invariant under $I_0$, i.e.,
\begin{equation}
\langle 0\vert T\big\{I_0 \psi_\alpha(x_1)I_0^{-1}, \overline {I_0\psi_\beta(x_2)I_0^{-1}}\,\big\}\vert 0\rangle = \langle 0\vert
T\big\{\psi_\alpha(x_1), \overline \psi_\beta(x_2)\big\}\vert 0\rangle\,. \nonumber
\end{equation}

Using Eq (\ref{conf5}) we find the adjoint transformation
$$
\bar\psi(x)= \psi^\dag(x)\gamma^0 \xrightarrow{\, t_0 \,} -(x^2)^{\delta-1/2} \psi^\dag(-x/x^2)
\,\slashed{x}^\dag\gamma^0\gamma^2\gamma^0 =(x^2)^{\delta-1/2} \bar\psi^\dag(-x/x^2)
\,\slashed{x}\gamma^2\gamma^0,
$$
where equalities $\gamma^2\gamma^0= -\gamma^0\gamma^2$ and $\gamma^0\slashed{x}^\dag \gamma^0 = \slashed{x}$
were used. Thus, the invariant condition takes the form
$$
(x_1^2 x_2^2)^{\delta - 1/2}\frac{\gamma^0\gamma^2 \slashed{x_1}(x^2_1\slashed{x}_2- \slashed{x}_1 x^2_2)
\slashed{x}_2\gamma^2\gamma^0}{[(x_1-x_2)^2+i\epsilon]^2} = (x_1^2 x_2^2)^{\delta + 3/2}
\frac{\slashed{x}_1- \slashed{x}_2}{[(x_1-x_2)^2+i\epsilon]^2}\,,
$$
which yields $\delta = -3/2$, i.e., the dimension of $\psi$. Therefore, Eq (\ref{I0delta}) can be
factorized as follows
\begin{equation} \label{Igammax2}
{\cal I}_0(x) =
\bigl(x^2\bigr)^{-3/2}\bigg(\frac{-i\,\slashed{x}}{\vert x\vert}\bigg)i\,\gamma^2\gamma^0=
(x^2)^D {\cal S}_0(x){\cal C}= (x^2)^D {\cal C}{\cal S}_0(x)\,,
\end{equation}
where $D$ is the dilation generator, $\cal C$ the charge conjugation matrix and ${\cal S}_0(x) =\pm
i\,\slashed{x}/\vert x \vert x$ is a spin matrix satisfying the equation $\bar{\cal S}_0(x)\,{\cal S}_0(x)
= {\cal S}_0(x)\,\bar{\cal S}_0(x)=1$.

Note that ${\cal S}_0(x)$ acts as a self--adjoint reflection operator, which transforms the
spinor field to its mirror image with respect to the 3D space orthogonal to time--like 4--vector
$x^\mu$ at $x=0$. In effect, we have
\begin{equation}
\label{S0yS0} {\cal S}_0(x)\,\slashed{y}\, \bar{\cal S}(x) = \slashed{y}-
2\,\frac{(xy)}{x^2}\slashed{x}\,,\quad {\cal S}_0(x)\,\gamma^\mu\, \bar{\cal
S}_0(x) = \gamma^\mu- 2\,\frac{x^\mu\slashed{x}}{x^2}\,.
\end{equation}

\subsection{Conformal transformations of tensors in 4D}
\label{I0onspinors} Since $\slashed{x} \slashed{y} = xy -ix^\mu y^\nu \sigma_{\mu\nu}$, where $\sigma_{\mu\nu}
= \frac{i}{2}[\gamma_\mu, \gamma_\nu]$, we obtain from Eqs  (\ref{ITI}) and (\ref{Igammax2})
\begin{equation}
\label{SXB} {\cal E}^\sigma_\rho(-b, x)= \frac{\big(1+bx -ix^\mu b^\nu \sigma_{\mu\nu}\big)^\sigma_\rho}
{\big(1+2bx +b^2x^2\big)^2}\,.
\end{equation}
Noting that $\big(x^\mu b^\nu \sigma_{\mu\nu}\big)^2 = x^2b^2-(bx)^2$, we define
$$
\sigma  = \frac{x^\mu b^\nu \sigma_{\mu\nu}}{\sqrt{x^2b^2-(bx)^2}}\,,
$$
so $\sigma^2=1$. We can then pose
$$
1+bx -ix^\mu b^\nu \sigma_{\mu\nu} = A\bigg(\!\cos \frac{\theta}{2} - i \sigma
\sin \frac{\theta}{2}\bigg) = A\,e^{-i\sigma\theta/2}\,,
$$
with
\begin{equation}
\label{TAN}A = \sqrt{1+2bx +b^2x^2}\quad\hbox{and}\quad \tan \frac{\theta}{2} =
\frac{\sqrt{x^2b^2-(bx)^2}}{1+2bx}\,.
\end{equation}
In conclusion, we can write
$$
{\cal E}(-b, x)= \big(1+2bx +b^2x^2\big)^{-3/2} e^{-i\sigma\theta/2}\,,
$$
showing that ${\cal E}(-b, x)$ is the product of a local dilation and Lorentz
rotation $e^{-i\sigma\theta/2}$ acting on the spinor space, both of which
depend on $x$.

Since the transformation properties of all possible tensors can be derived by reducing direct products of
spinor--field representations, we can infer the general form of inversion and elation spin--matrices for
fields $\Phi(x)$ of any spin in ${\cal M}_4$ as
\begin{equation}
\label{I0&E} {\cal I}_0(x) = (x^2)^D\,{\cal S}_0(x)\,{\cal C}\,,\quad {\cal
E}(-b,x)=(1+2bx +b^2x^2)^{D}\, {\cal R}_0(-b, x)\,,
\end{equation}
where $D$ is the dilation generator, ${\cal S}_0(x)$ is the spin--reflection matrix for $\Phi(x)$, ${\cal C}$
the charge conjugation matrix and ${\cal R}_0(-b,x) = e^{-i \Sigma \theta}$, with $\theta$ defined as in Eq
(\ref{TAN}) and
$$
\Sigma =\frac{x^\mu b^\nu \Sigma_{\mu\nu}}{\sqrt{b^2x^2-(xb)^2}}\,,
$$
as the spin--rotation matrix for $\Phi(x)$ at $x=0$.

As an application of the results so far achieved, let us study the transformation properties of fermionic bilinear
forms and corresponding boson fields under the action of the conformal group.

\subsection{Orthochronous inversions of tetrads, currents and gauge fields}
In the last two subsections, the behavior of the fields under orthochronous inversion was studied in the particular
case of flat or conformally flat spacetime (cf. \S\,7.1 of Part I), basing on the transformation properties of spinor
fields. As soon as we try to transfer the very same concepts to Dirac Lagrangian densities and equations, we encounter
immediately the problem of that Dirac's matrices $\gamma^\mu$, as dealt with so far, have no objective meaning and must be
replaced by expressions of the form $\gamma^\mu(x)= e^\mu_a(x)\, \gamma^a$, where $e^\mu_a(x)$ is the tetradic tensor
and $\gamma^a$ a standard representation of Dirac's matrices in 4D \cite{WEINBERG}.

Since $e^\mu_a(x)$ is related to the metric tensor by the equation $e^\mu_a(x)\,e^{a\mu}(x) = g^{\mu\nu}(x)$, it has
dimension $-1$ and is therefore transformed by $I_0$ as follows
$$
e^\mu_a(x) \xrightarrow{\,I_0\,} \frac{1}{x^2}\,\bigg[e^\mu_a(-x/x^2)-\frac{x^\mu
x_\nu}{x^2}\,e^\nu_a(-x/x^2)\bigg]\,.
$$
consistently with Eq (\ref{S0yS0}).

Let $\psi^a$ be a spinor field of dimension $\!-3/2$ satisfying canonical anti--commuta\-tion relations, where $a$
is some family superscript, and consider the hermitian bilinear forms
\begin{eqnarray}
\label{antibilinears} & & J^{(a,b)\mu}(x)=
\frac{1}{2}\,e^\mu_a(x)\,[\bar\psi^a(x), \gamma^a\, \psi^b(x)]\,,\quad
J^{5(a,b)\mu}(x)= \frac{1}{2}\,e^\mu_a(x)[\bar\psi^a(x), \gamma^a\,\gamma^5
\psi^b(x)], \nonumber\\
& & J^{a,b}(x)= \frac{1}{2}[\bar\psi^a(x), \psi^b(x)]\,,\quad J^{5a,b}(x)=
\frac{1}{2}[\bar\psi^a(x),\gamma^5 \psi^b(x)]\,.
\end{eqnarray}
These may be respectively envisaged as the currents and axial--vector currents of some local algebra, and scalar and
pseudo--scalar densities of $\psi^a$; all of which are expected to be coupled with appropriate boson fields in some
Langrangian density. We leave as an exercise the reader to prove that $I_0$ acts on them as follows:
\begin{eqnarray}
\label{I0veccurr} & & J^{(a,b)\mu}_\mu(x)\xrightarrow{\,I_0\,}
-\frac{1}{(x^2)^4}\bigg[J^{(a,b)\mu}(I_0x) - \frac{x^\mu x_\nu}{x^2}\,J^{(a,b)\nu}(I_0x) \bigg]\,;\\
\label{I0axveccurr} & & J^{5(a,b)\mu}_\mu(x)\xrightarrow{\,I_0\,}
-\frac{1}{(x^2)^4}\bigg[J^{5(a,b)\mu}(I_0x) - \frac{x^\mu
x_\nu}{x^2}\,J^{5(a,b)\nu}(I_0x) \bigg]\,;\\
\label{I0scaldens} & & J^{(a,b)}(x)\xrightarrow{\,I_0\,}
\frac{1}{(x^2)^3}J^{(a,b)}(I_0x)\,;\quad J^{5(a,b)}(x)\xrightarrow{\,I_0\,}
\frac{1}{(x^2)^3}J^{5(a,b)}(I_0x)\,.
\end{eqnarray}
where $I_0 x=-x/x^2$.

As explained at the beginning of \S\,3.1 Part I, covariant vector fields $A_\mu(x)$ and covariant axial--vector fields
$A^5_\mu(x)$ have dimension 0, whereas scalar fields $\varphi(x)$ and pseudoscalar fields $\pi(x)$ have dimension $-1$.
Therefore, for consistency with field equations, they are transformed by $I_0$ as follows
\begin{eqnarray}
\label{I0vec} & & A_\mu(x)\xrightarrow{\,I_0\,} -A_\mu(I_0
x) + \frac{x_\mu x^\nu}{x^2}\,A_\nu(I_0 x)\,;\\
\label{I0axvec} & & A^5_\mu(x)\xrightarrow{\,I_0\,} -A^5_\mu(I_0 x) +
\frac{x_\mu x^\nu}{x^2}\,A^5_\nu (I_0 x)\,;\\
\label{I0scalfields} & & \varphi(x)\xrightarrow{\,I_0\,}
\frac{1}{x^2}\,\varphi(I_0x)\,;\quad \pi(x)\xrightarrow{\,I_0\,}
\frac{1}{x^2}\,\pi(I_0 x)\,.
\end{eqnarray}
Combining Eqs  (\ref{I0veccurr})--(\ref{I0scaldens}) with Eqs
(\ref{I0vec})--(\ref{I0scalfields}) and using the replacements
$$g_{\mu\nu}(x)\xrightarrow{\,I_0\,} (x^2)^2 g_{\mu\nu}(I_0x)\,,\quad \sqrt{-g(x)}
\xrightarrow{\,I_0\,} (x^2)^4 \sqrt{-g(I_0x)}\,,$$
we obtain the transformation laws
\begin{eqnarray}
\label{I0currvec} & & \sqrt{-g(x)}\,J^{(a,b)}_\mu(x)\,A^\mu(x)\xrightarrow{\,I_0\,}
\sqrt{-g(I_0x)}\,J^{(a,b)}_\mu(I_0x)\,A^\mu(I_0x);\nonumber\\
\label{I0axcurraxvec} & & \sqrt{-g(x)}\,J^{5(a,b)}_\mu(x)\,A^{5\mu}(x)\xrightarrow{\,I_0\,}
\sqrt{-g(I_0x)}\,J^{5(a,b)}_\mu(I_0x)\,A^{(5)\mu}(I_0x)\,;\nonumber\\
\label{I0scaldens2} & & \sqrt{-g(x)}\,J^{(a,b)}(x)\,\varphi(x)\xrightarrow{\,I_0\,}
\sqrt{-g(I_0x)}\,J^{(a,b)}(I_0x)\,\varphi(I_0x)\,;\nonumber\\
\label{I0pscaldens} & & \sqrt{-g(x)}\,J^{5(a,b)}(x)\,\pi(x)\xrightarrow{\,I_0\,}
\sqrt{-g(I_0x)}\,J^{5(a,b)}(I_0 x)\,\pi(I_0x)\,;\nonumber
\end{eqnarray}
showing that all Lagrangian densities of interest are transformed by $I_0$ as
\begin{equation}
\label{IogL} \sqrt{-g(x)}\,L(x)\xrightarrow{\,I_0\,} \sqrt{-g(I_0 x)}\, L(I_0 x)\,.
\end{equation}

\subsection{Action--integral invariance under orthochronous inversion $I_0$}
\label{I0Actinv}
On the 3D Riemann manifold, the action of $I_0$ in the past and future cones $H^+_0\equiv H^+_0(+\infty)$ and
$H^-_0\equiv H^-_0(0^-)$ satisfies the self--mirroring linear properties described in \S\,\ref{OnIa}, provided
that the geometry is conformally flat. If the geometry is appreciably distorted by the gravitational field,
self--mirroring properties can still be found, provided that the points of cones are parameterized by polar
geodesic coordinates $\{\tau, \vec\rho\}$, as described in \S\,7.3 of Part I, with $\tau> 0$ for $H^+_0$
and $\tau <0$ for $H^-_0$. In this case, we can define $I_0$ as the operation which maps point
$x = \Sigma(\tau)\cap \Gamma(\vec\rho)\in H^+_0$ into point $x'=\Sigma(-1/\tau)\cap \Gamma(-\vec\rho)\in
H^-_0$, and vice versa.

These changes suggest that, in order for the motion equations to reflect appropriately the conditions for the
spontaneous breakdown of conformal symmetry, the action integral $A$ of the system must be restricted to the union of
$H^+_0$ and $H^-_0$.

Using Eq (\ref{IogL}), putting $x' = -x/x^2$ and renaming $x'$ as $x$, we can immediately establish the invariance under
$I_0$ of all action integrals of the form $A=A^-+A^+$, where
\begin{equation}
\label{AHplusAHminus} A^-= \int_{H^-_0}\!\!\!\!
\sqrt{-g(x)}\,L(x)\,d^4x\,,\quad A^+=
\int_{H^+_0}\!\!\!\!\sqrt{-g(x)}\,L(x)\,d^4x\,,
\end{equation}
because $H^\pm_0 \xrightarrow{\,I_0\,} H^\mp_0$ and $A^\pm \xrightarrow{\,I_0\,} A^\mp$.

\section{The spontaneous breakdown of conformal symmetry}
\label{spontbreakdown}
Assume that the action integral of a quantum field theory is invariant under a continuous non--Abelian group
$G$ of transformations, and let $\vert\Omega\rangle$ be the ground state of one of its field representations.
If $\vert\Omega\rangle$ is invariant under $G$, we say that the theory has the symmetry of $G$, otherwise,
the symmetry is spontaneously broken; in this case, the original symmetry is not simply lost,
but is replaced by a systematic rearrangement of the field representations, characterized by
the following facts  \cite{UMEZAWA1}: (1) residual invariance of $\vert\Omega\rangle$ under a subgroup
$S\subset G$, called the {\em stability subgroup} of symmetry breaking; (2) creation of one or more boson fields of
nonzero VEVs and a gapless energy spectrum, called Nambu--Goldstone (NG) fields, which are one--to--one with
the Lie generators of $G$ which do not annihilate $\vert\Omega\rangle$; (3) degeneration of these generators into
the generators of an Abelian group, called the {\em contraction subgroup}, which acts as the group of
NG--field amplitude translations and parametric rearrangements of field representations;
(4) if $\vert\Omega\rangle$ is invariant under spacetime translations, the energy spectrum of the Hamiltonian
exhibits zero--mass poles, which means that one or more NG fields are of massless particles with nonzero
VEVs; otherwise, the NG fields take the form of extended objects depending on spacetime coordinates.
As proved below, the NG fields generated by the spontaneous breakdown of conformal symmetry belong precisely to the second case.

\subsection{Possible spontaneous breakdowns of conformal symmetry}
\label{Fubini1}
The mechanism of the spontaneous breakdown of conformal symmetry was studied by Fubini in 1976 \cite{FUBINI}.
We report here his main results.

It is known that the 15--parameter Lie algebra of the conformal group $G\equiv{C}(1, 3)$, described by
Eqs (\ref{commut1})--(\ref{commut4}), is isomorphic with that of hyperbolic--rotation group $O(2,4)$ on the 6D
linear space $\{x^0, x^1, x^2, x^3, x^4, x^5\}$ of metric $(x^0)^2 + (x^5)^2 - (x^1)^2 - (x^2)^2 - (x^3)^2 - (x^4)^2$.
The spontaneous breakdown of conformal symmetry can occur only in three ways, corresponding to the
following stability subgroups of $G$:
\begin{itemize}
\item[--] $O(1,3)$: the {\em Poincar\'e group}, i.e, the 10--parameter Lie algebra generated by
$M_{\mu\nu}$ and $P_\mu$. With this choice, NG--boson VEVs are invariant under translations
and are therefore constant.

\item[--] $O(1,4)$: the {\em deSitter group} generated by the 10--parameter Lie
algebra which leaves invariant the quadric $(x^0)^2 - (x^1)^2 - (x^2)^2 - (x^3)^2 -
(x^4)^2$ \cite{MOSCHELLA}, which characterizes the class $dS_4$ of the deSitter
spacetimes as particular 4D--submanifolds, with constant positive curvature, of
the linear space $\{x^0, x^1, x^2, x^3, x^4\}$. Its generators are $M_{\mu\nu}$ and
$$
L_\mu=\frac{1}{2}\,\bigl(P_\mu - K_\mu\bigr)\,,
\vspace{-2mm}
$$
which anti--commute with orthochronous inversion $I_0$ and satisfy the commutation relations $[L_\mu, L_\nu]=-i\,M_{\mu\nu}\,;
\quad [M_{\mu\nu}, L_\rho]  = i \big(g_{\nu\rho}L_\mu -g_{\mu\rho}L_\nu\big)$. Since vacuum state $|\Omega\rangle$ is
invariant under this subgroup, the NG--field $\sigma_+(x)$ associated with the contraction subgroup of $G$ satisfies
equations
\vspace{-2mm}
\begin{equation}
\label{Lsigma-}
L_\mu\sigma_+(x)|\Omega\rangle=0\,, \quad M_{\mu\nu}\,\sigma_+(x)|\Omega\rangle\equiv
-i\big(x_\mu\partial_\nu-x_\nu\partial_\mu\big)\sigma_+(x)|\Omega\rangle=0\,;\nonumber
\end{equation}
the second of which implies that $\sigma_+(x)$ depends on $\tau^2\equiv x^2$ only.

\item[--] $O(2,3)$: the {\em anti--deSitter group} generated by the 10--parameter Lie
algebra which leaves invariant the quadric $(x^0)^2 + (x^4)^2 - (x^1)^2 - (x^2)^2
- (x^3)^2$, which characterizes the class  $AdS_4$ of the anti--deSitter spacetimes
as particular 4D--submanifolds, with constant negative curvature, of a 5D linear space.
Its generators are $M_{\mu\nu}$ and $$R_\mu=\frac{1}{2}\,\bigl(P_\mu+K_\mu\bigr)\,,$$
which commute with orthochronous inversion $I_0$ and satisfies the commutation relations
$[R_\mu, R_\nu]=i\,M_{\mu\nu}\,; \quad [M_{\mu\nu}, R_\rho]  = i (g_{\nu\rho}R_\mu
-g_{\mu\rho}R_\nu)$. Since $|\Omega\rangle$ is invariant under these transformations,
the NG--field $\sigma_-(x)$ associated with the contraction subgroup of $G$,
satisfies equations
\begin{equation}
\vspace{-2mm}
\label{Lsigma+}
R_\mu\sigma_-(x)|\Omega\rangle=0\,, \quad M_{\mu\nu}\,\sigma_-(x)|\Omega\rangle\equiv
-i\big(x_\mu\partial_\nu-x_\nu\partial_\mu\big)\,\sigma_-(x)|\Omega\rangle=0\,,\nonumber
\vspace{-2mm}
\end{equation}
the second of which implies that $\sigma_-(x)$  depends on $x^2=\tau^2$ only.
\end{itemize}

Comparing the results obtained for the de Sitter and anti--de Sitter groups, we note that $L_\mu$, $D$ are the
generators of the set--theoretical complement of $O(3,2)$ in $G$, and $R_\mu$, $D$ are those of the set--theoretical
complement of $O(1,4)$ in $G$. Thus, using commutation relations
$$
\vspace{-2mm}
[R_\mu, D]= i\,L_\mu\,,\quad [L_\mu, D]= i\,R_\mu\,,
\vspace{-2mm}
$$
we derive
\begin{equation}
\vspace{-2mm}
\label{LRsigma}
[R_\mu, D]\,\sigma_+(\tau)|\Omega\rangle = i\,L_\mu\,\sigma_+(\tau)|\Omega\rangle=0\,
\quad [L_\mu,D]\,\sigma_-(\tau)|\Omega\rangle = i\,R_\mu\,\sigma_-(\tau)|\Omega\rangle=0\,,
\end{equation}
showing that these set--theoretical complements act respectively on $\sigma_+(\tau)|\Omega\rangle$
and  $\sigma_-(\tau)|\Omega\rangle$ as Abelian subgroups of transformations.

Using Eqs (\ref{dergenf1}) and (\ref{dergenf2}), we obtain the explicit expressions of
Eqs (\ref{LRsigma}) for $\sigma_\pm(x)$ of dimension $-1$
\begin{eqnarray}
\vspace{-2mm}
L_\mu\sigma_+(\tau)|\Omega\rangle & \equiv & -i\bigg[\frac{1+x^2}{2}\,\partial_\mu-
x_\mu(x^\nu\partial_\nu +1)\bigg]\sigma_+(\tau)|\Omega\rangle=0\,,\nonumber\\
R_\mu\sigma_-(\tau)|\Omega\rangle& \equiv & -i\bigg[\frac{1-x^2}{2}\,\partial_\mu+
x_\mu(x^\nu\partial_\nu +1)\bigg]\sigma_-(\tau)|\Omega\rangle=0\,.\nonumber
\vspace{-2mm}
\end{eqnarray}
Contracting these equations with $x^\mu$, then putting $x^2\equiv\tau^2$ and $x^\mu\partial_\mu \equiv \tau\partial_\tau$,
we can easily verify that their solutions are satisfied for
\begin{equation}
\vspace{-2mm}
\label{sigmaplusminus0}
\sigma_+(\tau)= \frac{\sigma(0)}{1+\tau^2}\,,\quad \sigma_-(\tau)=
\frac{\sigma(0)}{1-\tau^2}
\end{equation}
and, which is particularly interesting, they satisfy the d'Alembert equations
\begin{equation}
\vspace{-2mm}
\label{squarefs}
\square\,\sigma_\pm(\tau)\pm
c^2\,\sigma^3_\pm(\tau)=0\quad \mbox{with } \sigma_\pm(x)=\sigma_\pm(\tau)\,,
\vspace{-2mm}
\end{equation}
where $c = 8/\sigma(0)^2$ and
$$
\vspace{-2mm}
\square f(\tau)\equiv \eta^{\mu\nu}\partial_\mu \partial_\nu f(\tau) = \Big(\partial_\tau^2
+\frac{3}{\tau}\partial_\tau \Big)f(\tau)\,.
$$

However, these expressions are not uniquely determined, since by applying the change of scale $\tau\rightarrow
\tau/\tau_0$,  $\tau_0 > 0$, Eqs (\ref{sigmaplusminus0}) become
\begin{equation}
\label{sigmaplusminus} \sigma_+(\tau)=
\frac{\sigma(0)}{1+(\tau/\tau_0)^2}\,,\quad \sigma_-(\tau)=
\frac{\sigma(0)}{1-(\tau/\tau_0)^2}\,,\,\, \mbox{where }
\sigma(0)=\frac{1}{\tau_0}\, \sqrt{\frac{8}{c}}\,.
\end{equation}
The energy spectra of these functions are gapless and free of poles; hence, they represent
time--dependent extended objects, as predicted by NG theory. In effect, we have
$$
\int_{-\infty}^{\infty}\frac{e^{i\,\omega\tau}}{1+\tau^2/\tau_0^2}d\tau = 2\pi\tau_0\cosh(\tau_0\,\omega)\,;\quad
\int_{-\infty}^{\infty}\frac{e^{i\,\omega\tau}}{1-(\tau-i\epsilon)^2/\tau_0^2}d\tau = 2\pi\tau_0\cos(\tau_0\,\omega)\,.
$$

\subsection{The NG fields of spontaneously broken conformal symmetry}
\label{Fubini2}
It is evident from Eq (\ref{squarefs}) that $\sigma_+(\tau)$ and $\sigma_-(\tau)$  are solutions of the motion equations
deducible from the classical conformal--invariant action integrals
$$
A_\pm = \int\bigg[ \pm \frac{1}{2} \eta^{\mu\nu}[\partial_\mu\sigma_\pm(x)]
\partial_\nu\sigma_\pm(x) - \frac{c^2_\pm}{4}\,\sigma_\pm(x)^4\bigg]\,d^4x\,.
$$

As classical fields,  $\sigma_+$ and $\sigma_-$ may be respectively interpreted as the VEVs $\langle\Omega|\varphi(x)
|\Omega\rangle$ and $\langle\Omega|\sigma(x)|\Omega\rangle$ of a massless physical scalar field $\varphi(x)$ and of a massless
ghost scalar field $\sigma(x)$, which enter the conformal--invariant action integrals, respectively,
$$
A^{\varphi} = \int\bigg[\frac{1}{2} \eta^{\mu\nu}(\partial_\mu\varphi)\,\partial_\nu\varphi -
\frac{c_\varphi^2}{4}\,\varphi^4\bigg]\,d^4x\,,\quad
A^{\sigma} = -\int\bigg[\frac{1}{2} \eta^{\mu\nu}(\partial_\mu\sigma)\partial_\nu\sigma
+ \frac{c_\sigma^2}{4}\,\sigma^4\bigg]\,d^4x\,,
$$
where $\eta_{\mu\nu}$ is the metric of a flat spacetime of any dimension. Instead, if spacetime
is a curved manifold, conformal invariance requires these action integrals to be replaced by
\begin{eqnarray}
\label{Avarphi}
& & A^{\varphi} = \int\bigg[\frac{\sqrt{-g}}{2} g^{\mu\nu}(\partial_\mu\varphi)\partial_\nu\varphi -
\frac{c_\varphi^2}{4}\,\varphi^4 +\frac{R}{6}\varphi^2\bigg]d^4x,\\
\label{Asigma}
& & A^{\sigma} = -\int\bigg[\frac{\sqrt{-g}}{2}g^{\mu\nu}(\partial_\mu\sigma)\partial_\nu\sigma
+ \frac{c_\sigma^2}{4}\sigma^4 +\frac{R}{6}\sigma^2\bigg]d^4x,
\end{eqnarray}
where $R\neq 0$ is the Ricci scalar of metric tensor $g_{\mu\nu}$. In this case, as proven in \S\,3.3 of Part I,
$A^{\varphi}$ and $A^{\sigma}$ are conformal invariant, but in 4D only, up to a harmless surface term.

Comparing these results with those described in Part I, we note that $\sigma_-(\tau)$ may be regarded
as the ``seed'' of the ghost scalar field $\sigma(x)$ introduced in the {\em geometric} Lagrangian
density described \S\,3.4 of Part I. Similarly, $\sigma_+(x)$ may be regarded as the ``seed'' of the
scalar field $\varphi$ of nonzero VEVs, described in \S\,6 of Part I. We shall prove that the mechanism of
cosmic inflation is precisely based on the interaction of these two fields.

Despite the appealing properties of this conformal--invariant picture, we cannot ignore two main problems:

1) The dynamics of $\sigma$ derived from Eq (\ref{Asigma}) is seriously questionable, since the negative
sign of the kinetic energy term of $A^\sigma$  leads to uncontrollable growth in $\sigma$ amplitude.
However, as we prove later, this inconvenience can be neutralized by adding an interaction of $\sigma$
with $\varphi$ which forces the energy density of the system to be bounded from below.

2) Despite the absence of dimensional constants, when we pass to the quantized version of the theory, the
conformal invariance  in the semiclassical approximation is expected to be destroyed by renormalization.
But here the quantum theory of CGR may have several surprises in store.  For instance, the zero--point
energy fluctuations of $\varphi(x)$ and $\sigma(x)$ cancel each other because of the opposite signs of
their kinetic--energy densities and, as we shall see  in Part III, the theory in the semiclassical
approximation predicts the precise values of the cosmological constant and other important cosmological
parameters, without invoking any contribution from zero--point energy fluctuations due to quantization.

\section{Relevant coordinate systems in CGR}
In Part I, Conformal General Relativity (CGR) has been described in two different ways:
one in which conformal symmetry is manifest; in this case, spacetime is a Riemann manifold of
metric tensor $g_{\mu\nu}(x)$ and the theory includes a ghost scalar--field $\sigma(x)$ which acts as the
inflationary factor of the universe. The other, in which conformal symmetry is hidden,
spacetime is a Cartan manifold of fundamental tensor $e^{\alpha(x)}g_{\mu\nu}(x)$,
$\sigma(x)$ is converted into a constant $\sigma_0$ of dimension $-1$, and $e^{\alpha(x)}$
becomes an internal degree of freedom of the conformal geometry.

Since we presume that the future cone created by the spontaneous breakdown of conformal symmetry
is imprinted by the symmetry of stability group $O(2,3)$, as described in \S\,\ref{Fubini1}, we are
naturally led to assume that it reflects, at least in the neighborhood of its origin, the structure and
optimal parameterization of an anti--deSitter spacetime.

In effect, in the Riemann--manifold representation, we are naturally led to adopt hyperbolic coordinates.
But we prefer to rename them as {\em kinematic--time coordinates}, as this was the name used by Brout {\em et.al.}
in their inspiring paper of 1979 \cite{BROUT2}.

In the Cartan manifold representation, in which the standard tensor calculus is replaced by its conformal
extension, as described in the Appendix to Part I, we are naturally led to adopt the {\em conformal
hyperbolic coordinates}. But we will convert them into {\em proper--time coordinates}, by a suitable
redefinition of the time parameter, so that they correspond to the proper--time coordinates used
in standard inflationary cosmology \cite{MUKHANOV}.

\subsection{Hyperbolic polar coordinates}
\label{polargeods}
All worldlines stemming from a point $V$ of a smooth Riemann manifold, and propagating within the future cone of origin
$V$, are called {\em polar geodesics} from $V$. By means of a suitable diffeomorphism of the manifold, we
can parameterize the cone in the neighborhood of $V$ in Minkowskian coordinates. The set of all polar
geodesics stemming from $V$ can then be used to implement a system of {\em hyperbolic polar coordinates}.

Since any polar geodesic is one--to--one with its direction $\vec\rho\,$ at $V$, we can denote it as $\Gamma(\vec\rho\,)$.
In particular, a polar geodesic, generally only one, say $\Gamma(\vec\rho_0)\equiv \Gamma(0)$, can be transformed
into a straight axis by means of a further diffeomorphism which does not alter the metric near $V$. We identify {\em kinematic time}
$\tau$ of an event $O\in\Gamma(\vec\rho\,)$ as the length of geodesic segment $VO$, as shown in Fig.\,2; then,
{\em hyperbolic angle} $\varrho\,$, ($-\infty\le\varrho\le +\infty$), as the derivative with respect to $\tau$, at $\tau=0$, of the
length of the hyperboloid arc between $\Gamma(0)$ and $\Gamma(\vec\rho\,)$. Lastly, we indicate by $\{\theta, \phi\}$
the Euler angles of the projection $\vec r$ of $\Gamma(\vec\rho\,)$ onto the 3D--plane orthogonal to $\Gamma(0)$
at $V$. Since the metric in the neighborhood of $V$ is Minkowskian, we can put $\vec\rho=\{\varrho, \theta, \phi\}$
and $\vec\rho_0=\{0, 0, 0\}$.
\begin{figure}[!ht]
\centering
\mbox{%
\begin{minipage}{0.35\textwidth}
\includegraphics[scale=0.73]{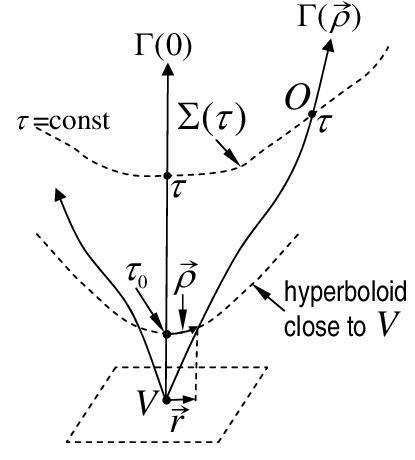}
\end{minipage}%
\quad
\begin{minipage}[c]{0.53\textwidth}
\caption{\small  Geodesics passing through a point $V$ of a spacetime manifold and spanning
the interior of the future cone of origin $V$ can be parameterized by {\em hyperbolic polar coordinates}
$\{\tau, \vec\rho\,\}$. This is possible because each geodesic $\Gamma(\vec\rho\,)$ depends uniquely on its
direction $\vec\rho\,=\{\varrho, \theta,\phi\}$ at $V$. Kinematic time $\tau$ of an event $O\in\Gamma(\vec\rho\,)$
can then be defined as the length of geodesic segment $VO$. 3D surface $\Sigma(\tau)$ is the locus of all
events which have the same time  $\tau$.}
\end{minipage}
}
\end{figure}

Since each line--element $ds$ of a polar geodesic has length $d\tau$ and $\vec\rho= \{\varrho, \theta,\phi\}\equiv
\{\rho^1, \rho^2,\rho^3\}$ is a constant triplet, we have $d\tau/ds =1$ and $d\rho^i/ds =0$. We can then cast any
squared line--element lying in the future cone in the general form
\begin{equation}
\label{pollinelement} ds^2 = d\tau^2 -\tau^2 \,\gamma_{ij}(\tau, \vec\rho\,)\,d\rho^i d\rho^j\,,\quad (i, j
=1,2,3)\,,
\end{equation}
where $\gamma_{ij}$ are adimensional functions of $\tau$ and $\vec\rho$  with initial conditions
\begin{eqnarray}
\lim_{\tau \rightarrow 0}\gamma_{11} = 1;\,\,\, \lim_{\tau \rightarrow 0}\gamma_{22}=(\sinh\varrho)^2;\,\,\,
\lim_{\tau \rightarrow 0}\gamma_{33} =(\sinh\varrho\,\sin\vartheta)^2;\,\,\,
\lim_{\tau \rightarrow 0}\gamma_{ij}=0\,\, (i\neq j).\nonumber
\end{eqnarray}

Therefore, the volume element of the interior of a future cone can be expressed as $\sqrt{-g(x)}\,d^4x =
\tau^3\sqrt{\gamma(\tau, \vec\rho\,)}\,d\tau d^3\rho$, where $\gamma(\tau, \vec\rho\,)$ is the determinant of
matrix $\big[\gamma_{ij}(\tau, \vec\rho\,)\big]$ and $d^3\rho\equiv d\rho^1d\rho^2d\rho^3 \equiv
d\varrho\,d\theta\,d\phi$ is the volume element of the hyperbolic--Euler angles.

If the metric is not so curved as to require a multi--chart representation, complete information
about the gravitational field is incorporated in functions $\gamma_{ij}(\tau, \vec\rho\,)$.

The set of all events of the future cone with the same kinematic time $\tau$ forms a 3D subspace
$\Sigma(\tau)$. Since a point $O$ running along a polar geodesic is presumed to represent an
observer comoving with the expanding universe and equipped with a kinematic--time clock, $\Sigma(\tau)$
represents the set of all possible comoving observers synchronized at time $\tau$.

Of course, as the gravitational field interacting with the evolving distribution of matter becomes more and more
complicated, the multi--chart representation and therefore also the description of $\Sigma(\tau)$
tend to evolve toward levels of indescribable complexity.

The importance of this hyperboloidal foliation of the future--cone is that the volume of each $\Sigma(\tau)$
is infinite at any $\tau$, whereas that of each 3D section orthogonal to the time axis, which is characteristic
of the Minkowskian foliation, is finite.

The reason for the importance is that the infinity of the hyperboloid volumes allows us to define the
{\em thermodynamic limit} of the evolving state of the universe at any kinematic time $\tau$ and to describe
the dispersion of the undetectable horde of infrared photons and gravitational waves at the infinite future,
which is necessary to represent and explain the macroscopic evolution of the universe as an irreversible
thermodynamic process.

In hyperbolic polar coordinates, standard tensor calculus undergoes an interesting simplification.
In particular, the set of general Christoffel symbols $\Gamma^\lambda_{\mu\nu} = \frac{1}{2} g^{\lambda\rho}
\big(\partial_\mu g_{\nu\rho} +\partial_\nu g_{\mu\rho} - \partial_\rho g_{\mu\nu}\big)$  simplifies as follows
\begin{equation}
\label{polarchrist}
\Gamma^\lambda_{00}\! = 0;\,\,\Gamma^0_{0i}\! = 0;\,\,\,\Gamma^j_{0i}\! =\! \frac{\gamma^{jk}\partial_\tau(\tau^2\gamma_{ki})}{2};
\,\,\,\Gamma^0_{ij}\!=\! \frac{\partial_\tau(\tau^2\gamma_{ij})}{2};\,\,\,\Gamma^k_{ij} = \frac{\gamma^{kh}\big[\partial_i \gamma_{jh}+\partial_j \gamma_{ih}-\partial_h \gamma_{ij}\big]}{2};
\end{equation}
where $\gamma^{ij}(x)$ are the elements of inverse matrix $[\gamma^{ij}(x)] = [\gamma_{ij}(x)]^{-1}$.

Therefore, the covariant differential operators acting on a scalar function $f(\tau, \vec\rho\,)$ are
\begin{eqnarray}
\label{polarconfders}
& & \hspace{-16mm} D_\mu f = \partial_\mu f \,;\quad D_\mu D_\nu f =
D_\mu\partial_\nu f =\partial_\mu\partial_\nu f- \Gamma^\lambda_{\mu\nu}\partial_\lambda f \,;
\quad D^2 f = D^\mu \partial_\mu f\,;  \\
\label{polarbeltdalemb}
& &  \hspace{-16mm} D^2f  = \frac{1}{\sqrt{-g}}\partial_\mu\big(\sqrt{-g}
\,g^{\mu\nu}\partial_\nu f\big) = \partial_\tau^2 f + \partial_\tau\ln\big(\tau^3\!\sqrt{\gamma\,}\big)\partial_\tau f -
\frac{1}{\tau^2\sqrt{\gamma}}\partial_i\big(\sqrt{\gamma\,}\gamma^{ij}\partial_j f\big).
\end{eqnarray}

\subsection{Conformal--time coordinates and proper--time coordinates}
\label{conft&propcoord}
The metric tensor $g_{\mu\nu}(\tau, \vec\rho\,)$  of a future cone embedded in a Riemann manifold $H^+$,
parameterized by the kinematic--time coordinates $x=\{\tau, \vec\rho\,\}$, is related through equations
$\hat g_{\mu\nu}(\tau, \vec\rho\,) = e^{2\alpha(\tau, \vec\rho)} g_{\mu\nu}(\tau, \vec\rho)$ to the
fundamental tensor $\hat g_{\mu\nu}(\tau, \vec\rho\,)$ of the corresponding future cone $\widehat H^+$ of the
Cartan manifold parameterized by the same coordinates.

Correspondingly, the squared line--element of $\widehat H^+$ and $\hat g_{\mu\nu}(\tau, \vec\rho\,)$
take the general forms
\begin{eqnarray}
\label{fundtensqline}
& &  \hspace{-5mm}d\hat s^2  =  e^{2\alpha(\tau, \vec\rho\,)}d^2s = e^{2\alpha(\tau, \vec\rho\,)}
\bigl[d\tau^2 -\tau^2 \,\gamma_{ij}(\tau, \vec\rho\,)\,d\rho^i d\rho^j\big],\,\,(i, j =1,2,3);\\
\label{conffundtens}
& &  \hspace{-5mm}\hat g_{00}(\tau, \vec\rho\,) = e^{2\alpha(\tau, \vec\rho\,)};\quad \hat g_{0i}(\tau, \vec\rho\,)=0;
\quad \hat g_{ij}(\tau, \vec\rho\,) = -\tau^2 e^{2\tilde\alpha(\tau, \vec\rho\,)}\gamma_{ij}(\tau, \vec\rho\,);
\end{eqnarray}
where spatial coefficients $\gamma_{ij}(\tau, \vec\rho\,)$ may depend on a gravitational field.
Consequently, the volume element $\sqrt{-\hat g(\tau, \vec\rho\,)}\,\tau^3 d\Omega(\vec\rho\,)$ of $H^+$ is
replaced by $e^{4\alpha(\tau, \vec\rho\,)}\sqrt{-g(x)}\,\tau^3 d\Omega(\vec\rho\,)$ of $\widehat H^+$,
where $d\Omega(\vec\rho\,)$ is the volume element of hyperbolic--Euler angles.

Since, in passing from $H^+$ to $\widehat H^+$, $d\tau$ is multiplied by Weyl scale--factor
$e^{\alpha(\tau, \vec\rho\,)}$, and also because in this form it becomes the analog of the
conformal--time element of standard inflationary cosmology, we find it suitable to rename $\{\tau, \vec\rho\,\}$
as the {\em conformal--time coordinates} of the Cartan manifold.

In passing from kinematic-time coordinates to conformal--time coordinates, the standard tensor calculus
of Riemannian geometry is replaced by the conformal tensor calculus of Cartan geometry (see Appendix to Part I).

Correspondingly, the covariant derivatives, which in kinematic--time coordinates act on $f(x)$ as
described by Eq (\ref{polarconfders}), now act on a scalar function $\hat f(x)$ of the Cartan manifold
$\widehat H^+$ as follows
\begin{equation}
\label{hatDmuDnuf}
\hat D_\mu f = \partial_\mu \hat f \,;\quad \hat D_\mu \hat D_\nu \hat f = \hat D_\mu\partial_\nu f =\partial_\mu\partial_\nu
\hat f-\hat \Gamma^\lambda_{\mu\nu}\partial_\lambda \hat f \,,
\end{equation}
where $\hat\Gamma^\lambda_{\mu\nu}= \frac{1}{2}g^{\lambda\rho}(\partial_\mu g_{\nu\rho}+\partial_\nu g_{\mu\rho}-
\partial_\rho g_{\mu\nu})+ \delta^\lambda_\nu \partial_\mu \alpha +\delta^\lambda_\mu \partial_\nu \alpha
-g_{\mu\nu}\partial^\lambda\alpha$ are the Christoffel symbols constructed from $\hat g_{\mu\nu}(x)$, as
shown in Eq (A-15) of Part I. Accordingly, the Beltrami--d'Alembert operator (\ref{polarbeltdalemb})
is changed to its conformal counterpart
\begin{equation}
\label{hatBeltdAlamb}
\hat D^2 \hat f = \frac{\partial_\tau^2 \hat f}{e^{2\alpha}}
+ \frac{\partial_\tau\big(\tau^3 e^{2\alpha}\sqrt{\gamma\,}\big)}{\tau^3 e^{4\alpha} \sqrt{\gamma\,}}\,\partial_\tau \hat f -
\frac{\partial_i\big(\tau\, e^{2\alpha}\sqrt{\gamma\,}\,\gamma^{ij}\partial_j \hat f\big)}{\tau^3e^{4\alpha}\sqrt{\gamma\,}}.
\end{equation}
It is evident that these coordinates are not of the hyperbolic polar type.

We now introduce another set of coordinates, which have the quality of being polar and correspond
to the proper--time representation of standard inflationary cosmology. Let us define
$d\tilde \tau = e^{\alpha(\tau, \vec\rho\,)}d\tau$ as the {\em proper--time element} corresponding to $d\tau$.
We then have
\begin{equation}
\label{tautotildetau}
\tilde \tau(x)\equiv \tilde \tau(\tau, \vec\rho\,) = \int_0^\tau e^{\alpha(\tau', \vec\rho\,)}d\tau'\,,
\end{equation}
where the integration is carried out at constant $\vec\rho$. For a homogeneous and isotropic spacetime, $\tilde\tau(x)$
depends only on $\tau$. This makes sense, because $\tau$ is by definition the running parameter of the geodesic
stemming from the origin $O$ of the future cone at constant hyperbolic--Euler angles $\vec\rho$, as described in \S\,\ref{polargeods}.
The set of parameters $\tilde x= \{\tilde \tau, \vec\rho\}$ is called the {\em proper time coordinates} and
the future--cone manifold spanned by these coordinates is called the proper--time manifold $\widetilde H^+$.

Vice versa, expressing $\tau$ as a function of $\{\tilde \tau, \vec\rho\}$ and defining
$x(\tilde x)\equiv\{\tau(\tilde x), \vec\rho\,\}$, we can write any function
$\hat f(x)$ grounded in $\widehat H^+$ as $\hat f[x(\tilde x)]=\tilde f(\tilde x)$. In
particular, putting $\tilde\alpha(\tilde x)=\alpha[x(\tilde x)]$ and $\tilde\gamma_{ij}(\tilde x)
\equiv \hat \gamma_{ij}\big[x(\tilde x)\big]$, we can write the squared line--element (\ref{fundtensqline}) and
the metric tensor (\ref{conffundtens}) respectively as
\begin{eqnarray}
\label{tildesqlinel2}
&&\hspace{-14mm} d\tilde s^2  = d\tilde\tau^2 -e^{2\tilde\alpha(\tilde \tau, \vec\rho\,)} \tau(\tilde\tau, \vec\rho\,)^2
\tilde\gamma_{ij}(\tilde \tau, \vec\rho\,)\,d\rho^i d\rho^j,\,\, (i,j=1,2,3)\,;\\
\label{tildemettens}
&&\hspace{-14mm} \tilde g_{00}(\tilde x) =  \tilde g_{00}(\tilde x)=1;\quad \tilde g_{0i}(\tilde x)=\tilde g^{0i}(\tilde x) =0;\quad
\tilde g_{ij}(\tilde x) = - \tau(\tilde\tau, \vec\rho\,)^2 e^{2\tilde\alpha(\tau, \vec\rho\,)}
\tilde \gamma_{ij}(\tilde \tau, \vec\rho\,);\\
&&\hspace{-14mm}\tilde g^{ij}(\tilde x) = -
\tau(\tilde\tau, \vec\rho\,)^{-2} e^{-2\tilde\alpha(\tau, \vec\rho\,)} \tilde \gamma^{ij}(\tilde \tau, \vec\rho\,);\quad
\sqrt{-\tilde g(\tilde x)}= \tau(\tilde\tau, \vec\rho\,)^3
e^{3\tilde \alpha(\tilde \tau, \vec\rho\,)}\sqrt{\tilde\gamma(\tilde x)}\,;
\end{eqnarray}
which are manifestly hyperbolic polar. Correspondingly, the Cartan manifold $\widehat H^+$
is replaced by a Riemann manifold $\widetilde H^+$, and the covariant derivatives act on a scalar
function $\tilde f(\tilde x)$ of proper--time coordinates $\tilde x \equiv \{\tilde \tau, \vec\rho\,\}$ as follows
\begin{equation}
\label{tildeDmuDnuf}
\tilde D_\mu \tilde f =
\tilde \partial_\mu \tilde f;\,\,\, \tilde D_\mu \tilde D_\nu \tilde f
=\tilde \partial_\mu\tilde \partial_\nu\tilde f-\tilde \Gamma^\lambda_{\mu\nu}\tilde \partial_\lambda \tilde f;
\,\,\, \tilde D_\mu \tilde f^\mu \equiv\frac{\tilde \partial_\mu \big(\sqrt{-\tilde g\,} \tilde  f^\mu\big) }{\sqrt{-\tilde g\,}}
= \big(\tilde \partial_\mu + \tilde \Gamma^\lambda_{\lambda\mu}\big)\tilde f^\mu;
\end{equation}
where $\tilde \partial_0 = \partial_{\tilde\tau}$, $\tilde \partial_i = \partial_i$, $\tilde\Gamma^\lambda_{\mu\nu}
= \frac{1}{2}\tilde g^{\lambda\rho}(\tilde \partial_\mu \tilde g_{\nu\rho}+\tilde \partial_\nu \tilde g_{\mu\rho}
-\tilde \partial_\rho \tilde g_{\mu\nu})$ are the Christoffel symbols constructed from $\tilde g_{\mu\nu}(x)$
and $\tilde \Gamma^\lambda_{\lambda\mu}= \tilde\partial_\mu \ln\sqrt{-\tilde g\,}$. Accordingly, the
conformal Beltrami--d'Alembert operator (\ref{hatBeltdAlamb}) is changed to its proper--time counterpart
\begin{equation}
\label{tildeBeltdAlamb}
\hspace{-3pt}\tilde D^2 \tilde f(\tilde x)\!=\!\partial_{\tilde \tau}^2 \tilde f(\tilde x)+\Big\{\!\partial_{\tilde \tau}
\ln\!\big[\tau(\tilde x)^3 e^{3\tilde\alpha(\tilde x)}\!\!\sqrt{\tilde\gamma(\tilde x)}\,\big]\!\Big\}
\partial_{\tilde \tau}\!\tilde f(\tilde x)-\frac{\partial_i\big[\tau(\tilde x)e^{\tilde\alpha(x)}\!
\sqrt{\tilde \gamma(\tilde x)}\tilde\gamma^{ij}(\tilde x)\partial_j \tilde f(\tilde x)\big]}
{\tau^3(\tilde x) e^{3\tilde\alpha(\tilde x)}\sqrt{\tilde\gamma(\tilde x)}}.
\end{equation}
It is then evident that these coordinates are of the hyperbolic polar type.

\subsection{Milne spacetime in kinematic--time coordinates}
\label{hyperbcoordinates}
If spacetime is flat, the most suitable parametrization of a future cone is provided by adimensional parameters $x^\mu$,
which optimally represent the symmetry of orthochronous inversion $I_0$, i.e., the {\em kinematic--time coordinates} centered at $x=0$.
These comprise {\em kinematic time} $\tau$, {\em hyperbolic angle} $\varrho$ ($0\leq \varrho\leq \infty$) and {\em Euler angles}
$\{\theta, \phi\}$, which are related to orthogonal Lorentzian coordinates $x=\{x^0, x^1, x^2, x^3\}$ by equations
\begin{eqnarray}
x^0 & = & \tau\,\cosh\varrho\,;\quad x^1 = \tau\,\sinh\varrho\,\sin\theta\,\cos\phi\,; \nonumber \\
x^2 & = & \tau\,\sinh\varrho\,\sin\theta\,\sin\phi\,;\quad x^3  = \tau\,\sinh\varrho\,\cos\theta\,. \nonumber
\end{eqnarray}
The squared line--element $ds^2 = (dx^0)^2 - (dx^1)^2-(dx^2)^2-(dx^3)^2$ is then
\begin{equation}
\label{hypsqlinel}
ds^2 = d\tau^2 - \tau^2\bigl[d\varrho^2 + \bigl(\sinh\varrho)^2 d\theta^2 +
\bigl(\sinh\varrho\,\sin\theta\bigr)^2d\phi^2 \bigr]\,.
\end{equation}

A future cone of the Minkowski spacetime parameterized by these coordinates is known
as the Milne universe (Mukhanov, 2005) \cite{WALKER}, or {\em Milne spacetime} $M^+_0$.

Since the covariant metric tensor and its contravariant companion, respectively,
\begin{eqnarray}
\label{hypcovmtensor}
& & g_{\mu\nu}(x) = \hbox{diag}\bigl[1, -\tau^2, -(\tau\sinh\varrho)^2, -(\tau\sinh\varrho\,\sin\theta)^2\bigr]\,,\\
\label{hypcontrmtensor}
& & g^{\mu\nu}(x) = \hbox{diag}\bigg[1, -\frac{1}{\tau^2}, -\frac{1}{(\tau\sinh\varrho)^2},
-\frac{1}{(\tau\sinh\varrho\,\sin\theta)^2}\bigg]\,,
\end{eqnarray}
are diagonal, the kinematic--time coordinates are also orthogonal. Therefore, the determinant and
spacetime volume element of this metric are, respectively,
$$
g(x) = - \tau^6(\sinh\varrho)^4(\sin\theta)^2\,\quad\mbox{and}\quad\sqrt{-g(x)}\,dx^4 \equiv
\tau^3(\sinh\varrho)^2\sin\theta\, d\tau\, d\rho\, d\theta\, d\phi\,.
$$

Defining $\vec\rho=\{\rho, \theta, \phi\}\equiv \{\rho^1, \rho^2, \rho^3\}$, we can write $x^\mu=x^\mu(\tau, \vec\rho)$ and
the 4--velocity along the direction of $x^\mu$ as $u^\mu\equiv \partial_\tau x^\mu = x^\mu/\tau$.
Thus we have $u^\mu u_\mu= \eta_{\mu\nu} u^\mu u^\nu =
(u^0)^2- \vert \vec u\vert^2 =1$.
\begin{figure}[!ht]
\centering
\mbox{%
\begin{minipage}{.42\textwidth}
\includegraphics[scale=0.43]{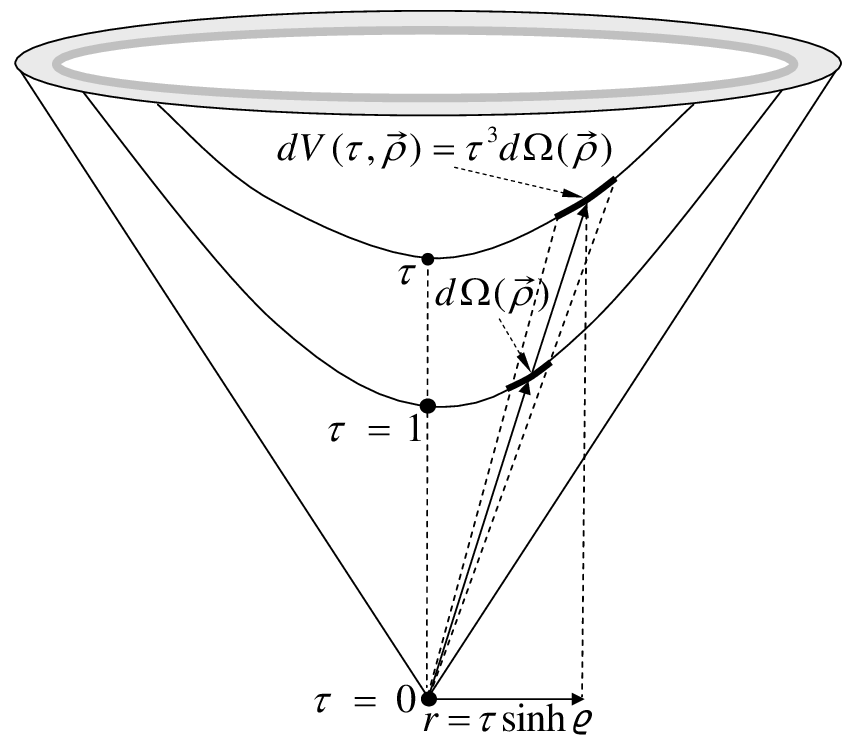}
\end{minipage}%
\quad
\begin{minipage}[c]{0.46\textwidth}
\caption{\small Milne spacetime $M^+_0$ in kinematic--time coordinates.
The position of a point on a hyperboloid is determined by kinematic time $\tau$ and
hyperbolic--Euler angles $\vec\rho= \{\varrho, \theta,  \phi\}$. The spacetime--volume element
at $x\equiv \{\tau, \vec\rho\,\}$ may be written as $d^4x  \equiv dV(\tau, \vec\rho\,)\,d\tau
=\tau^3 d\Omega(\vec\rho\,)\, d\tau$, where $d\Omega(\vec\rho\,)$ is the volume element of the
hyperbolic--Euler--angle space.}
\end{minipage}%
}
\end{figure}

Fig.\,3 illustrates the structure of Milne spacetime $M^+_0$ together with the profiles
of the unit hyperboloid and the hyperboloid at conformal time $\tau$. The
volume element of the hyperbolic--Euler--angle space and the 3D volume element of the
hyperboloid at conformal--time $\tau$ are respectively
\begin{equation}
\label{dV1&dVtau}
d\Omega(\vec\rho\,)\equiv (\sinh\varrho)^2\sin\theta\, d\varrho\,d\theta\, d\phi\,; \quad
d\Omega_\tau(\vec\rho\,)\equiv \tau^3 d\Omega(\vec\rho\,)\,.
\end{equation}
Therefore, the spacetime volume element is $\sqrt{-g(x)}\,d^4x = \tau^3
d\Omega(\vec\rho\,)\,d\tau \equiv  d\Omega_\tau(\vec\rho)\,d\tau$.
The squared gradient of a scalar function $f(\tau, \vec\rho\,)$ is
$g^{\mu\nu}(\partial_\mu f)(\partial_\nu f) =(\partial_\tau)^2 - \vert\vec \nabla_\Omega f\vert^2/\tau^2$, where
\begin{equation}
\label{hyperbsquaregrad} |\vec\nabla_\Omega f|^2 = (\partial_\varrho f)^2 +
\frac{(\partial_\theta f)^2}{(\sinh\varrho)^2} + \frac{(\partial_\phi
f)^2}{(\sinh\varrho\, \sin \theta)^2}\,.
\end{equation}

The only nonzero Christoffel symbols constructed from metric tensor (\ref{hypcovmtensor}) are
\begin{eqnarray}
&&\hspace{-5mm}\Gamma^0_{11} = \tau;\,\,\, \Gamma^0_{22} = \tau\sinh^2\!\rho;\,\,\, \Gamma^0_{33} =
\tau(\sinh\varrho\,\sin\theta)^2;\,\,\,\Gamma^1_{10}=\Gamma^2_{20}=\Gamma^3_{30} = \frac{1}{\tau};
\,\,  \Gamma^3_{32}= \cot\theta;\nonumber\\
&&\hspace{-5mm}\Gamma^2_{21} = \Gamma^3_{31}= \coth\rho;\,\,\,\Gamma^2_{33}=-\sin\!\theta
\cos\!\theta;\,\,\,\Gamma^1_{22} = -\sinh\!\rho \cosh\!\rho;\,\,\,\Gamma^1_{33} =
-\sinh\!\rho \cosh\!\rho\sin^2\!\theta\,; \nonumber
\end{eqnarray}
and the Beltrami--d'Alembert operator is $D^2 f=
\partial_\tau^2 f +3\tau^{-1}\partial_\tau f- \tau^{-2}\Delta_\Omega f$, where
\begin{equation}
\label{unitlaplacian}
\Delta_\Omega\,f \equiv \frac{1}{(\sinh\varrho)^2} \bigg\{\partial_\varrho
\big[(\sinh\varrho)^2\partial_\rho f\big]+\frac{1}{\sin\theta}\,\partial_\theta
(\sin\theta\, \partial_\theta f) +\frac{1}{(\sin\theta)^2}\,\partial^2_\phi f\bigg\}
\end{equation}
is the 3D Laplacian operator in the hyperbolic--Euler--angle space.

This operator has a complete set of orthonormalized eigenfunctions $\Phi_{\hat k}(\vec\rho\,)$
labeled by the {\em hyperbolic momentum eigenvalues}
$$
\hat k =\{k,l,m\}\,, \,\,\mbox{where } 0 \le k < +\infty\,;\,\,
l=0,1,2,\dots\,;\,\, -l\leq m \leq l\,;
$$
and a continuous spectrum of eigenvalues $-(k^2+1)$ \cite{GRIB}; precisely, we have:
\begin{eqnarray}
& & \Phi_{\hat k}(\vec\rho\,)\equiv \Phi_{\{k, l, m\}}(\vec\rho\,)
=\frac{1}{\sqrt{\sinh\varrho}}\bigg\vert\frac{\Gamma(ik+l+1)}{\Gamma(ik)}\bigg\vert\,
P^{-l-1/2}_{ik-1/2}(\cosh\varrho)\, Y_l^m(\theta, \phi)\,; \nonumber \\
& & \int_{K} \Phi^*_{\hat k}(\vec\rho_1)\,\Phi_{\hat k}
(\vec\rho_2)\,d^3\hat k \equiv  \sum_{l,m}\int_0^{+\infty} \Phi^*_{\{k,l,m\}}
(\vec\rho_1)\,\Phi_{\{k,l,m\}}(\vec\rho_2)\,d k = \delta_\Omega^3(\vec\rho_2-\vec\rho_1)\,;
\nonumber\\
& & \bigl(\Delta_\Omega+ k^2+1\bigr)\,\Phi_{\hat k}(\vec\rho\,)=0\,;\quad
\int_{\Omega}\Phi^*_{\hat k}(\vec\rho\,)\,\Phi_{\hat k'}(\vec\rho\,)\, d\Omega(\vec\rho\,)=
\delta^3_{K}(\hat k - \hat k')\,; \nonumber\\
& & \int_{\Omega}\big[\vec\nabla_\Omega\Phi^*_{\hat k}(\vec\rho\,)\big]\!\cdot\!
\big[\vec\nabla_\Omega\Phi_{\hat k'}(\vec\rho\,)\big] d\Omega(\vec\rho\,)= (k^2+1)\delta^3_K(\hat k
- \hat k'); \quad \sum_{l,m}\vert\Phi_{\{k, l,m\}}(\vec\rho\,)\vert^2=\frac{k^2}{2\pi^2}.\nonumber
\end{eqnarray}
where $P^\mu_\lambda(\cosh\varrho)$ are associated Legendre polynomials with
subscripts and superscripts on the complex domain, $ Y_l^m(\theta, \phi)$ are
the familiar 3D spherical harmonics, $K$ is the integration volume
of hyperbolic--momentum  vectors $\hat k$, $d^3\hat k$ its volume element,
$d\Omega(\vec\rho\,)= (\sinh\varrho)^2 \sin\theta\, d\varrho\,d\theta\,d\phi$
and the Dirac deltas are defined as follows
$$
\int_{\Omega}\!\!\!\delta^3_\Omega(\vec\rho_2-\vec\rho_1)\,d\Omega(\vec\rho\,)=1\,,
\quad\delta^3_K(\hat k - \hat k')= \delta_{l'l}\,\delta_{m'm}\,\delta(k-k')\,.
$$

\subsection{Accelerated Milne spacetime in kinematic--time coordinates}
\label{curvedhyperbcoord}
The evidence that the universe on the large scale is homogeneous and isotropic has been enriched by the
discovery that its expansion is slightly accelerated \cite{RIESS} \cite{PERLMUTTER}.
The simplest explanation of this fact is that the Milne spacetime is affected by a small
positive cosmological constant $\Lambda$ or, which is the same thing, by a spacetime curvature
$R= -4\Lambda$, where $R$ is the Ricci scalar. We shall call this {\em accelerated Milne spacetime} $M^+$.

To account for this property, we can give metric tensor (\ref{hypcovmtensor}) the
Robertson--Walker metric of an open universe \cite{WALD}; namely,
\begin{equation}
\label{FRWmet}
g_{\mu\nu}(x) = \hbox{diag}\bigl[1, -c(\tau)^2, -c(\tau)^2\sinh\varrho^2, - c(\tau)^2(\sinh\varrho\,\sin\theta)^2\bigr]\,,
\end{equation}
with the condition that $c(\tau)$ assures a negative constant curvature. We shall also require the coordinates of
this metric to be hyperbolic polar, i.e., they must satisfy the condition $\lim_{\tau\rightarrow 0}c(\tau)/\tau \rightarrow 1$,
so that the geodesics stemming from the origin of the future cone can be labeled by $\vec\rho$, as described in \S~\ref{polargeods}.

Since metric (\ref{FRWmet}) is diagonal, we can use Eqs (A-2) of Part I to obtain the only nonzero Christoffel symbols:
\begin{eqnarray}
\label{curvchristlist}
& & \Gamma_{01}^1=\Gamma_{10}^1=\Gamma_{02}^2 =\Gamma_{20}^2 =\Gamma_{03}^3 = \Gamma_{30}^3 =\frac{\dot c}{c};
\quad\Gamma^0_{1 1} = c\,\dot c; \quad \Gamma^0_{2 2} = c\,\dot c\, (\sinh\varrho)^2;\nonumber\\
& & \Gamma^1_{22} = -\sinh\varrho\cosh\varrho\,;\quad \Gamma^2_{21} = \Gamma^2_{12} = \Gamma^3_{31} = \Gamma^3_{13} =
\frac{\cosh\varrho}{\sinh\varrho}\,;
\quad\Gamma_{23}^3=\Gamma^3_{32}= \frac{\cos\theta}{\sin\theta}\,;
\nonumber\\
& & \Gamma^0_{3 3} = c\,\dot c\,(\sinh\varrho)^2 (\sin\theta)^2\,;\quad\Gamma^1_{33} = -\sinh\varrho\cosh\varrho\,(\sin\theta)^2\,;
\quad\Gamma^2_{33} = -\sin\theta\cos\theta\,;\nonumber\\
& & \Gamma^\rho_{0\rho} = \Gamma^\rho_{\rho\, 0}= 3\frac{\dot c}{c}\,; \quad \Gamma^\rho_{\rho 1} = \Gamma^\rho_{1\rho} =
2\frac{\cosh\varrho}{\sinh\varrho}\,;
\quad \Gamma^\rho_{2\rho} =\Gamma^\rho_{\rho 2} = \frac{\cos\theta}{\sin\theta}\,;\quad \Gamma^\rho_{3\rho} =\Gamma^\rho_{\rho 3} =0.
\nonumber\\
& & \Gamma_{0\rho}^\lambda \Gamma_{0\lambda}^\rho = 3\,\dot c^2;
\quad  \Gamma_{1\rho}^\lambda \Gamma_{1\lambda}^\rho  = 2\dot c^2;\quad\Gamma_{2\rho}^\lambda \Gamma_{2\lambda}^\rho  =
2\,\dot c^2(\sinh\varrho)^2-2(\cosh\varrho)^2 + \frac{(\cos\theta)^2}{(\sin\theta)^2};\nonumber\\
& & \Gamma_{3\rho}^\lambda \Gamma_{3\lambda}^\rho = 2\dot c^2 (\sinh\varrho)^2(\sin\theta)^2
-2(\cosh\varrho)^2(\sin\theta)^2-2(\cos\theta)^2\,.
\end{eqnarray}
Using $R^\mu_\nu= g^{\mu\lambda}(\partial_\rho\Gamma^\rho_{\lambda\nu} -\partial_\nu \Gamma^\rho_{\lambda\rho}+
\Gamma^\sigma_{\lambda\nu}\Gamma^\rho_{\sigma\rho}-\Gamma^\sigma_{\lambda\rho}\Gamma^\rho_{\sigma\nu})$,
we obtain the components of the mixed Ricci--tensor:
\begin{eqnarray}
& & \hspace{-16mm} R^0_0 = - 3\frac{\ddot c}{c};\quad R^i_j = -\delta^i_j\bigg(\frac{\ddot c}{c}
+ 2\,\frac{\dot c^2-1}{c^2}\bigg); \quad R^i_0 = 0,\,\,(i, j =1,2,3);\nonumber\\
\label{R&Rmunu}
& & \hspace{-16mm}R = -6\bigg(\frac{\ddot c}{c}+\frac{\dot c^2-1}{c^2}\bigg);\quad G^0_0 = 3\,\frac{\dot c^2-1}{c^2};
\quad G^i_i= 2\frac{\ddot c}{c}+\frac{\dot c^2-1}{c^2};
\end{eqnarray}
where, $\delta^i_j$ is the Kronecker delta, the repeated index $i$ is not summed and
$G^\mu_\nu = R^\mu_\nu- \delta^\mu_\nu R/2$ is the Einstein gravitational tensor.

The reader can easily verify that $R^\mu_\nu = 0$ if and only if $a(\tau)= \tau-\tau_0$, where $\tau_0$ is a constant,
and that the squared gradient, the  Beltrami--d'Alembert operator and the double covariant derivatives are,
respectively,
\begin{eqnarray}
\vspace{-4mm}
\label{curvsqrgrad}
& & \big(D^\mu f\big)\big(D_\mu f\big) = (\partial_\tau)^2 - \frac{1}{c^2}\vert \vec \nabla_\Omega f\vert^2\,;\\
\label{curvedD2}
\vspace{-2mm}
& & D^2 f = \partial^2_\tau f + 3\,\big(\partial_\tau\log c\big)\,\partial_\tau f- \frac{1}{c^2}\Delta_\Omega f\,;\\
\vspace{-2mm}
\label{curvedDmuDnu} & & D_\mu D_\nu f = D_\mu\partial_\nu f =\partial_\mu\partial_\nu f-
\Gamma^\lambda_{\mu\nu}\partial_\lambda f\,,
\vspace{-2mm}
\end{eqnarray}
with $\vec\nabla_\Omega f$ and $\Delta_\Omega f$ defined as in \S\,\ref{hyperbcoordinates} and $\Gamma^\lambda_{\mu\nu}$
given by Eqs (\ref{curvchristlist}).

For an empty universe of cosmological constant $\Lambda$, we have
\begin{equation}
\label{Rmunu&Lambda}
R= -4\,\Lambda, \quad R^\mu_\nu = -G^\mu_\nu= -\Lambda\,\delta^\mu_\nu= \frac{R}{4}\delta^\mu_\nu\,,\quad
\mbox{hence }\,\frac{\ddot c}{c} = \frac{\dot c^2-1}{c^2} = \frac{\Lambda}{3}.
\vspace{-2mm}
\end{equation}
The Riemann tensor is then $R_{\mu\nu\rho\sigma} = (R/12)\big(g_{\mu\rho}g_{\nu\sigma}-g_{\mu\nu}g_{\rho\sigma}\big)$
and the Weyl tensor $C_{\mu\nu\rho\sigma}$ vanishes (see end of Appendix to Part I).

As the reader can easily check, the solution to the last of Eqs (\ref{Rmunu&Lambda}) is $c(\tau) = \tau_\Lambda\sinh (\tau/\tau_\Lambda) =
\tau(1+ \tau^2/6\tau_\Lambda^2 +\dots)$ where $\tau_\Lambda = \sqrt{3/\Lambda}$. Since astronomic data suggest
$\Lambda \simeq 10^{-35}$\,s$^{-2}$, we obtain $\tau_\Lambda\simeq 2.74\times 10^{17}$s.
Quite surprisingly, this is about half the currently estimated age of universe $\tau_U \simeq 4.38 \times 10^{17}$s.

Therefore, during the short inflationary epoch, and as long as $\tau \ll  \tau_U$, the coordinates of metric (\ref{FRWmet})
are well approximated by the kinematic--time coordinates described in \S\,\ref{hyperbcoordinates}. The distortion
imparted by the negative curvature to Milne spacetime  $M^+_0$ shown in Fig.\,3 is qualitatively represented in Fig.\,4.
\begin{figure}[!ht]
\mbox{%
\begin{minipage}{.45\textwidth}
\includegraphics[scale=0.35]{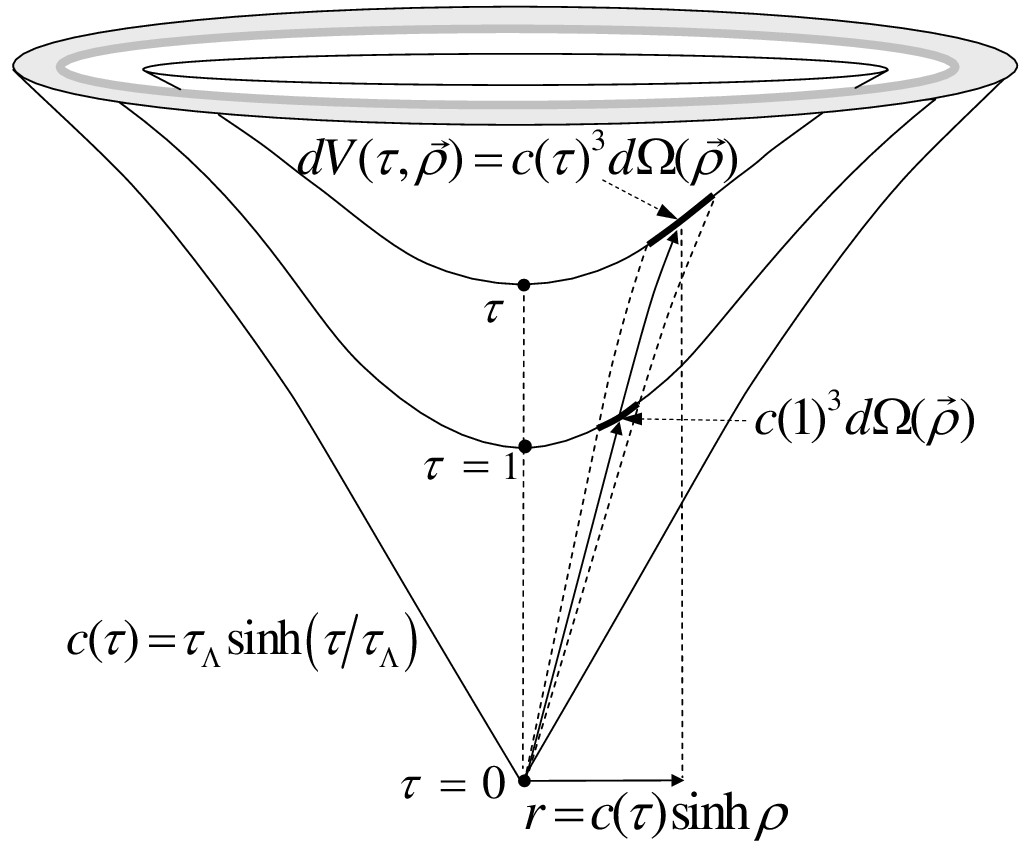}
\end{minipage}%
\begin{minipage}[c]{.52\textwidth}
\caption{\small
Accelerated Milne spacetime $M^+$ in kinematic--time coordinates. Moving along a polar geodesic from kinematic
time $1$ to kinematic time $\tau$, the volume element of the hyperboloid changes from $dV(1,\vec\rho\,)
=c(1)^3 d\Omega(\vec\rho\,)$ to $dV(\tau, \vec\rho\,)= c(\tau)^3 d\Omega(\vec\rho\,)$, where $d\Omega(\vec\rho\,) = (\sinh\varrho)^2\sin\theta\,d\varrho\,d\theta\,d\phi$ is the volume element of the hyperbolic--Euler angles and
$c(\tau)= \tau_\Lambda\sinh(\tau/\tau_\Lambda)$, with $\tau_\Lambda=\sqrt{3\,/\Lambda}$.}
\end{minipage}%
\vspace{-2mm}
}\end{figure}

To complete our analysis, we assume that, during most of the history of the universe all departures from homogeneity and
isotropy have been small, so that they can be treated as first--order perturbations. We can then represent the perturbed
metric of $M^+$ as $\bar g_{\mu\nu}=g_{\mu\nu} + h_{\mu\nu}$, where $g_{\mu\nu}$ is the unperturbed metric (\ref{FRWmet})
and $h_{\mu\nu}$ a small perturbation. Hence, the contravariant perturbation is $h^{\mu\nu}= g^{\mu\rho}g^{\nu\sigma}
h_{\rho\sigma}$. The Christoffel symbols then become perturbed, as follows $\delta \Gamma^\lambda_{\mu\nu} = \frac{1}{2}
g^{\lambda\rho}\big(\partial_\mu h_{\nu\rho} + \partial_\nu h_{\mu\rho} - \partial_\rho h_{\mu\nu} - 2\,\Gamma^\sigma_{\mu\nu}
h_{\rho\sigma}\big)$. Replacing in these the expressions for $\Gamma^\lambda_{\mu\nu}$ listed in Eqs (\ref{curvchristlist}),
we obtain the only nonzero perturbations:
\begin{eqnarray}
\label{ChristPerts}
&&\hspace{-5mm}\delta \Gamma^0_{00} = \frac{\dot h_{00}}{2};\quad \delta \Gamma^0_{0i} = \frac{\partial_i h_{00}-
2\,h_{0i}\,c\dot c }{2} ;\quad \delta \Gamma^1_{00} =\frac{\partial_1 h_{00} - 2\,\dot h_{01}}{2c^2};
\quad \delta \Gamma^2_{00} = \frac{\partial_2 h_{00} - 2\,\dot h_{02}}{2\,c^2 (\sinh\varrho)^2};\nonumber\\
&&\hspace{-5mm} \delta \Gamma^3_{00} =\frac{\partial_3 h_{00} -2\,\dot h_{03}}{2\,c^2 (\sinh\varrho \,\sin\theta)^2}\,;
\quad \delta \Gamma^1_{01} =-\frac{\dot h_{11}+ 2\,h_{11}\,\dot c/c}{2\,c^2};\quad  \delta \Gamma^2_{02} =
-\frac{\dot h_{22}+  2\,h_{22}\,\dot c/c}{2\,c^2(\sinh\varrho)^2};\nonumber\\
&&\hspace{-5mm}\delta \Gamma^2_{01} = \frac{\partial_2 h_{01}- \partial_1 h_{12}-\dot h_{12} + 2\,h_{12}\,\dot c /c}
{ 2\,c^2(\sinh\varrho)^2} \,;\quad\delta \Gamma^3_{01} = \frac{\partial_3 h_{01}- \partial_1 h_{13}-\dot h_{13}
+ 2\,h_{13}\,\dot c/c}{2\, c^2(\sinh\varrho\sin\theta)^2} \,;\nonumber\\
&&\hspace{-5mm}\delta \Gamma^1_{02} = \frac{\partial_1 h_{02}- \partial_2 h_{01}- \dot h_{12}+
2\,h_{12}\,\dot c/c}{2\,c^2};\quad\delta \Gamma^3_{02} = \frac{\partial_3 h_{02} - \partial_2 h_{03}-
\dot h_{23}+ 2\,h_{23}\,\dot c/c}{2\,c^2(\sinh\varrho\sin\theta)^2};\nonumber\\
&&\hspace{-5mm}\delta \Gamma^1_{03} = \frac{\partial_1 h_{03} - \dot h_{13} - \partial_3 h_{01} +
2\,h_{13}\,\dot c/c}{2\,c^2};\!\quad \delta \Gamma^2_{03} = \frac{\partial_2 h_{03} - \partial_3 h_{02}-
\dot h_{23} + 2h_{23}\,\dot c/c}{2\,c^2(\sinh\varrho)^2};\nonumber\\
&&\hspace{-5mm} \delta \Gamma^3_{03} = -\frac{\dot h_{33}+2h_{33}\,\dot c/c}{2\,c^2(\sinh\varrho\,\sin\theta)^2};\quad
\delta \Gamma^3_{13}=\frac{2\coth\!\varrho\,h_{33}-\partial_1 h_{33}}{2c^2(\sinh\varrho\sin \theta)^2};\quad
\!\delta \Gamma^3_{23}=\frac{2\coth\!\varrho\,h_{33}-\partial_2 h_{33}}{2c^2(\sinh\varrho\sin \theta)^2};\nonumber\\
&&\hspace{-5mm} \delta \Gamma^0_{11}  =\partial_1 h_{0 1} -\frac{\dot h_{11}}{2}- c\dot c\, h_{00};\quad\!
\delta \Gamma^0_{22} = \partial_2 h_{0 2} - \frac{\dot h_{22}}{2} - c\dot c\,h_{00}(\sinh\varrho)^2  +
h_{01}\sinh\varrho\,\cosh\rho; \nonumber\\
&&\hspace{-5mm}\delta \Gamma^0_{33}  = \partial_3 h_{0 3} - \frac{1}{2}\dot h_{33} - c\,\dot c\,
(\sinh\varrho\sin\theta)^2 h_{00} + \sinh\varrho\cosh\varrho(\sin\theta)^2\, h_{01} + \sin\theta\cos\theta h_{02}\,; \nonumber\\
&&\hspace{-5mm}\delta \Gamma^1_{11} = -\frac{\partial_1 h_{11} +2\,c\dot c\,h_{01}}{2\,c^2};\quad \delta \Gamma^2_{11} =
\frac{\partial_2 h_{11}-2\,\partial_1 h_{12} + 2\,c\dot c\,h_{01}}{2\,c^2(\sinh\varrho)^2};\nonumber\\
&&\hspace{-5mm}\delta \Gamma^3_{11} = \frac{\partial_3 h_{11}-2\,\partial_1 h_{13} +2\,c\dot c\,h_{01}}{2\,
c^2(\sinh\varrho\sin \theta)^2}\,;\quad \delta \Gamma^1_{12} =\frac{2\coth\varrho\,h_{12}-\partial_2 h_{11}}{2\,c^2}; \nonumber\\
&&\hspace{-5mm}\delta \Gamma^2_{12} = \frac{2\coth\varrho\, h_{22} -\partial_1 h_{22}}{2\,c^2(\sinh\varrho)^2}\,;\quad
\delta \Gamma^3_{12} = \frac{\partial_3 h_{12}-\partial_1 h_{23} - \partial_2 h_{13} +
2\,\coth\varrho\, h_{23}}{2\,c^2(\sinh\varrho\sin\theta)^2}\,;\nonumber\\
&&\hspace{-5mm}\delta \Gamma^1_{13} = \frac{2\coth\varrho\,h_{13}- \partial_3 h_{11}}{2\,c^2};
\quad \delta \Gamma^2_{13} =\frac{\partial_3 h_{12}-\partial_1 h_{23} - \partial_2 h_{13} +
2\,\coth\varrho\, h_{23}}{2\,c^2(\sinh\varrho)^2};\nonumber
\end{eqnarray}
Using Ricci's formula $\delta R_{\mu\nu}= \partial_\rho\delta \Gamma^\rho_{\lambda\nu} -\partial_\nu \delta \Gamma^\rho_{\lambda\rho}+
\Gamma^\sigma_{\lambda\nu}\delta\Gamma^\rho_{\sigma\rho}-\Gamma^\sigma_{\lambda\rho}\delta\Gamma^\rho_{\sigma\nu}$, we obtain
\vspace{-2mm}
\begin{equation}
\label{Ricci00A}
\delta R_{00} = \delta R^0_0= \frac{1}{2\,c^2} \bigg[\nabla^2_\Omega h_{00} + 3\,c\,\dot c\, \dot h_{00}-
2\,\partial^i\dot h_{0i}+\ddot h^i_i -\frac{2\,\dot c}{c}\,\dot h^i_i+ 2\bigg(\frac{\ddot c}{c} -
\frac{\dot c^2}{c^2}\bigg)h^i_i\bigg],
\vspace{-1mm}
\end{equation}
where $\partial_1 \equiv \partial_\rho$, $\partial_2 \equiv \partial_\theta$, $\partial_3 \equiv \partial_\phi$, $\partial^i h_{0i}
\equiv\sum_i |g^{ii}|\,\partial_ih_{0i}$, $\nabla^2_\Omega h_{00} \equiv \sum_i |g^{ii}|\,\partial_i^2 h_{00}$, and $h^i_i =
\sum_i |g^{ii}|\,h_{ii}$, where $g^{ii}= 1/g_{ii}$ and $i=1,2,3$.

A perturbation of particular interest is $h_{\mu\nu}(\tau, \vec\rho\,) = \mbox{diag }[2 \,c(\tau)^2 \Phi(\vec \rho\,), 0,0,0]$,
where $\Phi(\vec\rho\,)$ is a static Newtonian potential and $c(\tau)$ the acceleration factor. In this case, Eq (\ref{Ricci00A})
simplifies to Poisson equation $\delta R_{00}(\vec\rho\,) = \nabla^2_\Omega\Phi(\vec\rho\,)$.

\subsection{Milne spacetime in conformal-- and proper--time coordinates}
\label{confhyperbcoord}
Metric tensor $g_{\mu\nu}(\tau, \vec\rho\,)$  of the Milne spacetime $M_0^+$ in kinematic--time coordinates is
related to the fundamental tensor of the Milne Cartan spacetime $\widehat M_0^+$ in conformal--time coordinates
by equation $\hat g_{\mu\nu}(\tau, \vec\rho\,) = e^{2\alpha(\tau)}g_{\mu\nu}(\tau, \vec\rho\,)$, with $\alpha(x)$ only depending on $\tau$.

Therefore, the squared line--element and the fundamental tensor of $\widehat M_0^+$ are, respectively,
\begin{eqnarray}
\label{hatsqlinel1}
& & \hspace{-12mm} d\hat s^2 \!\!  =  e^{2\alpha(\tau)}d^2s = e^{2\alpha(\tau)}\bigl\{d\tau^2 -
\tau^2\bigl[d\varrho^2 \! + \bigl(\sinh\varrho\bigr)^2 d\theta^2\! + \bigl(\sinh\varrho\,\sin\theta\bigr)^2d\phi^2
\bigr]\bigr\};\\
\label{hatgmunu1}
& & \hspace{-12mm}\hat g_{\mu\nu}(x)=  e^{2\alpha(\tau)}\hbox{diag}\bigl[1, -\tau^2,
-\tau^2\sinh\varrho^2, -\tau^2\sinh\varrho^2\,\sin\theta^2\bigr];
\end{eqnarray}
Consequently, the squared gradient, Beltrami--d'Alembert operator and double covariant derivatives of a scalar function
$\hat f(\tau, \vec\rho\,)$ are, respectively,
\begin{eqnarray}
\label{hatsqrgrad}
& & \big(\hat D^\mu \hat f\big)\hat D_\mu \hat f = \frac{(\partial_\tau\hat f)^2}{e^{2\alpha(\tau)}} -
\frac{1}{\tau^2 e^{2\alpha(\tau)}}\vert\vec \nabla_\Omega \hat f\vert^2\,;\\
\label{hatdD2}
& & \hat D^2 \hat f = \frac{\partial^2_\tau \hat f}{e^{2\alpha(\tau)}}+ \frac{\partial_\tau\ln\big[\tau^3
e^{2\alpha(\tau)}\big]}{e^{2\alpha(\tau)}}\partial_\tau \hat f-
\frac{1}{\tau^2e^{2\alpha(\tau)}}\Delta_\Omega \hat f\,;\\
\label{hatDmuDnu}
& & \hat D_\mu \hat D_\nu \hat f = \hat D_\mu\partial_\nu \hat f =\partial_\mu\partial_\nu \hat f-
\hat \Gamma^\lambda_{\mu\nu}\partial_\lambda \hat f\,,
\end{eqnarray}
where $\hat \Gamma^\lambda_{\mu\nu}(x)$ are the conformal Christoffel symbols derived from $\hat g_{\mu\nu}(x)$.

In passing to proper--time coordinates, as indicated in \S\,\ref{conft&propcoord}, Eqs (\ref{hatsqlinel1}) (\ref{hatgmunu1}) become
\begin{eqnarray}
\label{tildesqlinel}
& & \hspace{-16mm}d\tilde s^2 = d\tilde\tau^2 -
\tau(\tilde \tau)^2 e^{2\tilde\alpha(\tilde \tau)}\bigl[d\varrho^2 \! +
\bigl(\sinh\varrho\bigr)^2 d\theta^2\! + \bigl(\sinh\varrho\,\sin\theta\bigr)^2d\phi^2
\bigr];\\
\label{tildegmunu}
& & \hspace{-16mm}\tilde g_{\mu\nu}(\tilde x)= \hbox{diag}\bigl[1, -\tau(\tilde\tau)^2e^{2\tilde\alpha(\tilde \tau)}
-\tau(\tilde \tau)^2e^{2\tilde\alpha(\tilde \tau)}\sinh\varrho^2,
-\tau(\tilde \tau)^2e^{2\tilde\alpha(\tilde \tau)}\sinh\varrho^2\,\sin\theta^2\bigr];
\end{eqnarray}
where, in virtue of the homogeneity and isotropy of $\widehat M_0^+$, $\tau(\tilde x)$ only depends on $\tilde \tau$.

Therefore, we also have
\begin{equation}
\label{tildedV1&dVtau}
\sqrt{-\tilde g(\tilde x)}=e^{3\tilde\alpha(\tilde \tau)}\sqrt{-g[\tau(\tilde\tau), \vec\rho\,]}\,;
\quad d^4 \tilde x \equiv \tau(\tilde\tau)^3\,d\Omega(\vec\rho\,)\, e^{\tilde\alpha(\tilde \tau)}d\tau
\equiv \tau(\tilde x)^3\,d\Omega(\vec\rho\,)\,d\tilde\tau\,;
\end{equation}
where $\tilde \partial_\mu= \{\partial_{\tilde\tau}, \partial_\varrho, \partial_\theta, \partial_\phi\}$ are
proper--time partial derivatives, $\tilde \partial^\mu = \tilde g^{\mu\nu}(\tilde\tau,\vec\rho\,) \tilde \partial_\nu$
the contravariant ones and,  in particular, $\partial_{\tilde\tau}=e^{-\tilde\alpha(\tilde\tau)}\partial_\tau$ and
$\partial^{\tilde \tau} \equiv \partial_{\tilde\tau}$. Therefore, covariant derivatives $\tilde D_{\mu}$ act on a
scalar function $\tilde f(\tilde x)\equiv\tilde f(\tilde\tau,\vec\rho\,)$ as $\tilde \partial_\mu\tilde \partial_\nu\tilde
f(\tilde x) - \tilde \Gamma^\lambda_{\mu\nu}(\tilde x)\tilde \partial_\lambda \tilde f(\tilde x)$.
Since metric tensor (\ref{tildegmunu}) has the same structure as (\ref{FRWmet}), but with
$\tau(\tilde \tau)e^{\tilde\alpha(\tilde \tau)}$ in place of $c(\tau)$ and $\tilde\tau$ in place of
$\tau$, $\tilde \Gamma^\lambda_{\mu\nu}(\tilde x)$ mimic the symbols listed in Eqs (\ref{curvchristlist}).
In particular, we obtain $\tilde \Gamma_{0i}^j= \delta_i^j\,\partial_{\tilde\tau}
\ln \big[\tau(\tilde \tau)\,e^{\tilde\alpha(\tilde \tau)}\big]$; $\tilde \Gamma^0_{11}=\frac{1}{2} \partial_{\tilde\tau}
\big[\tau(\tilde \tau)\,e^{\tilde\alpha(\tilde \tau)}\big]^2$; $\tilde \Gamma^0_{22}=\tilde \Gamma^0_{11}(\sinh\varrho)^2$;
$\tilde \Gamma^0_{33}=\tilde \Gamma^0_{22}(\sin\theta)^2$.

Hence, we have
\begin{equation}
\label{tildeDmuDnuf}
\tilde D_0 \tilde f = \partial_{\tilde\tau} \tilde f;\quad \tilde D_i \tilde f =\partial_i \tilde f;
\quad \tilde D_0 \tilde D_0 =\partial_{\tilde\tau}^2 \tilde f;
\quad \tilde D_i \tilde D_0 \tilde f = 0\,,
\end{equation}
and the squared gradient of $\tilde f(\tilde \tau)$ and Eq (\ref{tildeBeltdAlamb}) specialize to
\begin{eqnarray}
\label{tildeD2f}
& &\hspace{-8mm} \big(\tilde D^\mu \tilde f\big)\tilde D_\mu \tilde f =\tilde g^{\mu\nu}\bigl(\tilde \partial_\mu
\tilde f\bigr)\bigl(\tilde\partial_\nu \tilde f\bigr) = \bigl(\partial_{\tilde\tau} \tilde f\bigr)^2-
\frac{1}{\tau(\tilde\tau)^2 e^{2\tilde \alpha(\tilde\tau)}}\vert\vec \nabla_\Omega \tilde f\vert^2; \\
& & \hspace{-8mm}\tilde D^2 \tilde f =\partial_{\tilde \tau}^2\tilde f  + 3\big\{\partial_{\tilde\tau}\!\ln
\bigl[\tau(\tilde\tau)\,e^{\tilde\alpha(\tilde\tau)}\bigr]\big\}\partial_{\tilde\tau}\tilde f
- \frac{1}{\tau(\tilde\tau)^2e^{2\tilde\alpha(\tilde\tau)}}\Delta_\Omega \tilde f\,.
\end{eqnarray}

Since $e^{2\tilde\alpha(\tilde x)}$ is intended to represent the spatial inflation factor of the Milne spacetime
introduced in \ref{hyperbcoordinates}, we call it {\em inflated Milne spacetime} $\widetilde M^+_0$.

In this case, as illustrated in Fig.\,5,  the ratio between the lengths $\Delta \tilde L_1$, $\Delta \tilde L_2$
of two radial intervals, respectively comoving with the 3D--spaces of synchronized observers at proper times
$\tilde\tau_1$ and $\tilde\tau_2$, is $\Delta\tilde L_{2}/\Delta\tilde L_{1} =\tau(\tilde \tau_2)\,
e^{\tilde \alpha(\tilde\tau_2)}/\tau(\tilde \tau_1)\, e^{\tilde \alpha(\tilde\tau_1)}$. Instead,
the measurement units of all physical quantities are preserved.
\vspace{-2mm}
\begin{figure}[!ht]
\centering
\mbox{%
\begin{minipage}{.38\textwidth}
\includegraphics[scale=0.32]{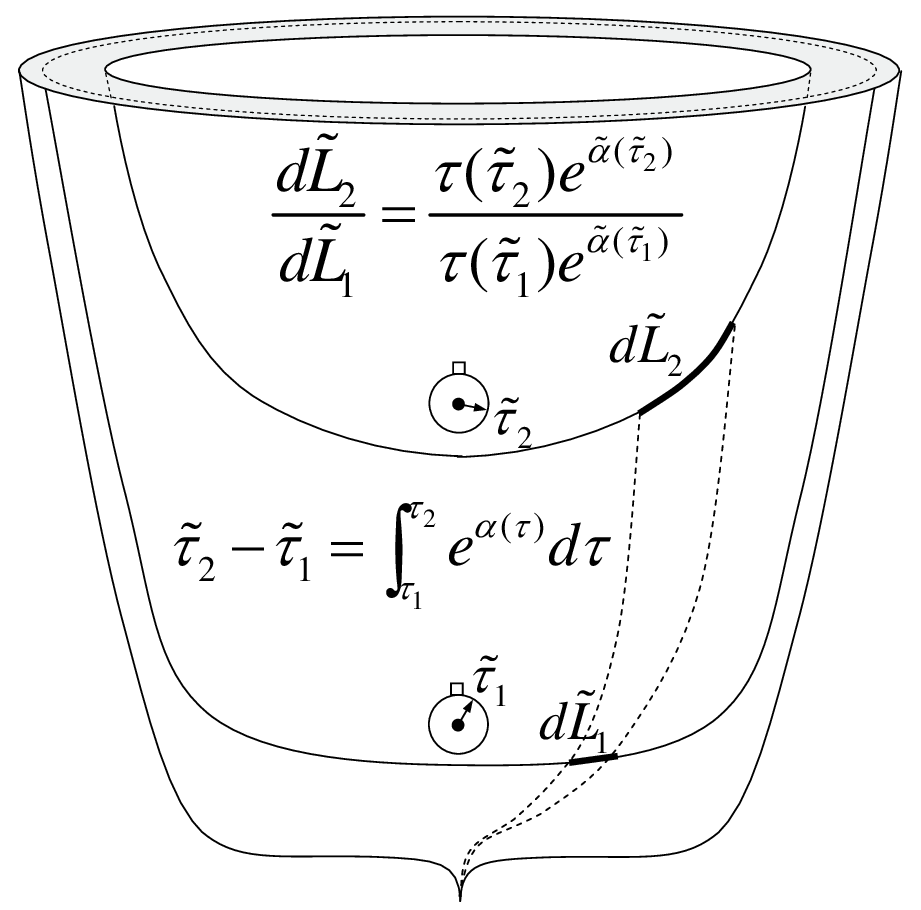}
\end{minipage}%
\begin{minipage}[c]{.55\textwidth}
\caption{\small Qualitative features of inflated Milne spacetime $\widetilde M^+_0$ in proper--time coordinates.
Passing from proper time $\tilde\tau_1$ to proper time $\tilde\tau_2$, the length $\Delta\tilde L$
of a comoving radial interval, as well as the measurement unit of length, increases by a factor of
$\tau(\tilde \tau_2) e^{\tilde \alpha(\tilde\tau_2)}/\tau(\tilde \tau_1)e^{\tilde \alpha(\tilde\tau_1)}$.
Since we expect the profile of $e^{\tilde \alpha(\tilde\tau)}$ to be sigmoidal, we also expect an
increasing flattening of hyperboloids in the apical region of the future cone.}
\end{minipage}
}
\end{figure}

Note that, instead, in the flat Milne spacetime represented in Fig.\,3, the lengths would increase by a
ratio of $\tau_2/\tau_1$, and the measurement unit of a physical quantity of dimension $n$ would increase
by a factor of $e^{n[\alpha(\tau_2)-\alpha(\tau_1)]}$.

By contrast, as shown in Fig.\,5, all worldlines of $\widetilde M^+_0$ are the geodesics stemming from
the future cone origin {\em O}. All comoving reference frames, which on the Riemann manifold are synchronized and at rest
in a 3D space labeled by conformal time $\tau$, are also synchronized and at rest in a 3D space
labeled by the proper time $\tilde\tau$ of $\widetilde M^+_0$.

If the fundamental tensor is not conformally flat, which may happen in the presence of
gravitational forces, the axial symmetry of Milne spacetime is generally lost, in both
conformal--time and proper--time representations.
\begin{figure}[!ht]
\centering
\mbox{%
\begin{minipage}{.35\textwidth}
\includegraphics[scale=0.3]{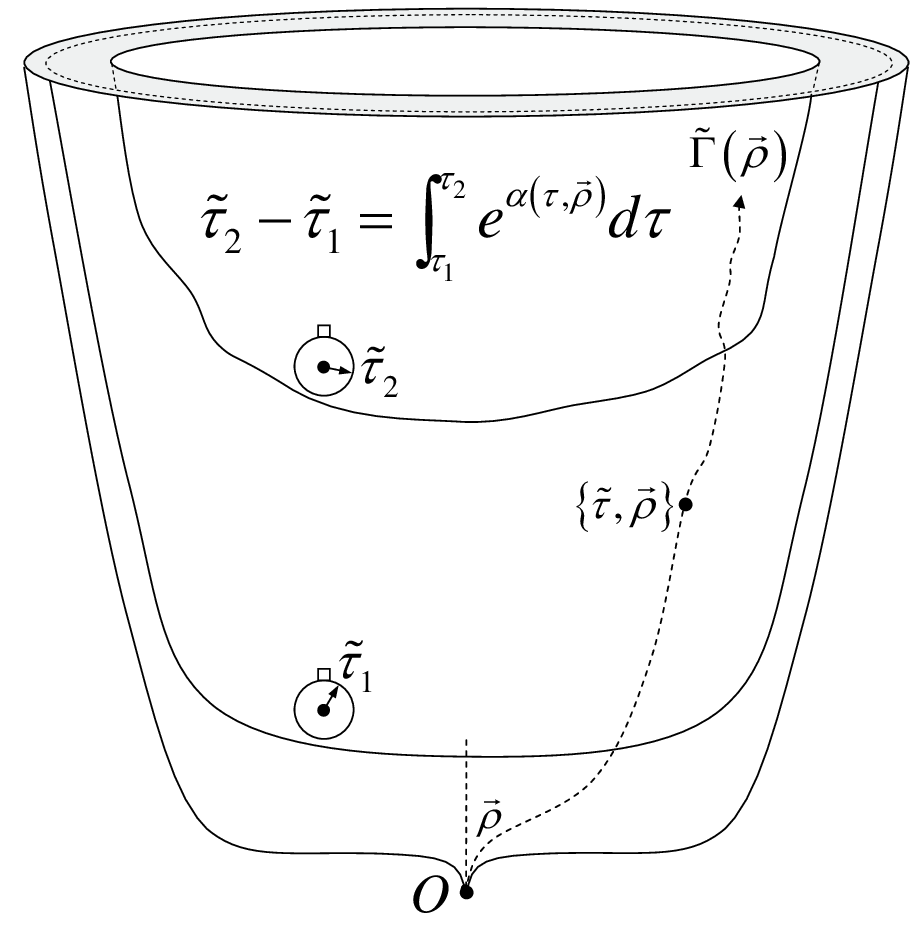}
\end{minipage}
\quad
\begin{minipage}[c]{.55\textwidth}
\caption{\small Generally future--cone manifold $\tilde H^+$ in proper--time coordinates $\{\tilde \tau, \vec\rho\,\}$.
The worldline of a particle comoving with the expanding universe is a geodesic $\tilde\Gamma(\vec\rho)$
stemming from the future--cone origin $O$, with running point $\{\tilde\tau, \vec\rho\,\}$. Each geodesic is uniquely
determined by hyperbolic--Euler angles $\vec\rho$ measured at $O$. Therefore, proper time $\tilde\tau$
labels the 3D spaces of synchronized comoving observers.}
\end{minipage}
}
\end{figure}

In Fig.\,6, the main features of a general future--cone manifold $\tilde H^+$ parameterized by proper--time coordinates
$\{\tilde\tau, \vec\rho\,\}$ are sketched. As the matter field collapses under the action of gravitational forces,
the geodesics, as well as the 3D spaces labeled by $\tilde \tau$, become more and more irregular and twisted in
the course of time.

\subsection{Inflated--accelerated Milne spacetime in proper--time coordinates}
\label{simpproptime}
In converting to proper--time coordinates the metric of the accelerated Milne spacetime represented in Fig.\,3,
as described in \S\,\ref{confhyperbcoord}, the squared line--element of a worldline and the metric tensor take
the inflated--accelerated form
\begin{eqnarray}
\label{dtildes2}
&&\hspace{-14mm}d\tilde s^2 = d\tilde\tau^2 -  e^{2\tilde\alpha(\tilde \tau)}\,\tilde c(\tilde\tau)^2\bigl[d\varrho^2 +
\bigl(\sinh\varrho)^2 d\theta^2 + \bigl(\sinh\varrho\,\sin\theta\bigr)^2d\phi^2
\bigr]\,,\\
\label{tildegmunutildex}
&&\hspace{-14mm}\tilde g_{\mu\nu}(\tilde x) = \hbox{diag}\Bigl[1, -e^{2\tilde\alpha(\tilde \tau)}
\tilde c(\tilde \tau)^2, -e^{2\tilde\alpha(\tilde \tau)}\tilde c(\tilde \tau)^2\!\sinh\varrho^2,
-e^{2\tilde\alpha(\tilde \tau)}\tilde c(\tilde \tau)^2\!\sinh\varrho^2\sin\theta^2\Bigr];
\end{eqnarray}
where $\tilde c(\tilde\tau) \equiv c[\tau(\tilde\tau)]$ represents the accelerated--expansion factor and
$e^{\tilde\alpha(\tilde \tau)} \equiv e^{\alpha[\tau(\tilde \tau)]}$ the inflation factor of the fundamental
tensor of the Cartan manifold in proper--time coordinates. Note that the scale factor now acts only
on spatial components.

Accordingly, the relevant differential operators are
\begin{eqnarray}
\label{curvconfderiv}
\hspace{-10mm}\big(\tilde D^\mu \tilde f\big)\tilde D_\mu \tilde f \!&=&\!
\bigl(\partial_{\tilde\tau} \tilde f\bigr)^2\! - \frac{1}
{[\tilde c(\tilde \tau)\,e^{\tilde\alpha(\tilde\tau)}]^2}\vert\vec \nabla_\Omega\tilde f\vert^2;\quad \tilde D_\mu\tilde D_\nu
\tilde f = \tilde\partial_\mu\tilde\partial_\nu \tilde f-
\tilde\Gamma_{\mu\nu}^\rho(\tilde\tau, \vec\rho\,)\,\tilde\partial_\rho \tilde f; \\
\label{curvconfD2}
\hspace{-10mm}\tilde D^2 \tilde f \!&=&\! \partial_{\tilde \tau}^2\tilde f+ 3\Big\{\partial_{\tilde \tau}
\ln\big[\tilde c(\tilde\tau)\,e^{\tilde\alpha(\tilde\tau)}\big]\!\Big\}
\partial_{\tilde \tau} \tilde f - \frac{1}{[\tilde c(\tilde\tau)\,
e^{\tilde\alpha(\tilde\tau)}]^2}\Delta_\Omega \tilde f\,;
\end{eqnarray}
where $\tilde f(\tilde x)$, $\tilde \partial_\mu \tilde f$, $\tilde D_\mu\tilde D_\nu \tilde f$ and
$\tilde\Gamma_{\mu\nu}^\rho$ are defined as in Eqs (\ref{tildeDmuDnuf}), $\big|\vec\nabla_\Omega \tilde f\big|^2$ as
in Eq (\ref{hyperbsquaregrad}) and $\Delta_\Omega \tilde f$  as in Eq (\ref{unitlaplacian}).

This further generalization $\widetilde M^+_0$ will be called the {\em inflated--accelerated Milne spacetime}
$\widetilde M^+$. We expect that it represents the geometry of our universe on the large scale.

As in \S\,\ref{confhyperbcoord}, the Christoffel symbols $\tilde\Gamma_{\mu\nu}^\rho(\tilde\tau,\vec\rho\,)$
in the second of Eqs (\ref{curvconfderiv}) can be obtained from those listed in Eqs (\ref{curvchristlist}) by replacing
$c(\tau)$, $\dot c(\tau)$ and $\ddot c(\tau)$, respectively, with $\tilde a(\tilde \tau)= \tilde c(\tilde \tau)
\,e^{\tilde\alpha(\tilde \tau)}$, $\dot{\tilde a}(\tilde \tau) = \partial_{\tilde\tau}[\tilde c(\tilde \tau)\,
e^{\tilde\alpha(\tilde \tau)}]$ and $\ddot{\tilde a}(\tilde \tau) = \partial^2_{\tilde\tau}[\tilde a(\tilde \tau)
\,e^{\tilde\alpha(\tilde \tau)}]$. In particular, we obtain $\tilde \Gamma_{0i}^j= \delta_i^j\,\partial_{\tilde\tau}
\ln \big[\tilde a(\tilde \tau)\,e^{\tilde\alpha(\tilde \tau)}\big]$; $\tilde \Gamma^0_{11}=\frac{1}{2} \partial_{\tilde\tau}
\big[\tilde a(\tilde \tau)\,e^{\tilde\alpha(\tilde \tau)}\big]^2$; $\tilde \Gamma^0_{22}=\tilde \Gamma^0_{11}(\sinh\varrho)^2$;
$\tilde \Gamma^0_{33}=\tilde \Gamma^0_{22}(\sin\theta)^2$.

Carrying out these replacements in Eqs (\ref{R&Rmunu}), we obtain the following mixed components
of the Ricci tensors in proper--time coordinates
\begin{eqnarray}
\label{tildericcitens}
&&\hspace{-8mm} \tilde R^0_0(\tilde \tau)=-3\frac{\ddot{\tilde a}(\tilde\tau)}{\tilde a(\tilde \tau)};
\quad \tilde R^i_j(\tilde \tau) = -\delta^i_j \Big[\frac{\ddot{\tilde a}(\tilde\tau)}{\tilde a(\tilde \tau)}
+ 2\frac{\dot{\tilde a}(\tilde\tau)^2\!\!-\!1}{\tilde a(\tilde \tau)}\Big];\quad \tilde R^i_0(\tilde \tau) = 0;
\quad (i,j = 1,2,3);\nonumber\\
&&\hspace{-8mm}\tilde R(\tilde \tau)\equiv\tilde R^\mu_\mu(\tilde \tau)= -6\Big\{\frac{\ddot{\tilde a}(\tilde\tau)}
{\tilde a(\tilde \tau)}\!+\!\frac{\dot{\tilde a}(\tilde\tau)^2\!\!-\!1}{\tilde a(\tilde \tau)^2}\Big\};\\
&&\hspace{-8mm}\tilde G^0_0(\tilde \tau) = 3\frac{\dot{\tilde a}(\tilde\tau)^2\!\!-\!1}
{\tilde a(\tilde \tau)^2};\quad\tilde G^i_j(\tilde \tau)= \delta^i_j \Big[2\frac{\ddot{\tilde a}(\tilde\tau)}{\tilde a(\tilde \tau)}\!+\!\frac{\dot{\tilde a}(\tilde\tau)^2\!\!-\!1} {\tilde a(\tilde \tau)^2}\Big],\,\,\mbox{where } \tilde G^\mu_\nu(\tilde \tau)\equiv \tilde R^\mu_\nu(\tilde \tau)\!-\!\frac{\delta^\mu_\nu}{2}\tilde R(\tilde \tau)\nonumber;
\vspace{-4mm}
\end{eqnarray}

Since in the following we prove that $e^{\tilde\alpha(\tilde \tau)}$ has a sigmoid--shaped profile, the deformation
of $M^+_0$ resulting from the combined action of inflation and accelerated expansion has approximately the
shape illustrated in Fig.\,7.
\begin{figure}[!ht]
\mbox{%
\begin{minipage}{.46\textwidth}
\includegraphics[scale=0.35]{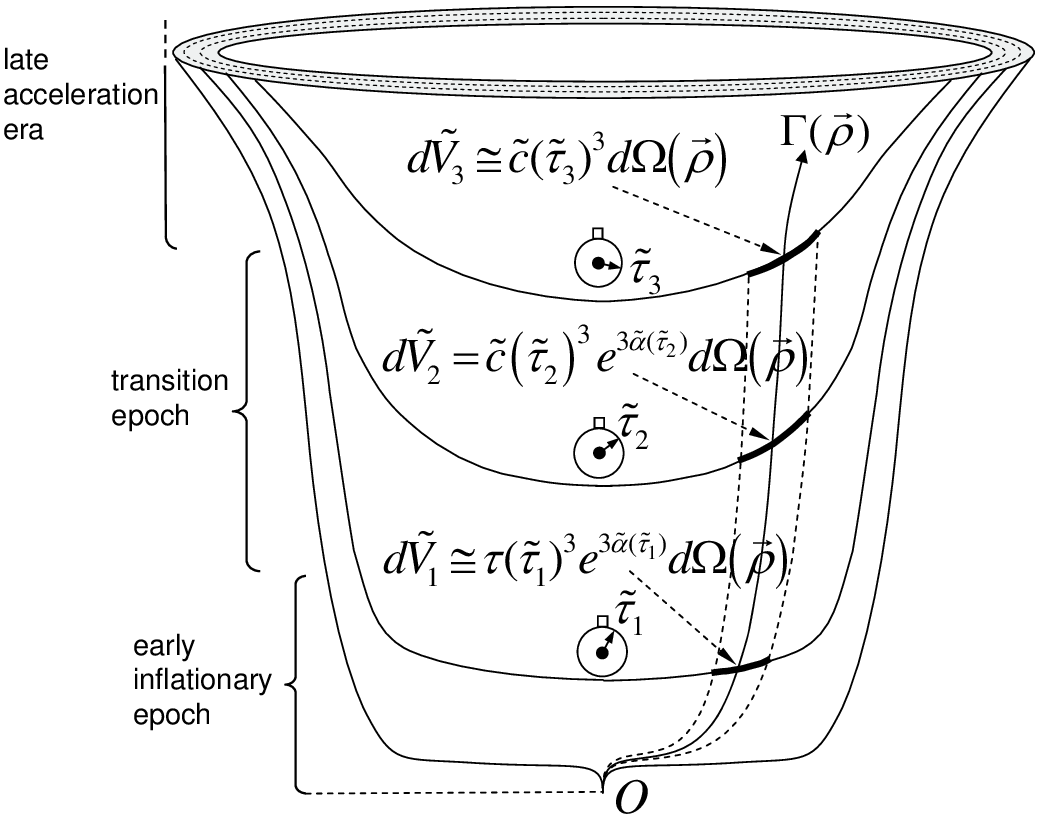}
\end{minipage}%
\quad
\begin{minipage}[c]{.48\textwidth}
\caption{\small Qualitative features of the inflated--accelerated Milne spacetime $\widetilde M^+$
in proper--time coordinates, showing the effects of inflation factor $e^{\tilde\alpha(\tilde \tau)}$
and acceleration factor $\tilde c(\tilde \tau)$. During the early--inflationary and transition epochs,
the deformation is similar to that of Fig.\,3 and, during the accelerated--expansion era, it
is similar to that of Fig.\,5.}
\end{minipage}%
}
\end{figure}

Quantity
\begin{equation}
\vspace{-2mm}
\label{tildeHubblepar}
\hspace{14mm} \tilde H(\tilde \tau)\equiv \partial_{\tilde \tau} \ln \tilde a(\tilde x)=
e^{-\alpha(\tau)}\partial_{\tau} \ln a(\tau)\big|_{\tau =\tau(\tilde\tau)}=\frac{\dot c(\tau)}
{c(\tau)e^{\alpha(\tau)}} + \frac{\dot\alpha(\tau)}{e^{\alpha(\tau)}}\bigg|_{\tau =\tau(\tilde\tau)},
\end{equation}
where equation $\partial_{\tilde\tau} = e^{\alpha(\tau)}\partial_{\tau}$ is used, is
the Hubble parameter of the universe on the large scale in proper-time coordinates  \cite{GORBUNOV}.
Carrying out the analysis of perturbations as in \S\,\ref{curvedhyperbcoord}, and replacing $c(\tau)$ with $\tilde a(\tilde \tau)$
and $\dot f \equiv \partial_\tau f$ with $\dot{\tilde f} \equiv \partial_{\tilde \tau} \tilde f$, we obtain from Eq (\ref{Ricci00A})
\begin{equation}
\label{Ricci00B}
\delta \tilde R_{00} = \delta \tilde R^0_0= \frac{1}{2\,\tilde a^2} \bigg[\nabla^2_\Omega\tilde h_{00} + 3\,\tilde a\dot{\tilde a}\,
\dot{\tilde h}_{00}-2\,\partial^i\dot{\tilde h}_{0i}+\ddot{\tilde h}^i_i -\frac{2\dot{\tilde a}}{\tilde a}\dot{\tilde h}^i_i+
2\bigg(\frac{\ddot{\tilde a}}{\tilde a} - \frac{\dot{\tilde a}^2}{\tilde a^2}\bigg)\tilde h^i_i\bigg].
\end{equation}

Thus, for a static Newtonian perturbation of the form $\tilde h_{\mu\nu}(\tilde\tau, \vec\rho\,) = \mbox{diag}[2\,\tilde a(\tilde\tau)^2
\Phi(\vec\rho\,), 0,0,0]$,  we obtain Poisson equation $\delta \tilde R_{00}(\vec\rho\,) = \nabla^2_\Omega \Phi(\vec\rho\,)$, as at
the end of \S\,\ref{curvedhyperbcoord}.

\section{Brief digression on standard inflationary cosmology}
\label{CGR&standtheor}
The metric of CGR spacetime may be called {\em conical} because the spontaneous breakdown of conformal symmetry
primes the opening of an inflated--accelerated future cone which contains the entire history of the universe.
This future cone is internally foliated by 3D hyperboloids and spanned by the worldlines which stem from the cone origin. The
metrics assumed in standard cosmology may be instead called {\em cylindrical}, as here the entire spacetime is foliated
into a set of 3D--spaces orthogonal to the time axis. Of this type are, for instance, the Robertson--Walker metrics
of general form
\vspace{-2mm}
\begin{equation}
\label{standardRWmetric}
ds^2 = dt^2 - a^2(t)\bigg[ \frac{dr^2}{1- K r^2} + r^2 d\theta^2 +r^2\big(\sin\theta\big)^2 d\phi^2 \bigg]\,.
\vspace{-2mm}
\end{equation}
Here, $r$ is not the radius of a polar--coordinate system but an adimensional parameter, $a(t)$ is a non--negative
scale factor with dimension of length, $\{\theta, \phi\}$ are Euler angles, and $K = 1, 0, -1$, according to whether
the 3D--space curvature is positive, zero or negative \cite{PEACOCK}; thus, we can regard $R(t)=a(t)\,r$ as
the evolving radius of curvature of the 3D--space.

For a homogeneous and isotropic universe described in comoving coordinates, the nonzero components of
matter  EM--tensor $T^\mu_\nu$ and gravitational tensor $G^\mu_\nu = R^\mu_\nu -\frac{1}{2}\,\delta^\mu_\nu R$
are, respectively,
\vspace{-3mm}
$$
T^0_0=\rho_E +\rho_{\hbox{\tiny vac}}, \quad T^i_i = -p +\rho_{\hbox{\tiny vac}}\quad\hbox{and}\quad
G^0_0 = 3\,\frac{\dot a^2+K}{a^2},\quad G^i_i = 2\,\frac{\ddot a}{a}+\frac{\dot a^2+K}{a^2}\,,
\vspace{-3mm}
$$
where $\rho_E>0$ is the matter energy density, $p$ the pressure of the matter field regarded as a homogeneous fluid,
and $\rho_{\hbox{\tiny vac}}$ the cosmological constant as nonnegative vacuum energy--density. Therefore,
Einstein's gravitational equations $G^\mu_\nu = \kappa\,T^\mu_\nu$, where $\kappa$ is the gravitational coupling
constant, takes the form of Friedmann--Lema\^itre equations \cite{FRIEDMANN} \cite{LEMAITRE}
\vspace{-2mm}
\begin{equation}
\label{FLRWeqs}
\frac{\ddot a}{a} = - \frac{\kappa}{6}\,\big(\rho_E  + 3\,p -
2\,\rho_{\hbox{\tiny vac}}\big)\,,\quad \frac{\dot a^2+K}{a^2} =
\frac{\kappa}{3}\,\big(\rho_E +\rho_{\hbox{\tiny vac}}\big)\,\Longrightarrow \frac{\dot a}{a}=-
\frac{\dot\rho_E}{3\big(\rho_E +p\big)}\,.
\vspace{-2mm}
\end{equation}
Since astronomic data corroborated by theoretical arguments \cite{LIDDLE} support the hypothesis that
spatial curvature is zero, we put $K=0$; therefore, Eqs \ref{FLRWeqs} simplify to
\vspace{-2mm}
\begin{equation}
\label{ddota}
\frac{\ddot a}{a} = - \frac{\kappa}{6}\big(\rho_E  + 3\,p - 2\,\rho_{\hbox{\tiny vac}}\big)\,;\quad \frac{\dot
a^2}{a^2} = \frac{\kappa}{3}\big(\rho_E + \rho_{\hbox{\tiny vac}}\big)\,
\Longrightarrow \frac{\dot a}{a}=- \frac{\dot\rho_E}{3\big(\rho_E +p\big)}\,.
\vspace{-2mm}
\end{equation}
Since $\rho_E + \rho_{\hbox{\tiny vac}}>0$, the universe must always be either expanding ($\dot a >0$) or contracting
($\dot a <0$). Astronomic data support the first assumption. Now let us assume $\rho_E  + 3\,p > 2\,\rho_{\hbox{\tiny vac}}$,
so that $\ddot a(t)<0$. Then, going backward in time, we find a time, say, $t=0$, at which $a(0)=0$. In this singular state,
all spacelike volumes shrink to zero, the EM--tensors of matter, $\Theta^M_{\mu\nu} \equiv T_{\mu\nu}$, and of geometry,
$\Theta^G_{\mu\nu}\equiv -\kappa^{-1}G_{\mu\nu}$, converge respectively to $+\infty$ and $-\infty$. These singular limits
are unavoidable as, although related by Einstein's gravitational equation $\Theta^M_{\mu\nu} +\Theta^G_{\mu\nu}=0$,
the two tensors are separately conserved.

This difficulty can be bypassed by dating the birth of the universe at Planck time $t_{Pl}\approx 10^{-43}$s, when matter
consisted of a gas of free relativistic particles in thermal equilibrium at Planck temperature $T_{Pl}\approx 10^{18}$GeV
and all spherical regions of Planck radius $r_{Pl} = c\, t_{Pl} \approx 1.6 \times 10^{-35}$m (where $c$ is the
speed of light) were homogeneous and isotropic. It is reasonable to assume that regions separated
by more than $2\,r_{Pl}$ should evolve independently of each other.

In these extreme initial conditions, the pressure is mostly due to zero--mass and nearly zero--mass particles (electromagnetic
radiation and neutrinos), and is therefore related to the energy density by equation $p = \rho_E/3$. Therefore, the solution to
the third of Eqs (\ref{ddota}) is simply $\rho_E= C/ a^4$, where $C$ is a positive constant.

On the other hand, thermodynamics teaches us that $\rho_E$ is related to temperature $T$ by equation
$\rho_E = (\pi^2 g_*/30)\,T^4$ and to entropy density $s = (\rho_E +p)/T$ by equation $s = (2\pi^2 g_{s*}/45)\,T^3$,
where $g_*$ and $g_{s*}$ are the effective degree--of--freedom degeneracies of $\rho_E$ and $s$, respectively \cite{KOLB};
of note, the same value of $g_{s*}$ is entitled to the entropy density of today's cosmic radiation at temperature
$T_{BK}\simeq 2.726 \,\mbox{$^{\mbox{\tiny o}}$K}$. Hence, we have $a\, T =$ const.

Since during homogeneous and isotropic expansion neither heat nor work can be exchanged between adjacent portions of matter,
the amount of entropy $\Delta S=s(t)\,a(t)^3 \Delta\Omega$ of the matter contained in any expanding volume element
$\Delta V = a(t)^3 \Delta\Omega$, where $\Delta\Omega$ is a comoving solid angle, is conserved. Hence, we have
$a(t_1)/a(t_2) = T(t_2)/T(t_1)$.

We are now in a position to evaluate the size $r_0$ of the Planckian sphere collection at $t =t_{PL}$, which has
evolved to the homogeneous and isotropic domain of the universe that we admire at present time $t_U\!$; a domain
which is at least as large as the present horizon scale $r_U=c\,t_U \ll 10^{28}$ cm. Since $r_0$ is smaller than
$r_U$ by scale--factor ratio $a(t_{PL})/a(t_U)$, we obtain $r_0 = c\,t_U\,T(t_U)/T(t_{Pl}) \equiv r_{Pl}\, Z$,
where $Z =r_0/r_{Pl}= 10^{28}$. Thus, according to this simple description of the universe in decelerated
expansion, the number of causally disconnected regions, from which our universe originated, was about $Z^3 = 10^{84}$.

However, the question still remains of how and why the initial chaotic distribution of zones now described could become as
homogeneous and isotropic as it is today.

The answer provided by standard cosmology is that the decelerated expansion was preceded by a period of accelerated expansion, 
called {\em inflation} \cite{GUTH}, which is possible only if $\rho_E+3p-2\rho_{\hbox{\tiny vac}}\!<\! 0$. According to this view, 
a single Planck sphere at $t=t_{Pl}$, first expanded adiabatically in acceleration up to time $t_c-\!dt$ by a factor of
$a(t_c-dt)/a(t_{Pl}) = T(t_{Pl})/T(t_c\!-dt)$, so as to produce a causally connected domain of radius $r_f =
r_{Pl}\,T(t_{Pl})/T(t_c\!-\!dt)$; then, at a critical time $t_c$, it underwent an explosive phase transition
(the {\em big bang}) which raised the temperature to $T(t_c\!+dt)= Z\,T(t_c\!-\!dt)$, and, afterward, continued
to expand adiabatically in deceleration to $t=t_U$. So, in the end, we obtain $r_0 \approx r_{Pl}$, as desired.

Unfortunately, this {\em ad hoc} conjecture is unable to resolve other stringent problems. In particular, it does not explain
why the cosmic microwave background (CMB) of the celestial sphere appears to be spotted by temperature anisotropies
in the order of $\epsilon = |\delta T/T|\approx10^{-5}$, which, because of entropy conservation, are presumed to
preserve the original pattern imprinted at big bang time $t_f$ until the formation of large--scale structures.

To explain this phenomenology, some cosmologists \cite{CHIBISOV} \cite{LINDE} \cite{LYTH} have postulated the
existence of a sort of super--Higgs field of mass $M\simeq \epsilon\, T_{Pl}\approx 10^{13}$ GeV, i.e., $\approx 10^{11}$
times  larger than the mass $\mu_H\simeq 126$GeV of the Higgs boson detected by LHC experiments, and have ascribed
the CMB anisotropies to primordial quantum fluctuations of this field.

In CGR, there is no need to invoke a super--Higgs field as the factor of inflation and CMB anisotropies.
Here, a ghost scalar field $\sigma(x)$, working as the agent of inflation, and a Higgs field $\varphi(x)$, with mass
parameter proportional to $\sigma(x)$, arise naturally as NG bosons of conformal--symmetry breakdown, making it
possible that, for a suitable interaction and initial conditions, a huge transfer of energy from geometry to matter
occurs. Also, the conformal extensions of $\Theta^M_{\mu\nu}$ and $\Theta^G_{\mu\nu}$ are not separately
conserved and remain finite as the $\varphi$--$\sigma$ interaction promotes the creation of a huge number of
Higgs bosons; this lasts until $\sigma(x)$ reaches its maximum $\sigma_0=\sqrt{6\,\kappa}$ and the mass
of $\varphi(x)$ converges to $\mu_H$. In the conical spacetime of CGR, although all finite volume elements of
3D--hyperboloids shrink to zero as $\tau\rightarrow 0$, the energies of matter and geometry in these volumes remain finite.

In addition, CGR explains the CMB anisotropies as effects of Jeans gravitational instabilities \cite{JEANS},
occurring in the Higgs boson gas at big bang time $t_c$, which were enormously amplified by the Weyl
scale factor of fundamental tensor $\tilde g_{\mu\nu}$ because, as we shall see, at $t_c$, the strength
of the gravitational field is multiplied by a factor of $Z$.

\section{Three equivalent pictures of CGR}
\label{twoequivreps}
The inflationary process created by the spontaneous breakdown of conformal symmetry has been so far described in
three different but equivalent ways:
\begin{itemize}
\item[{\em i})] As a manifestly conformal--invariant field theory including a ghost scalar--field $\sigma(x)$ grounded
on a Riemann manifold $H^+$, the squared--line--element of which has the general form $ds^2 =d\tau^2 -
\gamma_{ij}(x)\, d\rho^i d\rho^{\,j}$, where $x\equiv \{\tau,\vec\rho\,\}$ are the kinematic--time coordinates described
in \S\,\ref{polargeods}. The metric tensor of $H^+$ is then
\begin{equation}
\label{conftimematrix}
\big[\,g_{\mu\nu}(x)\big] \equiv  \left|\begin{array}{cccc}
1 & 0 & 0 & 0\\
0 & -\gamma_{11}(x) & -\gamma_{12}(x) & -\gamma_{13}(x)\\
0 & -\gamma_{12}(x) & -\gamma_{22}(x) & -\gamma_{32}(x)\\
0 & -\gamma_{13}(x) & -\gamma_{23}(x) & -\gamma_{33}(x)
\end{array}\right|.
\end{equation}
\item[{\em ii})] As a non--manifestly conformal--invariant field theory grounded in a conformally connected
Cartan manifold $\hat H^+$, the squared line--element of which has the general form $d\hat s^2 =  e^{2\alpha(x)}ds^2$,
where $d s^2$ is as in {\em i}); $\alpha(x)$ is a function of $x$ determined by the dynamics of the inflationary process,
which starts at $\tau=0$ with a negative value independent of $\vec\rho$ and zero slope, then increases with
$\tau$ and vanishes at $\tau \rightarrow \infty$, as explained in \S\,3.4 of Part I. The fundamental tensor
of $\hat H^+$ is then
\begin{equation}
\label{cartanmatrix} \big[\,\hat g_{\mu\nu}(x)\big] \equiv  \left|\begin{array}{cccc}
e^{2\alpha(x)} & 0 & 0 & 0\\
0 & - e^{2\alpha(x)}\gamma_{11}(x) & - e^{2\alpha(x)}\gamma_{12}(x) & - e^{2\alpha(x)}\gamma_{13}(x)\\
0 & - e^{2\alpha(x)}\gamma_{12}(x) & - e^{2\alpha(x)}\gamma_{22}(x) & - e^{2\alpha(x)}\gamma_{32}(x)\\
0 & - e^{2\alpha(x)}\gamma_{13}(x) & - e^{2\alpha(x)}\gamma_{23}(x) & - e^{2\alpha(x)}\gamma_{33}(x)
\end{array}\right|.
\end{equation}
\item[{\em iii})] As a non--manifestly conformal--invariant field theory grounded in Riemann manifold $\tilde H^+$ obtained by
re--parameterizing {\em ii}) in the form a conical Robertson--Walker metric in proper--time coordinates, any squared
line--element of which therefore takes the general form $d\tilde s^2 =  d\tilde \tau^2 - e^{2\tilde \alpha(\tilde x)}\tilde \gamma_{ij}(\tilde x) d\rho^i
d\rho^{\,j}$, where $\tilde\tau=\int_0^\tau e^{\alpha(\bar\tau, \vec\rho\,)}d\bar\tau$, $\tilde x \equiv \{\tilde \tau, \vec\rho\,\}$,
$\tilde \alpha(\tilde x) \equiv \alpha[x(\tilde x)]$ and $\tilde \gamma_{ij}(\tilde x) \equiv \gamma_{ij}[x(\tilde
x)]$. The metric tensor of $\tilde H^+$ is then
\begin{equation}
\label{propertimematrix} \big[\,\tilde g_{\mu\nu}(\tilde x)\big] \equiv \left|\begin{array}{cccc}
1 & 0 & 0 & 0\\
0 & - e^{2\tilde\alpha(\tilde x)}\tilde\gamma_{11}(\tilde x)& - e^{2\tilde\alpha(\tilde x)}\tilde\gamma_{12}(\tilde
x) & - e^{2\tilde\alpha(\tilde x)}\tilde \gamma_{13}(\tilde x)\\
0 & - e^{2\tilde \alpha(\tilde x)}\gamma_{12}(\tilde x) & - e^{2\alpha(\tilde x)}\tilde\gamma_{22}(\tilde x)
& - e^{2\tilde\alpha(\tilde x)}\tilde\gamma_{32}(\tilde x)\\
0 & - e^{2\tilde\alpha(\tilde x)}\tilde \gamma_{13}(\tilde x)& - e^{2\tilde \alpha(\tilde x)}\tilde\gamma_{23}(x) & -
e^{2\tilde\alpha(\tilde x)}\tilde \gamma_{33}(\tilde x)
\end{array}\right|.
\end{equation}
\end{itemize}

The first way will be called here the {\em kinematic--time picture}, as it is the analog of the kinematic--time
representation used by Brout {\em et.al.} (1979) in their theory of the causal universe; the second way, which
is characteristic of the Cartan manifold representation, will be called the {\em conformal--time picture}, as
it is the analog of the conformal--time representation of standard cosmology; the third way is called the
{\em proper--time picture}, as it is the analog of the proper--time representation of standard cosmology.

The kinematic--time picture represents the universe as a conformal--symmetric system, in a state of
spontaneously broken conformal symmetry, as may be described by comoving synchronized observers living
in the post--inflation era. Looking back to the past, these observers interpret all the events occurring
during the inflationary epoch as subject to the action of the dilation field $\sigma(x)\equiv \sqrt{6/\kappa}
\, e^{\alpha(x)}$, where $\kappa$ is the gravitational coupling constant. In particular, all quantities
of dimension $n$, including $\kappa$, which has $n=2$, are imagined to undergo scale changes of
magnitude proportional to $e^{n\alpha(x)}$. Possible anisotropies of this process are explained as metric distortions
caused by the gravitational field. If this is negligible, or can be regarded as a slight perturbation of the metric,
the optimal description is obtained in terms of the kinematic--time coordinates described in \S\,\ref{curvedhyperbcoord}.
Otherwise, the spatial components of the metric tensor must be replaced with others, $\gamma_{ij}(x)$, depending
on the gravitational field, as described in point {\em i}).

The conformal--time picture is a variant of the kinematic--time picture, in which the action of $\sigma(x)$
takes the form of a geometric effect of the conformally connected Cartan manifold. Since in this
case $\sigma(x)$ is replaced by $\sqrt{6/\kappa}$, the underlying conformal symmetry is not manifest,
so that the action integral assumes the appearance of an Einstein action integral.

The proper--time picture describes the universe as it might have been seen during the
inflationary epoch, and seen still today, by comoving and coeval observers equipped with co--scaling
rulers but non--co--scaling synchronized clocks. With respect to these observers, all physical
quantities preserve their size but, during the inflationary epoch, their time scale becomes highly
compressed. If the gravitational field is negligible or small, the best description is obtained
in terms of the proper--time coordinates introduced in \S\,\ref{simpproptime}. Otherwise, the spatial
components of the fundamental tensor must be replaced with $\hat\gamma_{ij}(\hat x)$, as described
in point {\em ii}).

These pictures fully express the concept of relativity with respect to scale changes. We can move from one
to another by replacing an action integral $A$, grounded in manifold $H^+$, with its ``hat" counterpart $\hat A$,
grounded in manifold $\hat H^+$, or with its ``tilde" counterpart $\tilde A$, grounded in manifold $\tilde H^+$.
As we are about to prove, these action integrals are functionally equivalent because they differ at most by a
harmless surface term. In the post--inflationary era, as $\alpha(x)\rightarrow 0$, all of them converge to the gravitational
action integral of Einstein.

To prove this, let us consider the relation between the conformal--time picture and the proper--time picture with regard to
conformal invariance. Let
\begin{equation}
\label{basicaction} A= \int_{H^+}\!\! \sqrt{-g(x)}\,L_{\sigma}(x)\,d^4x
\end{equation}
be the conformal-invariant action--integral of CGR grounded in the Riemann manifold ${\cal M}_R$ equipped
with the metric tensor $g_{\mu\nu}(x)$ described by Eq (\ref{conftimematrix}). The Lagrangian density
$L_{\sigma}(x)$ depends explicitly on the ghost scalar--field $\sigma(x)$ and includes the
gravitational part in the form of the conformal--invariant term $-\sigma^2(x)\, R(x)/12$, as described
in \S\,3.4 of Part I. As discussed in \S\S~2.3 and 2.4 of Part I, a Weyl transformation with scale factor
$e^{\alpha(x)}=\sigma(x)/\sigma_0$ acts on the local quantities of the theory as follows:
\begin{eqnarray}
& & g_{\mu\nu}(x) \rightarrow \hat g_{\mu\nu}(x) = e^{2\alpha(x)}g_{\mu\nu}(x)\,;\quad g^{\mu\nu}(x)\rightarrow \hat
g^{\mu\nu}(x)= e^{-2\alpha(x)}g^{\mu\nu}(x);\nonumber\\
& & \sqrt{-g(x)} \rightarrow \sqrt{-\hat g(x)} = e^{4\alpha(x)} \sqrt{-g(x)};\quad Q_n(x) \rightarrow \hat Q_n(x)=
e^{n\alpha(x)}Q_n(x);\nonumber\\
& &\varphi(x)\rightarrow \hat \varphi(x)= e^{-\alpha(x)}\varphi(x);\quad \sigma(x)\rightarrow \hat \sigma(x)=
e^{-\alpha(x)}\sigma(x) \equiv \sigma_0;\nonumber\\
& & D_\mu \rightarrow \hat D_\mu\,;\quad  \Gamma^\lambda_{\mu\nu}(x) \rightarrow \hat\Gamma^\lambda_{\mu\nu}(x)\,;
\quad R_{\mu\nu}(x) \rightarrow \hat R_{\mu\nu}(x); \quad R(x) \rightarrow \hat R(x)\,; \nonumber
\end{eqnarray}
where $Q_n(x)$ is any local quantity of dimension $n$, $\sigma(x)$ is the ghost scalar field, $\varphi(x)$ the
Higgs--boson field of mass proportional to $\sigma(x)$, $D_\mu$ are respectively the covariant derivatives and
$\Gamma^\lambda_{\mu\nu}(x)$ the Christoffel symbols derived from the metric tensor (\ref{conftimematrix}) of $H^+$,
$R_{\mu\nu}(x)$ are the Ricci tensor and $R(x)$ the Ricci scalar. The relations between the last three
quantities of the above transformations and their respective "hat" counterparts are listed in Eqs (A-15), (A-23), (A-24) of the Appendix to Part I.

Correspondingly, $A$ is transformed to a functionally equivalent action integral
\begin{equation}
\label{hataction}
\hat A=\int_{\hat H^+}\!\!\sqrt{-\hat g(x)}\,\hat L_{\sigma_0}(x)\, d^4 x\,,
\end{equation}
where $\hat L_{\sigma_0}(x)$ is the total Lagrangian density grounded in the Cartan manifold $\hat H^+$ equipped
with the fundamental tensor (\ref{cartanmatrix}). Since under the Weyl--transformation $\sigma(x)$ becomes $\sigma_0$,
all terms of $L_{\sigma}(x)$ depending on $\partial_\mu\sigma$, in particular, the negative kinetic--energy term
$-g^{\mu\nu}(\partial_\mu\sigma) (\partial_\nu\sigma)/2$, goes to zero and the gravitational term $-\sigma^2(x)\, R(x)/12$
goes  to $-\sigma^2_0\, \hat R(x)/12\equiv -\hat R(x)/2\kappa$, so that $\hat L_{\sigma_0}(x)$
becomes formally similar, but not substantially similar, to a Lagrangian density of GR.

To prove the functional equivalence of action integrals (\ref{basicaction}) and (\ref{hataction}), consider that
the conformal Ricci scalar $\hat R(x)$ appearing in $\hat L_{\sigma_0}(x)$
is related to $R(x)$ and $e^{\alpha(x)}\equiv \sigma(x)/\sigma_0$ by Eq (A-24) of Part I, which we rewrite
in the form
\begin{equation}
\label{tildeR2R}
\hat R(x)  =  e^{-2\alpha(x)}\bigl[R(x) - 6\,\sigma(x)^{-1}D^2 \sigma(x)\bigr]\equiv
\frac{e^{-4\alpha(x)}}{\sigma_0^2}\big[\sigma(x)^2 R(x) -6\,\sigma(x) D^2 \sigma(x)\big],
\end{equation}
where $D^2$ is the covariant Beltrami--d'Alembert operator associated with metric tensor
(\ref{conftimematrix}). The importance of Eq (\ref{tildeR2R}) becomes evident when we go back
from $\hat A$ to $A$, in which case $\sigma_0$ must be replaced by $\sigma(x)$,
$\sqrt{-\hat g(x)}$ by $e^{4\alpha(x)}\sqrt{-g(x)}$ and therefore $\sqrt{-\hat g(x)}\,\sigma^2_0 \hat R(x)/12$
by $\sqrt{-g(x)}\,\sigma^2(x) R(x)/12 - \sqrt{-g(x)}\,\sigma(x) D^2 \sigma(x)/2$. Now, by virtue of
the well--known "transparency" properties of covariant derivatives with respect
to arbitrary functions of $g_{\mu\nu}(x)$, described by Eqs (A-5) and (A-6) of Part I, we have
\begin{equation}
\label{sigd2sig}
\sigma(x) D^2 \sigma(x) \equiv D_\mu\big[g^{\mu\nu}(x)\,\sigma(x)\,\partial_\nu\sigma(x)\big] -
g^{\mu\nu}(x)\bigl[\partial_\mu\sigma(x)]\,\partial_\nu\sigma(x)\,.
\end{equation}
Therefore, the first term on the right--hand side of Eq (\ref{sigd2sig}),
being a surface term, can be removed from the action integral and we are left with only
$-g^{\mu\nu}\bigl(\partial_\mu\sigma)\,\partial_\nu\sigma/2$, which is exactly the negative
kinetic--energy term of $\sigma(x)$ contained in $L_{\sigma}(x)$.

Lastly, to pass from the conformal--time picture to the proper--time picture, we only need to express the
conformal--time coordinates $x$ of the Cartan manifold $\hat H^+$, of fundamental tensor (\ref{cartanmatrix}),
as functions of the proper--time coordinates $\tilde x$ of the polar Robertson--Walker manifold
$\tilde H^+$, of metric tensor (\ref{propertimematrix}), so that $\hat A$ is transformed to
action integral,
$$
\tilde A= \int_{\tilde H^+}\!\!\sqrt{-\tilde g(\tilde x)}\,\tilde L_{\sigma_0}(\tilde x)\, d^4\tilde x\,,
$$
where $\tilde L_{\sigma_0}(\tilde x)$ is the Lagrangian density as a function of "tilde" local quantities,
grounded in $\tilde H^+$. All these results condense into the equivalence relationships
\begin{equation}
\label{the3A}
A\sim \hat A\,,\quad \hat A = \tilde A\,,\quad A\sim \tilde A\,.
\end{equation}

Note that $A$ and $\tilde A$, although grounded in different Riemann manifolds, look very different
as regards conformal invariance. In fact, although the conformal invariance of $A$ is manifest in the
absence of dimensional constants and the vanishing of the EM tensor trace, instead, the conformal invariance
of $\tilde A$ is  obscured by elimination of the surface terms in the transition $A\rightarrow \hat A$
and the presence of the dimensional terms containing $\sigma_0$; in particular, of the term $-\sigma^2_0\,
\tilde R(\tilde x)/12\equiv -\tilde R(\tilde x)/2\,\kappa$, which is formally equal to the Einstein--Hilbert
Lagrangian density of the gravitational field in GR.

As noted in \S\S~\ref{Fubini1} and \ref{Fubini2}, this explicit lack of conformal symmetry, characteristic
of the Cartan and proper--time pictures, can be interpreted as the effect of the spontaneous breakdown
of local conformal symmetry consequent on the fact that, in the conformal--time picture, the vacuum
expectation value of $\sigma(x)$ is a positive function of $x$.

As explained in \S\,\ref{I0Actinv}, this spontaneous breakdown confines the geometry to
the interior of a Milne spacetime. We can therefore restate the action--integral equivalence in the form
$$
A\equiv \int_{M^+} \!\!\!\!\sqrt{-g(x)}\,L_{\sigma(x)}(x)d^4x \sim
\hat A\equiv\int_{\widehat M^+} \!\! \!\!\sqrt{-\hat g(x)}\,\hat L_{\sigma_0}(x) d^4 x=
\tilde A\equiv\int_{\widetilde M^+} \!\! \!\!\sqrt{-\tilde g(\tilde x)}\,\tilde L_{\sigma_0}
(\tilde x)d^4\tilde x,
$$
where $M^+$, $\widehat M^+$  and $\widetilde M^+$ stand for the generalized Milne spacetimes,
respectively in kinematic--time, Cartan and proper--time pictures.

Since we presume that matter distribution remains homogeneous and isotropic during the
inflationary epoch, we infer that, in these conditions, the spacetime curvature $R$ of
the Milne spacetime $M^+$ is constant. If $R=0$, the spacetime--volume elements of
Milne spacetimes $M^+$, $\widehat M^+$ and $\widetilde M^+$, respectively, are
\begin{eqnarray}
\label{Rvolelement}
& & \sqrt{-g(\tau,\vec\rho\,)}\,d^4x = dV(\tau, \vec\rho\,)\,d\tau \equiv \tau^3\,d\Omega(\vec\rho\,)\,d\tau\,;\\
\label{Cvolelement}
& &\sqrt{-\hat g(\tau,\vec\rho\,)}\,d^4 x = d\hat V(\tau, \vec\rho\,)
\,d\tau \equiv e^{4\alpha(\tau)}\tau^3\,d\Omega(\vec\rho\,)\,d\tau\,;\\
\label{Propervolelement}
& &\sqrt{-\tilde g(\tilde \tau,\vec\rho\,)}\,d^4\tilde x = d\tilde V(\tilde\tau, \vec\rho\,)
\,d\tilde\tau \equiv e^{3\tilde\alpha(\tilde \tau)} \tau(\tilde \tau)^3\,d\Omega(\vec\rho\,)\,d\tilde\tau\,;
\end{eqnarray}
where $d\Omega(\vec\rho\,)= (\sinh\varrho)^2\sin\theta\,d\varrho\,d\theta\,d\phi$,
as described by Eqs (\ref{dV1&dVtau}), and $\tilde \tau(\tau) = \int_0^{\bar\tau} e^{\alpha(\bar\tau)}d\bar\tau$,
as described by Eq (\ref{tautotildetau}). Otherwise, we have
\begin{eqnarray}
\label{Rvolelement2}
& & \sqrt{-g(\tau,\vec\rho\,)}\,d^4x = dV(\tau, \vec\rho\,)d\tau \equiv c(\tau)^3\,d\Omega(\vec\rho\,)\,d\tau\,;\\
\label{Cvolelement2}
& &\sqrt{-\hat g(\tau,\vec\rho\,)}\,d^4 x = d\hat V(\tau, \vec\rho\,)
\,d\tau \equiv  c(\tau)^3 e^{4\alpha(\tau)}\,d\Omega(\vec\rho\,)\,d\tau\,;\\
\label{Propervolelement2}
& &\sqrt{-\tilde g(\tilde \tau,\vec\rho\,)}\,d^4\tilde x = d\tilde V(\tilde\tau, \vec\rho\,)
\,d\tilde\tau \equiv  \tilde c(\tilde \tau)^3e^{3\tilde\alpha(\tilde \tau)}\,d\Omega(\vec\rho\,)\,d\tilde\tau\,;
\end{eqnarray}
where $c(\tau) = \sqrt{2\,\tau_\Lambda^2\,(\cosh\tau/\tau_\Lambda - 1)}$ and $\tilde c(\tilde\tau)=
c[\tau(\tilde\tau)]$, with $\Lambda$ the effective cosmological constant, are the
accelerated--expansion factors respectively described in \S\,\ref{curvedhyperbcoord} and \S~\ref{simpproptime}.

\section{The Higgs field in CGR}
In this section we introduce the Lagrangian formalism which is necessary to describe in three different
but equivalent ways the basic mechanism of spacetime inflation and matter generation: first, the
interaction of a massless ghost scalar field $\sigma$ with a physical massless scalar field $\varphi$,
which thereby becomes a Higgs field of dynamically varying mass. This simple system is capable
of representing the process of cosmic inflation as a huge transfer of energy from geometry to matter.

\subsection{The Higgs field in the kinematic--time picture}
\label{HiggsInR}
As anticipated in \S\,6 of Part I, the conformal--invariant interaction of a Higgs field
$\varphi$ with the dilation field $\sigma$ is described by the action integral $A = A^{M} + A^{G}$, where
\begin{eqnarray}
\label{AM0}
&&\hspace{-16mm} A^{M}\! = \!\!\int_{H^+}\!\!\!\!\sqrt{-g}\, L^M(x) d^4 x \! = \!\!\int_{H^+}\!\!\!\!\frac{\sqrt{-g}}{2}\bigg[g^{\mu\nu}\!\big(\partial_\mu\varphi\big)
\partial_\nu\varphi \!- \!\frac{\lambda}{2} \bigg(\varphi^2 \!-\!\frac{\mu^2}{\lambda}
\,\frac{\sigma^2}{\sigma_0^2} \bigg)^2\!+\!\frac{R}{6}\,\varphi^2\bigg]d^4x,\\
\label{AG0}
&&\hspace{-16mm} A^{G} = \!\!\int_{H^+}\!\!\!\!\sqrt{-g}\, L^G(x) d^4 x\! = - \!\!\int_{H^+}\!\!\!\!\frac{\sqrt{-g}}{2}
\bigg[g^{\mu\nu}\big(\partial_\mu\sigma\big)\partial_\nu\sigma +\frac{R}{6}\,\sigma^2 \bigg]d^4x
\end{eqnarray}
are the action integrals, respectively of matter and geometry, over a future cone $H^+$
parameterized by the hyperbolic polar coordinates $x=\{\tau, \vec\rho\,\}$
described in \S~\ref{polargeods}; $L^M(x)$ and $L^G(x)$ are the Lagrangian densities of
$A^{M}$ and $A^{G}$; $g^{\mu\nu}(x)$ is the metric tensor described by Eq (\ref{propertimematrix});
$R(x)$ is the Ricci scalar accounting for spacetime curvature and gravitational forces;
$\mu_H = \mu/\sqrt{2}\cong 126$ GeV is the Higgs--boson mass in the post--inflation era,
when $\sigma(x)$ converges to $\sigma_0$ (cf. \S\,6 of Part I); $\sigma_0 =\sqrt{6/\kappa} =
5.9654\times 10^{18}$ GeV, where $\kappa$ is the gravitational coupling constant; $\lambda$
is the self--coupling constant of $\varphi$ which, according to the Standard Model,
is related to the Fermi coupling constant $G_F \simeq  1.16637\times 10^{-5}$ GeV$^{-2}$
by equation $\lambda =\sqrt{2}\,\mu_H^2 G_F \simeq 0.131$.

From Eqs (\ref{AM0}) and (\ref{AG0}), we derive the motion equations for $\varphi$ and $\sigma$
\begin{equation}
\label{varphisigmamoeteq}
D^2\varphi + \lambda\,\bigg(\varphi^2 -\frac{\mu^2}{\lambda}\,\frac{\sigma^2}{\sigma_0^2}\bigg)\,\varphi
- \frac{R}{6}\,\varphi=0\,,\quad D^2\sigma +\frac{\mu^2}{\sigma_0^2}\,\bigg(\varphi^2 -\frac{\mu^2}{\lambda}
\,\frac{\sigma^2}{\sigma_0^2}\bigg)\,\sigma -\frac{R}{6}\,\sigma=0\,,
\end{equation}
where
\begin{equation}
\label{Beltropingencoord}
D^2 f\equiv\frac{1}{\sqrt{-g}}\,\partial_\mu\big(\sqrt{-g}\,g^{\mu\nu}\partial_\nu f\big)=
\partial_\mu\big(g^{\mu\nu} \partial_\nu f\big)+  \big(\partial_\mu \ln \sqrt{-g}\big)\,
g^{\mu\nu}\partial_\nu f
\end{equation}
is the Beltrami--d'Alembert operator of metric $g_{\mu\nu}(x)$ acting on a scalar field $f(x)$.

By functional variation of $2L^M(x)$ and $2L^G(x)$ with respect to $g^{\mu\nu}(x)$,  we obtain the Hilbert--Einstein EM
tensors of matter and geometry,  respectively
\begin{eqnarray}\label{tetamunuM}
&&\hspace{-12mm}\Theta^{M}_{\mu\nu}= (\partial_\mu\varphi)\partial_\nu\varphi -
\frac{g_{\mu\nu}}{2} \bigg[(\partial^\rho\varphi)\partial_\rho\varphi
-\frac{\lambda}{2}\bigg(\varphi^2 - \frac{\mu^2}{\lambda}
\frac{\sigma^2}{\sigma_0^2}\bigg)^{\!2} \bigg]+\nonumber \\
& & \frac{1}{6}\,\bigl(g_{\mu\nu}D^2 - D_\mu D_\nu\bigr)\,\varphi^2 + \frac{\varphi^2}{6}G_{\mu\nu}\,,\\
\label{tetamunuG}
&&\hspace{-12mm}\Theta^{G}_{\mu\nu} = -\big(\partial_\mu\sigma\big)\partial_\nu\sigma +
\frac{g_{\mu\nu}}{2}(\partial^\rho\sigma)\partial_\rho\sigma-\frac{1}{6}\big(g_{\mu\nu}D^2-D_\mu
D_\nu\big)\sigma^2 -\frac{\sigma^2}{6} G_{\mu\nu},
\end{eqnarray}
where $G_{\mu\nu}(x) = R_{\mu\nu}(x) - g_{\mu\nu}(x)R(x)/2$ and $D_\mu D_\nu f = D_\mu \partial_\nu f
=\partial_\mu\partial_\nu f-\Gamma^\lambda_{\mu\nu}\partial_\lambda f$, with
$\Gamma^\lambda_{\mu\nu}$ the Christoffel symbols constructed out of $g_{\mu\nu}(x)$. Therefore,
matching \S\,2.1 of Part I, the gravitational equation is
\begin{equation}
\label{complRiemGraveq}
\Theta_{\mu\nu}(x)= \Theta^{M}_{\mu\nu}(x) + \Theta^{G}_{\mu\nu}(x) = 0\,.
\end{equation}

Contracting the indices of this tensor by $g^{\mu\nu}(x)$, exploiting the identity $D^2f^2 =
2\,(\partial^\mu f)\,\partial_\mu f+ 2\,fD^2 f$, then using motion equations (\ref{varphisigmamoeteq}), we obtain the EM--traces
\begin{equation}
\label{thetatraces}
\Theta^{M}  = \frac{\mu^2\sigma^2}{\sigma_0^2}\bigg(\frac{\mu^2}{\lambda}
\frac{\sigma^2}{\sigma_0^2}-\varphi^2\bigg)\,,\quad \Theta^{G} = -\frac{\mu^2\sigma^2}{\sigma_0^2} \bigg(\frac{\mu^2}{\lambda}\frac{\sigma^2}{\sigma_0^2}-\varphi^2\bigg)\,,
\end{equation}
as well as the trace equation $\Theta = \Theta^M + \Theta^G =0$.

Defining the mixed energy--momentum tensor of $\varphi$ and $\sigma$ alone as
\begin{eqnarray}
\label{barthetamunu}
& & \bar\Theta^\mu_\nu = \big(\partial^\mu\varphi\bigr)\partial_\nu\varphi -
\big(\partial^\mu\sigma\big)\partial_\nu\sigma -\frac{\delta^\mu_\nu}{2}
\Big[(\partial^\rho\varphi\big)\partial_\rho\varphi
-(\partial^\rho\sigma\bigr)\partial_\rho\sigma\Big]+ \nonumber\\
& &  \frac{\lambda\,\delta^\mu_\nu}{4}\bigg(\varphi^2 - \frac{\mu^2}{\lambda}\,
\frac{\sigma^2}{\sigma_0^2}\bigg)^2 + \frac{1}{6}\,\bigl(D^2 - D^\mu D_\nu\bigr)
\big(\varphi^2-\sigma^2\big)\,,
\end{eqnarray}
we can rewrite the gravitational equation as
\begin{equation}
\label{shortgraveq}
G_{\mu\nu}(x)  = \frac{6\,\bar\Theta_{\mu\nu}(x)}{\sigma^2(x)-\varphi^2(x)}\,.
\end{equation}
Then, exploiting the second Bianchi identity  $D_\mu G^\mu_\nu(x)=0$, we obtain
\begin{equation}
\label{Labelle}
\big[\sigma(x)^2-\varphi(x)^2\big]D_\mu \frac{\bar\Theta^\mu_\nu(x)}{\sigma(x)^2-\varphi(x)^2} =
D_\mu\bar\Theta^\mu_\nu(x)-\bar\Theta^\mu_\nu(x)\, \partial_\mu \ln \big[\sigma(x)^2-\varphi(x)^2\big]= 0,
\end{equation}
showing that $\bar\Theta^\mu_\nu(x)$ in not conserved unless $\varphi(x)$ and $\sigma(x)$ are constant
or $\sigma(x)^2=\varphi(x)^2$.

All these equations simplify considerably if we assume that the universe is homogeneous
and isotropic, which means that $\varphi$, $\sigma$ and $R$  depend only on $\tau$.
In this case the metric--tensor is
\begin{equation}
\label{FRWmetmetrictens}
\big[g_{\mu\nu}(\tau, \vec\rho\,)\big] = \mbox{diag}\big[1,  -c(\tau)^2, -c(\tau)^2\bigl(\sinh\varrho)^2,
-c(\tau)^2\bigl(\sinh\varrho\,\sin\theta\bigr)^2\bigr]\,,
\end{equation}
where $c(\tau)$ is a scale factor accounting for the accelerated expansion. As explained in
\S~\ref{curvedhyperbcoord}, $c(\tau)$ must behave as $\tau$ near the future--cone origin.

Hence, we have $\sqrt{-g(\tau, \vec\rho\,)} = c(\tau)^3\bigl(\sinh\varrho\big)^2\!\sin\theta$ and Eqs (\ref{AM0}) (\ref{AG0}) simplify to
\begin{eqnarray}
\label{AMappr}
& & \hspace{-16mm}A^{M} = \!\int_{H^+}\!\!\!\!\sqrt{-g}\,L^M\!(x)\,d^4x =
\Omega\!\!\int_0^{\infty} \frac{c^3}{2}\bigg[\bigl(\partial_\tau\varphi\bigr)^2-\frac{\lambda}{2}\bigg(\varphi^2 -\frac{\mu^2}{\lambda}\frac{\sigma^2}{\sigma_0^2}\bigg)^2+\frac{R}{6}\,\varphi^2\bigg] d\tau,\\
\label{AGappr}
& & \hspace{-16mm}A^{G} = \!\int_{H^+}\!\!\!\!\sqrt{-g}\,L^M\!(x)\,d^4x =
- \Omega\!\!\int_0^{\infty}  \frac{c^3}{2}\bigg[\bigl(\partial_\tau\sigma\bigr)^2 +\frac{R}{6}\,\sigma^2 \bigg] d\tau,
\end{eqnarray}
where $H^+$ indicates the Riemann manifold of the future cone and $\Omega$ the infinite volume of the hyperbolic--Euler--angle space
represented in Fig.\,4.

From these action integrals, we derive the reduced motion equations of $\varphi(\tau)$ and $\sigma(\tau)$, which
are like those of Eqs (\ref{varphisigmamoeteq}) but with $D^2 f(\tau) =\partial^2_{\tau} f(\tau) +
3\,\big[\partial_\tau\ln c(\tau)\big]\,\partial_\tau f(\tau)$.
Using equations $D_\mu\partial_\tau f(\tau)= \partial_\mu \partial_\tau f(\tau)- \Gamma_{\mu\tau}^\tau(x)\,
\partial_\tau f(\tau)= \delta_{\mu\tau}\partial_\tau^2 f(\tau)$, $D_\mu f(\tau)= \delta_{\mu\tau}\partial_\tau f(\tau)$ and
\begin{equation}
\label{simplecovdiv}
D_\mu \big[a(\tau)\,\partial^\mu f(\tau)\big] = a(\tau)\,D^2 f(\tau) + \big[\partial_\mu a(\tau)\big]\partial^\mu f(\tau)\,,
\end{equation}
where $a(\tau)$ is any scalar function of $\tau$, we obtain the reduced gravitational equations
\begin{eqnarray}
\label{simpleteta00}
&&\hspace{-14mm}\Theta^0_0 = \frac{1}{2}\big(\dot\varphi^2
-\dot\sigma^2\big)+ \frac{\dot c}{c}\big(\varphi\dot\varphi-\sigma\dot\sigma\big)+ \frac{\lambda}{4}\bigg(\varphi^2-\frac{\mu^2}{\lambda}\frac{\sigma^2}{\sigma_0^2}\bigg)^2
 - \frac{\sigma^2-\varphi^2}{6}G^0_0=0;\\
\label{simpletetaii}
&&\hspace{-14mm}\Theta^i_i = -\frac{1}{2}\big(\dot\varphi^2
-\dot\sigma^2\big)+\frac{1}{6}D^2(\varphi^2-\sigma^2)+ \frac{\lambda}{4}\bigg(\varphi^2
- \frac{\mu^2}{\lambda}\frac{\sigma^2}{\sigma_0^2}\bigg)^2  - \frac{\sigma^2-\varphi^2}{6}G^i_i=0.
\end{eqnarray}
Exploiting the identity $D^2 \varphi^2 = 2 \dot\varphi^2 + 2 \varphi D^2 \varphi$ and
a similar identity for $D^2 \sigma^2$, then, using Eqs (\ref{varphisigmamoeteq}),
we find $\Theta^i_i =-\Theta^0_0/3$, showing that Eqs (\ref{simpleteta00}) and (\ref{simpletetaii})
condense into the single gravitational equation
\begin{equation}
\label{singleq}
G^0_0(\tau) =\frac{6\,\rho_E(\tau) }{\sigma(\tau)^2-\varphi(\tau)^2}
\end{equation}
where
\begin{equation}
\label{rhoE}
\rho_E(\tau) = \frac{1}{2}\dot\varphi(\tau)^2 -\frac{1}{2}\dot\sigma(\tau)^2
+\frac{\lambda}{4}\bigg[\varphi(\tau)^2 - \frac{\mu^2}{\lambda}\frac{\sigma(\tau)^2}{\sigma_0^2}\bigg]^2
\end{equation}
can be regarded as the homogeneous energy density of the $\{\varphi(\tau), \sigma(\tau)\}$ system.

Considering that one of Eqs (\ref{R&Rmunu}) gives
\begin{equation}
\label{G00vsdota}
G^0_0(\tau)= 3\,\frac{\dot c(\tau)^2-1}{c(\tau)^2},
\end{equation}
we can easily see that gravitational equation (\ref{singleq}) determines $\dot c(\tau)$ but not $\ddot c(\tau)$.
It is then clear that assuming $\ddot c(\tau)= \big[\dot c(\tau)^2-1\big]/c(\tau)$  does not contradict Eq (\ref{simpleteta00}).
As shown in \S~\ref{curvedhyperbcoord}, this condition, which is exactly equivalent to Eq (\ref{Rmunu&Lambda}),
is fulfilled provided that $G_0^0(\tau) = - R/4 =$ constant.

In summary, gravitational equations (\ref{simpleteta00}) and (\ref{simpletetaii}) condense into equation
\begin{equation}
\label{homgraveqriem}
\Theta^\tau_\tau\equiv \Theta^0_0=\frac{1}{2}\big(\dot\varphi^2- \dot\sigma^2\big) + \frac{\dot c}{c}
\big(\varphi\dot\varphi-\sigma\dot\sigma\big) +\frac{\lambda}{4}
\bigg(\varphi^2-\frac{\mu^2}{\lambda}\frac{\sigma^2}{\sigma_0^2}\bigg)^2 +\frac{\sigma^2-\varphi^2}{24}R =0\,.
\end{equation}

\subsection{The Higgs field in the conformal--time picture}
In the conformal--time picture, the action integral $\hat A=\hat A^{M}+ \hat A^{G}$ of the
Higgs field interacting with the dilation field can be directly obtained from action integral
$A= A^M+A^G$ by carrying out the following replacements in Eqs (\ref{AM0}) and (\ref{AG0}):
\begin{eqnarray}
\label{hatWeyltrans}
&&\hspace{-16mm} g_{\mu\nu}(x)\rightarrow \hat g_{\mu\nu}(x) =
e^{2\alpha(x)}g_{\mu\nu}(x);\quad g^{\mu\nu}(x)\rightarrow \hat g^{\mu\nu}(x) =
e^{-2\alpha(x)}g^{\mu\nu}(x);\nonumber  \\
&&\hspace{-16mm}\sqrt{-g(x)}\rightarrow  \sqrt{-\hat g(x)}=
e^{4\alpha(x)}\sqrt{-g(x)};\quad \varphi(x) \rightarrow \hat \varphi(x)= e^{-\alpha(x)}\varphi(x);\nonumber\\
&&\hspace{-16mm}\sigma(x)\rightarrow \sigma_0;\quad R(x)\rightarrow\hat R(x)=
e^{-2\alpha(x)}\bigl[R(x) - 6\, e^{-\alpha(x)}D^2  e^{\alpha(x)}\bigr];
\end{eqnarray}
where $x=\{\tau, \vec\rho\,\}$ are the hyperbolic polar coordinates described in \S~\ref{polargeods},
$\hat g_{\mu\nu}(x)$ is the fundamental tensor described by Eq (\ref{cartanmatrix}),
$e^{\alpha(x)}$ is the Weyl scale factor introduced in \S~2.2 of Part I and $\hat R(x)$
is the Ricci scalar constructed out of $\hat g_{\mu\nu}(x)$.

We thereby  obtain the following action integrals of matter and geometry
over the future cone $\hat H^+$ of the Cartan manifold
\begin{eqnarray}
\label{hatAM}
\hspace{-12mm}\hat A^{M}\!\!\! &=&\!\!\!\int_{\hat H^+}\!\!\!\sqrt{-\hat g}\,\hat L^Md^4x
=\int_{\hat H^+}\!\!\!\frac{\sqrt{-\hat g}}{2} \bigg[\hat g^{\mu\nu}
\big(\partial_\mu\hat\varphi\big) \partial_\nu\hat\varphi
-\frac{\lambda}{2}\bigg(\hat \varphi^2 - \frac{\mu^2}{\lambda}\bigg)^2
+\frac{\hat R}{6}\,\hat\varphi^2\bigg]d^4 x,\\
\label{hatAG}
\hspace{-12mm}\hat A^{G} \!\!\!&=&\!\!\!\int_{\hat H^+}\!\!\!\sqrt{-\hat g}\,\hat L^G d^4x =
-\int_{\hat H^+}\!\!\!\frac{\sqrt{-\hat g}}{12}\hat R \,\sigma_0^2
\,d^4 x \equiv -\frac{1}{\kappa}\int_{\hat H^+}\!\!\!\frac{\sqrt{-\hat g}}{2}\,\hat R
\,d^4 x\,,
\end{eqnarray}
where $\hat L^M(x)$ and $\hat L^G(x)$ are their respective Lagrangian densities. Contrary to $A$,
the minimum of potential--energy density falls at $\hat\varphi= \mu/\sqrt{\lambda}$,
but, in compensation, the conformal symmetry appears to be explicitly broken by the dimensional constants $\kappa$ and $\mu$.

Action integrals $\hat A$ and $A$, although different in measure, are functionally equivalent, in the sense that they
differ by a surface term. In fact, computing their differences we find
\begin{eqnarray}
&&\hspace{-10mm} \Delta A^M \equiv \hat A^{M} - A^{M}  = \int_{H^+} \frac{\sqrt{-g}}{2}\Big\{\varphi^2\big[g^{\mu\nu}
(\partial_\mu \alpha)\partial_\nu\alpha-e^{-\alpha}D^2 e^{\alpha}\big]-g^{\mu\nu} (\partial_\mu \alpha)
\,\partial_\nu\varphi^2\Big\} d^4 x;\nonumber\\
&&\hspace{-10mm}\Delta A^G \equiv\hat A^{G} - A^{G}  = -\int_{H^+} \frac{\sqrt{-g}}{2}\Big\{\sigma^2\big[g^{\mu\nu}
(\partial_\mu \alpha)\partial_\nu\alpha-e^{-\alpha}D^2 e^{\alpha}\big]-g^{\mu\nu} (\partial_\mu \alpha)\,
\partial_\nu\sigma^2\Big\} d^4 x;\nonumber
\end{eqnarray}
in which the expressions in curly brackets are pure covariant divergences, as the following identities clearly show
\begin{eqnarray}
\label{surftermvarphi2}
D_\mu\big(g^{\mu\nu}\varphi^2e^{-\alpha}\partial_\mu e^\alpha\big)&\equiv & \varphi^2 \big[e^{-\alpha}D^2 e^\alpha
- g^{\mu\nu}(\partial_\mu \alpha)\partial_\nu \alpha\big]+g^{\mu\nu}(\partial_\nu \alpha)\partial_\mu\varphi^2;\\
\label{surftermsigma2}
D_\mu\big(g^{\mu\nu}\sigma^2 e^{-\alpha}\partial_\mu e^\alpha\big) & \equiv & \sigma^2\big[ e^{-\alpha}D^2
e^\alpha- g^{\mu\nu}(\partial_\mu \alpha)\partial_\mu \alpha\big]+g^{\mu\nu}(\partial_\nu \alpha)\partial_\mu \sigma^2.
\end{eqnarray}
Remarkably, equations like these only hold in 4D spacetime. Using the covariant--divergence property $\sqrt{-g(x)}D_\mu f^\mu(x) =\partial_\mu \big[\sqrt{-g(x)}f^\mu(x)\big]$ and Eqs (\ref{surftermvarphi2}) (\ref{surftermsigma2}), we obtain
\begin{equation}
\label{hatA-A}
\sqrt{-g}\,\Delta \hat L = \frac{\sqrt{-g}}{2}\,D_\mu\big[g^{\mu\nu}(\varphi^2-\sigma^2)\,e^{-\alpha}\partial_\mu e^\alpha\big]=
\frac{1}{2}\,\partial_\mu \big[\sqrt{-g}\,g^{\mu\nu}(\varphi^2-\sigma^2)\,\sigma^{-1}\!\partial_\nu \sigma\big],
\end{equation}
where $\Delta L = \hat L^M -L^M + \hat L^G - L^G$, showing that $\hat A$ differs from $A$ by a mere surface term.

In passing from $A$ to $\hat A$, the conformal--invariance of the Lagrangian density ceases to be manifest.
The original conformal invariance of $A$ does appear explicitly broken  in $\hat A$ by
the dimensional constants $\sigma^2_0= \sqrt{6}/\kappa$ and $\mu$.

By equating to zero the functional variation of $2 \hat L(x) \equiv 2\big[\hat L^M(x) +\hat L^G(x)\big]$
with respect to $\hat g^{\mu\nu}(x)$,  we obtain the gravitational equation
\begin{eqnarray}
\label{hattetamunu}
\hat\Theta_{\mu\nu} & = &(\partial_\mu\hat \varphi)\,\partial_\nu\hat\varphi -
\frac{\hat g_{\mu\nu}}{2} \bigg[\hat g^{\rho\sigma}(\partial_\rho\hat\varphi)\partial_\sigma\hat\varphi
-\frac{\lambda}{2}\bigg(\hat\varphi^2 - \frac{\mu^2}{\lambda}\bigg)^{\!2}\bigg]+\nonumber \\
& & \frac{1}{6}\,\big(\hat g_{\mu\nu} \hat D^2 -\hat D_\mu \hat D_\nu\bigr)\,\hat\varphi^2
+ \frac{\hat\varphi^2-\sigma_0^2}{6}\,\hat G_{\mu\nu}=0\,
\end{eqnarray}
where $\hat\Theta_{\mu\nu}$ is the total EM--tensor of matter and geometry, and
$\hat G_{\mu\nu}(x)$ is the conformal gravitational. As shown by Eqs (A-23)--(A-25)
of Part I, this tensor satisfies the equality
\begin{eqnarray}
\label{hatgmunu}
\hat G_{\mu\nu}(x) & \equiv &  \hat R_{\mu\nu}(x)- \frac{1}{2}\,\hat g_{\mu\nu}\hat R(x) =  R_{\mu\nu} - \frac{1}{2}\,g_{\mu\nu}R  +
2\,\sigma^{-1}(g_{\mu\nu}D^2-D_\mu\partial_\nu)\,\sigma +\nonumber\\
& & \sigma^{-2}\Bigl[4\,(\partial_\mu\sigma)(\partial_\nu\sigma) -
g_{\mu\nu}(\partial^\rho \sigma)\,\partial_\rho\sigma\Bigl],
\end{eqnarray}
where $\hat R_{\mu\nu}(x)$ and $\hat R(x)$ are the conformal Ricci tensor and Ricci scalar.

Here are the expressions of the differential operators involved in Eq (\ref{hatgmunu})
\begin{eqnarray}
\label{hatcovderiv}
& & \hat D_\mu \hat f  =  \partial_\mu \hat f; \quad \hat D^\mu \hat f =  \partial^\mu \hat f  = e^{2\alpha(x)}\partial_\mu \hat f;\quad
\hat D_\mu \hat f_\nu \equiv \partial_\mu \hat f_\nu - \hat \Gamma_{\mu\nu}^\lambda \hat f_\lambda\,; \\
\label{hatBeltrop}
& & \hat D^2 \hat f = \frac{1}{\sqrt{-\hat g}}\,\partial_\mu\big(\sqrt{-\hat g}\,\hat g^{\mu\nu}\,\partial_\nu \hat f\,\big)=
\frac{e^{-4\alpha}}{\sqrt{-g}}\,\partial_\mu\big(e^{2\alpha}\sqrt{-g}\,g^{\mu\nu}\,\partial_\nu \hat f\,\big)=\nonumber\\
& & \qquad\quad e^{-2\alpha}\big[\partial_\mu (g^{\mu\nu}\partial_\nu \hat f) +
\big(2\,\partial_\mu\alpha+ \partial_\mu\ln\sqrt{-g}\,\big)\partial^\mu \hat f\,\big];\\
& & \hat D_\mu \hat D_\nu \hat f =\hat D_\mu \partial_\nu\hat f  =  \partial_\mu\partial_\nu \hat f - \hat \Gamma_{\mu\nu}^\lambda \partial_\lambda\hat f=
\partial_\mu\partial_\nu \hat f - \Gamma_{\mu\nu}^\lambda\partial_\lambda\hat f  - \nonumber \\
& & \qquad\qquad\,\,\,(\partial_\mu \alpha)\partial_\nu\hat f -(\partial_\nu \alpha)\partial_\mu\hat f
+g_{\mu\nu}(\partial^\lambda\alpha)\partial_\lambda\hat f.
\end{eqnarray}

The first row describes the action of the conformal covariant derivatives $\hat D_\mu$, respectively on a scalar field
$\hat f(x)$ and a vector field $\hat f_\nu(x)$, involving the conformal Christoffel symbols $\hat \Gamma_{\mu\nu}^\lambda(x)$
constructed from $\hat g_{\mu\nu}(x)$. The second row describes the action of the Beltrami--d'Alembert operator
$\hat D^2$ on $\hat f(x)$. The third row describes the action of operator $\hat D_\mu\hat D_\nu$ on a scalar field $\hat f(x)$,
where use has been made of the identity introduced by Eq (A-15) of Part I, namely. $\hat \Gamma_{\mu\nu}^\lambda(x) = \Gamma^\lambda_{\mu\nu}(x) + \delta^\lambda_\nu \partial_\mu \alpha(x) +\delta^\lambda_\mu \partial_\nu \alpha(x) -g_{\mu\nu}(x)\partial^\lambda\alpha(x)$,
where $\delta_\mu^\nu \equiv g_\mu^\nu(x)$ is the Kronecker delta.

From $\hat A$, we also obtain the motion equation
\begin{equation}
\label{hatvarphimoeteq}
\hat D^2\hat\varphi + \lambda\,\bigg(\hat\varphi^2 -\frac{\mu^2}{\lambda}\bigg)\,\hat\varphi - \frac{\hat R}{6}\,\hat\varphi=0\,,
\end{equation}

Of note, it may seem that, in passing from the kinematic--time picture to the conformal--time one, the motion equation
of the dilation field has disappeared. This is not in fact so, as it is already contained in the gravitational--trace
equation $\hat \Theta(x)  = \hat g^{\mu\nu}(x)\hat\Theta_{\mu\nu}(x)=0$. To prove this, let us first consider the trace
equation
\begin{eqnarray}
\label{hatEMtrace}
\hat \Theta & = & -\hat g^{\mu\nu}(\partial_\mu\hat \varphi)\,\partial_\nu\hat\varphi
+\lambda\bigg(\hat\varphi^2 - \frac{\mu^2}{\lambda}\bigg)^2+\frac{1}{2}\hat D^2\,\hat\varphi^2
- \frac{\hat\varphi^2-\sigma_0^2}{6}\,\hat R=\nonumber\\
& & \hat\varphi \hat D^2\hat\varphi +\lambda\bigg(\hat\varphi^2 - \frac{\mu^2}{\lambda}\bigg)\hat \varphi^2
-\frac{\hat R}{6}\,\hat\varphi^2 +\frac{\hat R}{6}\,\sigma_0^2 - \mu^2\bigg(\hat\varphi^2  -
\frac{\mu^2}{\lambda}\bigg)=\nonumber\\
& & \frac{\hat R}{6}\,\sigma_0^2 - \mu^2\bigg(\hat\varphi^2  - \frac{\mu^2}{\lambda}\bigg) =0,
\end{eqnarray}
where we exploited identity  $\hat D^2\,\hat\varphi^2 = 2\,\hat g^{\mu\nu} (\partial_\mu\hat \varphi)
\,\partial_\nu\hat\varphi + 2\,\hat\varphi \hat D^2\hat\varphi$ in the second step and Eq (\ref{hatvarphimoeteq})
in the third step. Using the last of Eqs (\ref{hatWeyltrans}), we obtain
$$
\hat \Theta=\frac{\hat R}{6}\,\sigma_0^2 - \mu^2\bigg(\hat\varphi^2  - \frac{\mu^2}{\lambda}\bigg)
= - e^{-4\alpha}\sigma\bigg[ D^2\sigma +\frac{\mu^2}{\sigma_0^2}\bigg(\varphi^2 -
\frac{\mu^2}{\lambda} \frac{\sigma^2}{\sigma_0^2}\bigg)\sigma -\frac{R}{6}\sigma\bigg] =0\,,
$$
which is exactly equivalent to the second of Eqs (\ref{varphisigmamoeteq}) because $ e^{-4\alpha}\sigma >0$.

We clearly see that the scale--expansion factor, which in the kinematic--time picture has the form of the
ghost scalar field $\sigma(x)=\sigma_0\, e^{\alpha(x)}$, in the conformal--time picture takes
the form of an adimensional degree of freedom $s(x)= e^{\alpha(x)}$ of the conformal gravitational field.

All equations so far considered simplify considerably if we assume that the Higgs field is homogeneous and
the rate of spacetime expansion is uniformly accelerated or constant. This means that $\hat\varphi$ and $\hat R$ only depend
on $\tau$ and that $R$ is zero or a negative constant.

Since in this case the squared line--element is given by Eq (\ref{dtildes2}),
we can write the fundamental tensor of the Cartan manifold and its inverse as
\begin{eqnarray}
& & \hat g_{\mu\nu}(x) = \hbox{diag}\Bigl[s(\tau)^2, - s(\tau)^2 c(\tau)^2,
-s(\tau)^2 c(\tau)^2\!\sinh\varrho^2,-s(\tau)^2 c(\tau)^2\!\sinh\varrho^2 \sin\theta^2\Bigr],\nonumber\\
& & \hat g^{\mu\nu}(x) = \hbox{diag}\bigg[\frac{1}{s(\tau)^2}, \frac{-1}{s(\tau)^2 c(\tau)^2},
\frac{-1}{s(\tau)^2 c(\tau)^2\!\sinh\varrho^2},\frac{-1}{s(\tau)^2 c(\tau)^2\!\sinh\varrho^2 \sin\theta^2}\bigg].\nonumber
\end{eqnarray}
where $x\equiv \{\tau, \vec\rho\,\}$ and $c(\tau)$ is the spatial scale--factor accounting
for accelerated expansion.

Hence, we have $\sqrt{-\hat g(x)}=s(\tau)^4 c(\tau)^3 \sinh\varrho^2\sin\theta$ and Eqs (\ref{hatAM})
(\ref{hatAG}) simplify to
\begin{eqnarray}
\label{hatAMappr}
&&\hspace{-16mm}\hat A^{M}\! = \!\!\int_{\hat H^+}\!\!\!\!\sqrt{-\hat g}\,\hat L^M\!(x)\,d^4x =
\Omega\!\!\int_0^{\infty} \frac{s^4 c^3}{2}\bigg[\bigl(\partial_\tau\hat\varphi\bigr)^2-\frac{\lambda}{2}\bigg(\hat\varphi^2 -\frac{\mu^2}{\lambda}\bigg)^2+\frac{\hat R}{6}\,\hat \varphi^2\bigg] d\tau,\\
\label{hatAGappr}
&&\hspace{-16mm}\hat A^{G}\! = \!\!\int_{\hat H^+}\!\!\!\!\sqrt{-\hat g}\,
\hat L^M\!(x)\,d^4x = - \Omega\!\!\int_0^{\infty}\!\!\!s^4 c^3\frac{\sigma^2_0}{12}\hat R\,d\tau
\equiv -\frac{\Omega}{2\kappa}\!\int_0^{\infty}\!\!\!s^4 c^3\hat R\,d\tau ,
\end{eqnarray}
where $\hat H^+$ indicates the Cartan manifold and $\Omega$ the infinite volume
of the hyperbolic--Euler--angle space (do not quibble about this mathematical license).

Now, the covariant operators $\hat D^2$ and $\hat D_\mu$ act on a scalar function $\hat f(\tau)$, depending only
on $\tau$, as follows: $\hat D_\mu \hat D_\nu \hat f(\tau) = 0\,\,(\mu\neq \nu)$ and
\begin{eqnarray}
\label{simplehatD2}
&&\hspace{-16mm}\hat D^2 \hat f(\tau) = \frac{\partial_\tau\big[s(\tau)^2 c(\tau)^3
\partial_\tau\hat f(\tau)\big]}{s(\tau)^4 c(\tau)^3}=\frac{\partial_\tau^2 \hat f(\tau)}{s(\tau)^2} +
\bigg[\frac{2\,\dot s(\tau)}{s(\tau)^3} +\frac{3\,\dot c(\tau)}{s(\tau)^2c(\tau)}\bigg] \partial_\tau \hat f(\tau)\,;\\
\label{Do2}
&&\hspace{-16mm}\hat D_0\hat D_0 \hat f(\tau) =\partial_\tau^2\hat f(\tau)-
\big[\Gamma^0_{00}(\tau)+ \dot\alpha(\tau)\big]\partial_\tau\hat f(\tau)=\partial_\tau^2\hat f(\tau)- \frac{\dot s(\tau)}{s(\tau)}\partial_\tau\hat f(\tau);\\
\label{D2-Do2}
&&\hspace{-16mm}\big[\hat g_{00}(x)\hat D^2 - \hat D_0 \hat D_0\big]\hat f(\tau) \equiv
\big[s(\tau)^2\hat D^2 - \hat D_\tau \hat D_\tau\big]\hat f(\tau) =
3\bigg[\frac{\dot s(\tau)}{s(\tau)} +\frac{\dot c(\tau)}{c(\tau)}\bigg] \partial_\tau \hat f(\tau)\,;
\end{eqnarray}
as $\hat g_{00}(x)=s(\tau)^2$, $\hat g_{0 i}(x)=0$ and $\Gamma^0_{00}(\tau)=0$, with dot superscripts denoting $\partial_\tau$.

Consequently, the gravitational equations (\ref{hattetamunu}) condense into the single equation
\begin{equation}
\label{simplehattetamunu}
\hat\Theta_{\tau\tau} = \frac{1}{2}(\partial_\tau\hat \varphi)^2 +\frac{s^2\lambda}{4}\bigg(\hat\varphi^2 -
\frac{\mu^2}{\lambda}\bigg)^2\!+\frac{1}{2}\,\bigg(\frac{\dot s}{s} +\frac{\dot c}{c}\bigg)\partial_\tau\hat\varphi^2
+ \frac{\hat\varphi^2-\sigma_0^2}{6}\,\hat G_{\tau\tau}=0.
\end{equation}

Using in this equation the equalities
\begin{equation}
\label{hatreplacement}
\hat\varphi=\frac{\varphi}{s};\quad \partial_\tau\hat\varphi = \frac{\dot\varphi}{s} - \varphi\frac{\dot s}{s^2};\quad
\partial_\tau\hat\varphi^2 = 2\frac{\varphi\dot\varphi}{s^2} - 2\varphi^2\frac{\dot s}{s^3};\quad
\hat G_{\tau\tau}= -\frac{R}{4} + 6\frac{\dot s\dot c}{sc} +3\frac{\dot s^2}{s^2};
\end{equation}
we find
\begin{equation}
\label{hattheta00equiv}
\hat\Theta_{\tau\tau}(\tau) = \frac{\Theta_{\tau\tau}(\tau)}{s(\tau)^2}=0\,;\quad \hat\Theta_\tau^\tau(\tau)
= \hat g^{\tau\tau}(\tau) \,\hat\Theta_{\tau\tau}(\tau) =\frac{\Theta_\tau^\tau(\tau)}{s(\tau)^4}=0\,;
\end{equation}
where $\hat g^{\tau\tau}(\tau)\equiv \hat g^{00}(x)$, clearly showing the equivalence of
Eqs (\ref{homgraveqriem}) and (\ref{simplehattetamunu}), as expected.

\subsection{The Higgs field in the proper--time picture}
In the proper--time picture, the action integral $\tilde A=\tilde A^{M}+ \tilde A^{G}$ of the
Higgs field interacting with the dilation field can be directly obtained from action integral
$\hat A= \hat A^M+\hat A^G$ of the conformal--time picture by means of the following substitutions
\begin{eqnarray}
\label{cart2proptime}
&&\hspace{-16mm}\tau \rightarrow \tilde\tau = \int_0^\tau e^{\alpha(\bar \tau, \vec\rho\,)}d\bar\tau;\quad \partial_\tau
\rightarrow  \partial_{\tilde \tau}\equiv \tilde \partial_0 = e^{-\alpha(x)}\partial_\tau;
\quad \hat\varphi(x) \rightarrow\tilde \varphi(\tilde x)= \hat\varphi[x(\tilde x)]; \nonumber\\
&&\hspace{-16mm}\hat g^{\mu\nu}(x) \rightarrow \tilde g^{\mu\nu}(\tilde x);\quad \sqrt{-\hat g(x)} \rightarrow
\sqrt{-\tilde g(\tilde x)};\quad\hat R(x)\rightarrow \tilde R(\tilde x)= \hat R[x(\tilde x)];
\end{eqnarray}
where $\tilde x= \{\tilde\tau, \vec\rho\,\}$ are the proper--time coordinates defined in \S\,\ref{confhyperbcoord},
$\tilde \partial_\mu$ are partial derivatives with respect to $\tilde x^\mu$, $\tilde g_{\mu\nu}(\tilde x)$ is
the polar metric described by Eq (\ref{conftimematrix}) and $\tilde R(\tilde x)$ is the Ricci scalar constructed from
$\tilde g^{\mu\nu}(\tilde x)$.

We therefore obtain the proper--time action integral of matter and geometry  $\tilde A = \tilde A^{M}+ \tilde A^{G}$, where
\begin{eqnarray}
\label{tildeAM}
\hspace{-8mm}\tilde A^{M} &=&\int_{\tilde H^+}\!\!\!\frac{\sqrt{-\tilde g}}{2} \bigg[\tilde g^{\mu\nu}\bigl(\tilde
\partial_\mu\tilde\varphi\bigr) \tilde \partial_\nu\tilde\varphi
-\frac{\lambda}{2}\bigg(\tilde \varphi^2 - \frac{\mu^2}{\lambda} \bigg)^2
+\frac{\tilde R}{6}\,\tilde\varphi^2\bigg]d^4\tilde x\,;\\
\label{tildeAG}
\hspace{-8mm}\tilde A^{G} &=& -\frac{1}{\kappa}\int_{\tilde H^+}\!\!\!\frac{\sqrt{-\tilde g}}{2}\,\tilde R
\,d^4\tilde x\,,
\end{eqnarray}
where $\tilde H^+$ is the inflated and accelerated future cone embedded in the Riemann manifold
of metric $\tilde g_{\mu\nu}(\tilde x)$ and $\kappa \equiv 6/\sigma_0^2$ is the gravitational
coupling constant of GR. As in the conformal--time picture, the minimum of potential--energy density falls
at $\tilde \varphi= \mu/\sqrt{\lambda}$ and the conformal symmetry appears to be explicitly broken by the
dependence of $\tilde A$ on $\kappa$ and $\mu^2$. The main difference is now that the metric tensor has
the hyperbolic--polar form
$$
\tilde g_{00}(\tilde \tau, \vec\rho\,) = 1;\quad \tilde g_{0i}(\tilde \tau, \vec\rho\,) = 0\,\, (i=1,2,3);\quad
\tilde g_{ij}(\tilde \tau, \vec\rho\,)= -e^{2\tilde\alpha(\tilde\tau, \vec\rho\,)}\tilde\gamma_{ij}(\tilde \tau, \vec\rho\,);
$$
described by Eq (\ref{propertimematrix}), so that the manifold returns to a Riemann structure.

Since $\tilde A$ is obtained from $\hat A$ by a simple redefinition of the conformal time $\tau$, it is
equal in measure to $\hat A$, but functionally equivalent to $A$, as $\tilde A$ inherits the surface term from
$\hat A$.

The motion equation of $\tilde\varphi$ is
\begin{equation}
\label{moteqtildevarphi} \tilde D^2\tilde \varphi + \lambda\bigg(\tilde \varphi^2 -
\frac{\mu^2}{\lambda}\bigg)\tilde\varphi- \frac{\tilde R}{6}\,\tilde\varphi=0\,,
\end{equation}
where
\begin{equation}
\label{tildeD2}
\tilde D^2\!\tilde f(\tilde x)\!=\!\partial_{\tilde \tau}^2\!\tilde f(\tilde x) +\bigg\{\!\partial_{\tilde \tau}
\ln\!\!\Big[\tau(\tilde x)^3\!e^{3\tilde\alpha(\tilde x)}\!\sqrt{\tilde\gamma(\tilde x)}\,\Big]\!\!\bigg\}
\partial_{\tilde \tau}\!\tilde f(\tilde x)-\frac{\partial_i\!\big[e^{2\tilde\alpha(\tilde x)}\!
\sqrt{\tilde \gamma(\tilde x)}\,\tilde\gamma^{ij}(\tilde x)\partial_j \tilde f(\tilde x)\big]}
{\tau^2(\tilde x)\, e^{4\tilde\alpha(\tilde x)}\sqrt{\tilde\gamma(\tilde x)}}
\end{equation}
is the Beltrami--d'Alembert operator constructed from $\tilde g^{\mu\nu}(\tilde x)$.

The EM tensors derived from (\ref{tildeAM}) and (\ref{tildeAM}) are respectively
\begin{eqnarray}
\label{tildeTetamunu}
\hspace{-8mm}\tilde \Theta^M_{\mu\nu} & = & \bigl(\tilde
\partial_\mu\tilde\varphi\bigr)\tilde \partial_\nu\tilde\varphi- \frac{\tilde g_{\mu\nu}}{2}
\tilde g^{\rho\sigma}\bigl(\tilde \partial_\rho\tilde\varphi\bigr)
\tilde \partial_\sigma\tilde\varphi + \tilde
g_{\mu\nu}\frac{\lambda}{4} \bigg(\tilde \varphi^2- \frac{\mu^2}{\lambda}\bigg)^2 +\nonumber\\
\hspace{-8mm}& & \frac{1}{6}\bigl( \tilde g_{\mu\nu}\tilde D^2 - \tilde
D_\mu \tilde \partial_\nu\bigr)\tilde \varphi^2 +
\frac{\tilde\varphi^2}{6}\tilde G_{\mu\nu}\,; \quad \tilde\Theta^G_{\mu\nu} = -\frac{\tilde G_{\mu\nu}}{\kappa} \equiv -\frac{\sigma_0^2}{6}
\tilde G_{\mu\nu}\,;\\
\label{tildeGmumu}
& & \hspace{-16mm}\mbox{where }  \tilde G_{\mu\nu}= \tilde R_{\mu\nu}- \frac{1}{2}\tilde g_{\mu\nu}\tilde R,
\,\,\mbox{which gives }\,\tilde G^0_0 =  -\tilde R\,\, \mbox{and }\, \tilde G^\mu_\mu =  -\tilde R\,.
\end{eqnarray}
from which we obtain the gravitational equation $\tilde\Theta_{\mu\nu}= \tilde\Theta^M_{\mu\nu}+\tilde\Theta^G_{\mu\nu}=0$.
Contracting Eqs (\ref{tildeTetamunu}) with $\tilde g^{\mu\nu}$, and then using identity $\tilde D^2\tilde \varphi^2 \equiv 2\,\tilde g^{\rho\sigma}\bigl(\tilde \partial_\rho\tilde\varphi\bigr) \bigl(\tilde \partial_\sigma\tilde\varphi\bigr) +
2\,\tilde\varphi\tilde D^2\tilde\varphi$ and motion equation (\ref{moteqtildevarphi}), we obtain the EM traces and trace equations
\begin{equation}
\label{tildetraces}
\tilde \Theta^M = \mu^2\bigg(\frac{\mu^2}{\lambda}-\tilde \varphi^2\bigg)\,; \quad
\tilde \Theta^G =  \frac{1}{\kappa}\tilde R\,;\quad \hbox{then } \tilde R =
\kappa\,\mu^2\bigg(\tilde \varphi^2- \frac{\mu^2}{\lambda}\bigg)\,.
\end{equation}
Replacing the expression for $\tilde R$ given by Eq (\ref{tildeR2R}) in the last of Eqs (\ref{tildetraces}), we obtain
$$
\tilde R(\tilde x) = e^{-2\alpha(x)} \big[R(x) - 6\, e^{-\alpha(x)} D^2  e^{\alpha(x)}\big]=
\kappa\,\mu^2\bigg(\tilde \varphi^2- \frac{\mu^2}{\lambda}\bigg).
$$

An interesting point is worth noting here: inserting the last of Eqs (\ref{tildetraces}) into Eq (\ref{moteqtildevarphi}),
we obtain
\begin{equation}
\label{moteqtildevarphi2}
\tilde D^2\tilde \varphi + \bigg(\lambda -  \frac{\mu^2}{\sigma_0^2}\bigg)\bigg(\tilde
\varphi^2 - \frac{\mu^2}{\lambda} \bigg)\tilde\varphi=0\,,
\end{equation}
showing that, in proper--time coordinates, the dependence of the dynamics of $\tilde\varphi$ on
$\tilde R$ simply results in the self--coupling--constant change
\begin{equation}
\label{tildelambda}
\lambda\rightarrow \tilde\lambda = \lambda-\frac{\mu^2}{\sigma_0^2}\,.
\end{equation}
Since $\tilde\lambda/\lambda \simeq 1- 1.7\times 10^{-33}$, the change is absolutely negligible.

All these equations simplify considerably if we assume that the universe is homogeneous
and isotropic, meaning that $\tilde\varphi$ and $\tilde R$ depend only on $\tilde\tau$.
Since in this case the squared line--element is given by Eq (\ref{dtildes2}), the metric tensor
simplifies to that of the inflated--accelerated Milne spacetime $\tilde M^+$ represented in Fig.\,7, i.e..
$$
\tilde g_{\mu\nu}(\tilde \tau, \vec\rho\,) = \hbox{diag}\Bigl[1, - \tilde a(\tilde \tau)^2,
-\tilde a(\tilde \tau)^2\!\sinh\varrho^2,-\tilde a(\tilde \tau)^2\!\sinh\varrho^2 \sin\theta^2\Bigr],
$$
where $\tilde a(\tilde \tau) = \tilde c(\tilde \tau)\, \tilde s(\tilde \tau)$; $\tilde c(\tilde \tau)$
and $\tilde s(\tilde \tau)=e^{\tilde\alpha(\tilde \tau)}$ are obtained respectively from $c(x)$
and $s(\tau)\equiv e^{\alpha(\tau)}$, according to the rule $f(\tau)\rightarrow
\tilde f(\tilde \tau) = f[\tau(\tilde \tau)]$. Therefore, we have
\begin{eqnarray}
\label{tildeterm}
&&\hspace{-8mm}\sqrt{-g(x)} = c(x)^3 (\sinh\varrho)^2\sin\theta \rightarrow
\sqrt{-\tilde g(\tilde x)}= \big[\tilde s(\tilde \tau)\,\tilde c(\tilde \tau)\big]^3(\sinh\varrho)^2\sin\theta\,;\nonumber\\
\label{tildeDVol}
&&\hspace{-8mm}d^4x = d\Omega(\vec\rho)\,d\tau \rightarrow d^4\tilde x= d\Omega(\vec\rho)\,d\tilde\tau\,;\nonumber
\end{eqnarray}
where $d\Omega(\vec\rho)$, as usual, is the volume element  of the hyperbolic--Euler--angle space $\Omega$.

With these changes, the motion equation (\ref{moteqtildevarphi2}) simplifies to
\begin{equation}
\label{simpmoteqtildevarphi} \partial_{\tilde\tau}^2\tilde\varphi(\tilde \tau)+
3\bigg[\frac{\partial_{\tilde\tau}\tilde c(\tau)}{\tilde c(\tau)} + \frac{\partial_{\tilde\tau}\tilde s(\tau)}{\tilde s(\tau)}\bigg]
\partial_{\tilde\tau}\,\tilde\varphi(\tilde \tau)+ \tilde\lambda\bigg(\tilde \varphi^2 - \frac{\mu^2}{\lambda}\bigg)\tilde\varphi=0\,,
\end{equation}
where Eq (\ref{tildelambda}) was used. The expression in squared brackets is the Hubble parameter in proper--time coordinates.

The homogeneous gravitational equation is therefore
\begin{equation}
\label{CartEinst}
\tilde \Theta_0^0 = \frac{1}{2}\bigl(\partial_{\tilde\tau}\tilde\varphi\bigr)^2
+\frac{\lambda}{4} \bigg(\tilde \varphi^2- \frac{\mu^2}{\lambda}\bigg)^2
+\frac{1}{2}\bigg(\frac{\partial_{\tilde\tau}\tilde c}{\tilde c}+ \frac{\partial_{\tilde\tau}\tilde s}
{\tilde s}\bigg)\partial_{\tilde \tau}\tilde\varphi^2- \frac{\sigma_0^2- \tilde\varphi^2}{6}\,\tilde G^0_0=0\,.
\end{equation}
Since $\tilde\varphi^2/\sigma_0^2$ is negligible relative to 1, this equation has the form of
the standard gravitational equation of Einstein for a homogeneous Higgs field.

To prove the equivalence of this equation with Eq (\ref{homgraveqriem}), let us retrieve equation
$$
\tilde G^0_0(\tilde \tau)=3\frac{[\partial_{\tilde\tau}{\tilde a}(\tilde\tau)]^2\!-\!1}
{\tilde a(\tilde\tau)^2}=3\bigg\{\!\frac{[\partial_{\tilde\tau}{\tilde c}(\tilde\tau)]^2}{{\tilde c}(\tilde\tau)^2} +
\frac{[\partial_{\tilde\tau}{\tilde s}(\tilde\tau)]^2}{{\tilde s}(\tilde\tau)^2}+2\frac{[\partial_{\tilde\tau}{\tilde c}
(\tilde\tau)][\partial_{\tilde\tau}{\tilde s}(\tilde\tau)]}{\tilde c(\tilde\tau)\tilde s(\tilde\tau)} -
\frac{1}{[\tilde c(\tilde\tau)\tilde s(\tilde\tau)]^2}\!\bigg\}
$$
from Eqs (\ref{tildericcitens}). Since in going back to the kinematic--time picture $\tilde f(\tilde\tau)$
becomes $f(\tau)$ and $\partial_{\tilde\tau}{\tilde f}(\tilde\tau)$ becomes $\dot f(\tau)/s(\tau)$, then
$\sigma_0^2- \tilde\varphi^2$ becomes $[\sigma(\tau)^2- \varphi(\tau)^2]/s(\tau)^2$
and $\tilde G^0_0(\tilde \tau)/6$ becomes
$$
\frac{1}{2 s(\tau)^2}\bigg[\frac{\dot c(\tau)^2}{c(\tau)^2} +\frac{\dot s(\tau)^2}{s(\tau)^2}+
2\frac{\dot c(\tau)\,\dot s(\tilde\tau)}{c(\tau)\,s(\tau)} -\frac{1}{c(\tau)^2}\bigg] =  \frac{1}{s(\tau)^2}
\bigg[\frac{1}{2}\frac{\dot s(\tau)^2}{s(\tau)^2}+ \frac{\dot c(\tau)}{c(\tau)}\frac{\dot s(\tau)}{s(\tau)}-
\frac{R}{24}\bigg],
$$
where Eq (\ref{G00vsdota}) and the last of Eqs (\ref{Rmunu&Lambda}) were used.

As $\tilde \tau$ becomes increasingly larger, $\tilde \tau$ approaches $\tau$, $\tilde s(\tilde \tau)$
approaches $1$, $\tilde c(\tilde \tau)$ approaches $c(\tau)$, $\tilde R^0_0$ approaches $R^0_0$ and $\tilde R$ 
approaches $R$. Consequently, Eq (\ref{CartEinst}) approaches the standard gravitational equation of a 
homogeneous Higgs field:
$$
\frac{1}{2}\big(\partial_\tau\varphi\big)^2 +\frac{\lambda}{4}\bigg(\varphi^2-\frac{\mu^2}{\lambda}\bigg)^2
+ \frac{1}{2}\frac{\partial_\tau c}{c}\,\partial_\tau\varphi^2= \frac{1}{\kappa}\bigg(R^0_0- \frac{1}{2}R\bigg)=
-\frac{R}{4\,\kappa}\,.
$$

The last step is clearly due to the fact that the mixed Ricci tensor of a homogeneous and isotropic open universe
has the form $R^\mu_\nu = \delta^\mu_\nu\,R/4$, where $R$ is a negative constant.

\end{document}